\input epsf
\documentclass[12pt]{article}
\usepackage{latexsym,amsmath,amssymb,graphicx}
\renewcommand{\thefootnote}{\fnsymbol{footnote}}

\usepackage{amsmath}
\usepackage{amsfonts}

\numberwithin{equation}{section}

\def\doubleset#1#2{\bgroup%
\def\doit#1#2{%
\setbox\dblsetbox=\hbox{$\cstyle #1$}%
\raise#2\ht\dblsetbox\copy\dblsetbox%
\hskip-\wd\dblsetbox%
\raise-#2\ht\dblsetbox\box\dblsetbox}%
\mathchoice%
{\def\cstyle{\displaystyle}\doit#1#2}%
{\def\cstyle{\textstyle}\doit#1#2}%
{\def\cstyle{\scriptstyle}\doit#1#2}%
{\def\cstyle{\scriptscriptstyle}\doit#1#2}\egroup}
\def\underarrow#1{\vbox{\ialign{##\crcr$\hfil\displaystyle
 {#1}\hfil$\crcr\noalign{\kern1pt\nointerlineskip}$\longrightarrow$\crcr}}}

\newbox\dblsetbox

\newlength{\extraspace}
\newlength{\extraspaces}
\newcommand{\be}{\begin{equation}
\addtolength{\abovedisplayskip}{\extraspaces}
\addtolength{\belowdisplayskip}{\extraspaces}
\addtolength{\abovedisplayshortskip}{\extraspace}
\addtolength{\belowdisplayshortskip}{\extraspace}}
\newcommand{\ee}{\end{equation}}

\newcommand{\ba}{\begin{eqnarray}
\addtolength{\abovedisplayskip}{\extraspaces}
\addtolength{\belowdisplayskip}{\extraspaces}
\addtolength{\abovedisplayshortskip}{\extraspace}
\addtolength{\belowdisplayshortskip}{\extraspace}}
\newcommand{\ea}{\end{eqnarray}}

\newcommand{\bd}{\begin{displaymath}
\addtolength{\abovedisplayskip}{\extraspaces}
\addtolength{\belowdisplayskip}{\extraspaces}
\addtolength{\abovedisplayshortskip}{\extraspace}
\addtolength{\belowdisplayshortskip}{\extraspace}}
\newcommand{\ed}{\end{displaymath}}

\newcounter{saveeqn}

\newcommand{\newsection}[1]{
\vspace{12mm}
\pagebreak[3]
\addtocounter{section}{1}
\setcounter{equation}{0}
\setcounter{subsection}{0}
\noindent{\bf \thesection. #1}
\nopagebreak
\medskip
\nopagebreak}

\newcommand{\newsubsection}[1]{
\vspace{0.8cm}
\pagebreak[3]
\addtocounter{subsection}{1}
\noindent{\it \thesubsection. #1}
\nopagebreak
\vspace{2mm}
\nopagebreak}

\begin{document}
\addtolength{\baselineskip}{1.5mm}

\thispagestyle{empty}
\begin{flushright}
hep-th/   \\
\end{flushright}
\vbox{}
\vspace{1.0cm}

\begin{center}
\centerline{\LARGE{Two-Dimensional Twisted Sigma Models,}}
\bigskip
\centerline{\LARGE{the Mirror Chiral de Rham Complex, and}}      
\bigskip       
\centerline{\LARGE{Twisted Generalised Mirror Symmetry}}       

\vspace{1.0cm}

{Meng-Chwan~Tan\footnote{E-mail: g0306155@nus.edu.sg}}
\\[2mm]
{\it Department of Physics\\
National University of Singapore \\
Singapore 119260} \\[1mm] 
\end{center}     
               
\vspace{0.5 cm}                         

\centerline{\bf Abstract}\smallskip \noindent                   

In this paper, we study the perturbative aspects of a  ``B-twisted" two-dimensional $(0,2)$ heterotic sigma model on a holomorphic gauge bundle $\mathcal E$ over a complex, hermitian manifold $X$. We show that the model can be naturally described in terms of the mathematical theory of ``Chiral Differential Operators". In particular, the physical anomalies of the sigma model can be reinterpreted as an obstruction to a global definition of the associated sheaf of vertex superalgebras derived from the free conformal field theory describing the model locally on $X$. In addition, one can  also obtain a novel understanding of  the sigma model one-loop beta function solely in terms of holomorphic data.  At the $(2,2)$ locus, one can describe the resulting half-twisted variant of the topological B-model   in terms of a $\it{mirror}$ ``Chiral de Rham complex" (or CDR) defined by Malikov et al. in \cite{GMS1}. Via mirror symmetry, one can also derive various conjectural expressions relating the sheaf cohomology of the mirror CDR to that of the original CDR on pairs of Calabi-Yau mirror manifolds. An analysis of the half-twisted  model on a non-K\"ahler group manifold with torsion also allows one to draw conclusions about the corresponding sheaves of CDR (and its mirror) that are consistent with mathematically established results by Ben-Bassat in \cite{ben} on the mirror symmetry of generalised complex manifolds. These conclusions therefore suggest an interesting relevance of the sheaf of CDR in the recent study of generalised mirror symmetry.

\newpage

\renewcommand{\thefootnote}{\arabic{footnote}}
\setcounter{footnote}{0}

\newsection{Introduction}  

The mathematical theory of ``Chiral Differential Operators'' or CDO's is a fairly well-developed subject that aims to provide a rigorous mathematical construction of conformal fields theories, possibly associated with sigma models in two-dimensions, without resorting to mathematically non-rigorous methods such as the path integral. It was first introduced and studied in a series of seminal papers by Malikov et al. \cite{MSV1, MSV2, GMS1, GMS2, GMS3}, and in \cite{BD} by Beilinson and Drinfeld, whereby a more algebraic approach to this construction was taken in the latter. These developments have found interesting applications in various fields of geometry and representation theory such as mirror symmetry \cite{Bo} and the study of elliptic genera \cite{BL, BL1, BL2} etc. However, the explicit interpretation of the theory of CDO's, in terms of the physical models it is supposed to describe, has been somewhat unclear, that is until recently.   
    
In the pioneering papers of Kapustin \cite{Ka} and Witten \cite{CDO}, initial steps were taken to provide a physical interpretation of some of the mathematical results in the general theory of CDO's. In \cite{Ka}, it was argued that on a Calabi-Yau manifold $X$, the mathematical theory of a CDO  known as the chiral de Rham complex or CDR for short, can be identified with the infinite-volume limit of a half-twisted variant of the topological A-model. In \cite{CDO}, the perturbative limit of a half-twisted $(0,2)$ sigma model with right-moving fermions was studied, where its interpretation in terms of the theory of a CDO that is a purely bosonic version of the CDR was elucidated. An explicit computation (on ${\mathbb P}^1$) was also carried out by Frenkel et al. in \cite {Frenkel} to verify mathematically, the identification of the CDR as the half-twisted sigma model in perturbation theory. 
              
Shortly thereafter, a generalisation of the model in \cite{CDO} to include left-moving worldsheet fermions valued in a holomorphic gauge bundle over the target space, was considered by the present author in \cite{MC}, that is, the perturbative aspects of a twisted $\it{heterotic}$ $(0,2)$ sigma model were being studied in \cite{MC}.  The objective of \cite{MC} was to seek a physical interpretation of the mathematical theory of a $\it{general}$ class of CDO's (constructed from generic vertex superalgebras) that has been formally defined by Malikov et al. in \cite{GMS1,GMS3}.  It was then shown in \cite{MC} that the physical anomalies of the sigma model can be reinterpreted as an obstruction to a global definition of an associated sheaf of vertex superalgebras derived from the free conformal field theory describing the model locally on $X$. It was also shown that one can obtain a novel understanding of  the sigma model one-loop beta function solely in terms of holomorphic data. In addition, at the $(2,2)$ locus, the interpretation of the resulting half-twisted variant of the A-model in terms of a sheaf of CDR, was also made manifest on an arbitrary (not necessarily Calabi-Yau) smooth manifold. The results in \cite{MC} therefore serve as an alternative verification and generalisation of the specific findings established earlier in \cite{Ka}.                     
       
In this paper, we shall study the $\it{perturbative}$ aspects of a $(0,2)$ heterotic sigma model with a $\it{different}$ twist - at the $(2,2)$ locus, the twisted heterotic sigma model actually specialises to a half-twisted variant of the topological B-model instead.  Our main objective is to furnish a purely physical interpretation of the $\it{mirror}$ chiral de Rham complex defined by Malikov et al. in \cite{GMS1}. In doing so, we will be able to derive several important mathematical results that have not been computed anywhere in the literature before - an example of particular importance is the set of automorphism relations of this sheaf of mirror CDR, and by considering the equivalence of elliptic genera under mirror symmetry of the underlying untwisted $(2,2)$ models on Calabi-Yau manifolds, one can also derive various conjectural expressions relating the sheaf cohomology of the original CDR to that of its     mirror, on mirror pairs of Calabi-Yau manifolds. Moreover, by analysing an explicit example of the half-twisted B-model  on a non-K\"ahler, parallelisable group manifold with torsion, one can also draw conclusions about the corresponding sheaves of CDR (and its mirror) that are consistent with the mathematics of mirror symmetry on generalised complex manifolds established by Ben-Bassat in \cite{ben}. These conclusions therefore suggest an interesting relevance of the sheaf of CDR in the study of generalised mirror symmetry.  Like in \cite{MC}, we will be able to obtain, at various points in the paper, novel insights into the physics of the twisted models via a reinterpretation of some established mathematical results in the theory of CDO's, and vice-versa.                         
              
Additional motivation for this work also come from the fact that the twisted heterotic sigma model $\it{is}$ relevant to the heterotic string - the twisted correlation functions are related to the actual string correlation functions via spectral flow. Another relevant point to note is that an isomorphic model has been considered by Sharpe in \cite{other}. In \cite{other}, Sharpe analyses the quantum correlation functions of the twisted model and derives an interestingly new anomaly cancellation condition for the $(2,2)$ B-model from a $(0,2)$ perspective. Perhaps certain results in \cite{other} can be understood in the context of the mirror CDR as well.       

\vspace{0.1cm}

Towards our end, we shall follow closely the approach taken in \cite{MC}.

\smallskip\noindent{\it A Brief Summary and Plan of the Paper}            

A brief summary and  plan of the paper is as follows. First, in Section 2, we will review the two-dimensional heterotic sigma model with $(0,2)$ supersymmetry on a rank-$r$ holomorphic gauge bundle ${\mathcal E}$ over a K\"ahler manifold $X$. We will then perform a certain twist on the model which will serve to redefine the spins of the relevant worldsheet fields such that the resulting Lagrangian will specialise to a topological B-model Lagrangian at the $(2,2)$ locus.              

Next, in Section 3, we will focus on the space of physical operators of this twisted heterotic sigma model. In particular, we will study the properties of the chiral algebra furnished by these operators. In addition, we will show how the moduli of the chiral algebra arises when we include a non-K\"ahler deformation of $X$. The geometrical properties of a specific non-K\"ahler group manifold relevant to our analysis later in the paper, will also be elaborated in this section.  

In Section 4, we will discuss, from a purely physical perspective, the anomalies of this particular model. The main aim in doing so is to prepare for the observations and results that we will make and find in the next section.   

In Section 5, we will introduce the notion of a sheaf of perturbative observables.  An alternative description of the chiral algebra of physical operators in terms of the elements of a Cech cohomology group will also be presented. Thereafter, we will show that the twisted model on a local patch of the target space can be described in terms of a free $\it{hybrid}$ $bc$-$\beta\gamma$ system, where in order to give a complete description of the model on the $\it{entire}$  target space,  it will first be necessary to study its local symmetries. Using the local symmetries, one can then glue together the free conformal field theories (each defined on a local patch of the target space by the free hybrid $bc$-$\beta\gamma$ system) to obtain a globally-defined sheaf of CDO's or vertex superalgebras which span a subset of the chiral algebra of the model. It is at this juncture that one observes the mathematical obstruction to a global definition of the sheaf (and hence the existence of the underlying theory)  to be the physical anomaly of the model itself. Via an example, we will be able to obtain a novel understanding of the non-zero one-loop beta function of the twisted heterotic sigma model solely in terms of holomorphic data.

In Section 6, we will study the twisted model at the $(2,2)$ locus where $\mathcal E = TX$, such that the obstruction to a global definition of the sheaf of vertex superalgebras vanishes for $\it{any}$ smooth manifold $X$ if one works locally on the worldsheet $\Sigma$. In doing so, we obtain a purely mathematical description of the half-twisted variant of the topological B-model in terms of the theory of the mirror CDR, that for a target space with vanishing first Chern class such as a Calabi-Yau manifold, acquires the  structure of a topological vertex algebra. Using the CFT state-operator correspondence in the Calabi-Yau case, one can express the elliptic genus in terms of the Cech cohomology of the mirror CDR. Consequently, from the equivalence of elliptic genera under mirror symmetry of the underlying untwisted $(2,2)$ models on Calabi-Yau manifolds, one can also derive various conjectural expressions relating the Cech cohomology of the original CDR on $\widetilde X$, to that of its mirror on $X$, where $X$ and $\widetilde X$ are a pair of mirror Calabi-Yau's.               
   
In Section 7, we will analyse, as examples, sheaves of mirror CDR that describe the physics of the half-twisted B-model on two different smooth manifolds. The main aim is to illustrate the rather abstract discussion in the preceding sections. By studying the sheaves of mirror CDR on $\mathbb {CP}^1$, we find that a subset of the infinite-dimensional space of physical operators furnishes an underlying superaffine Lie-algebra. As will be explained, this observation is consistent  with the definition of the mirror CDR as a sheaf of vertex algebras that is isomorphic to the sheaf of  CDR. Similar to Section 5, we will be able to obtain a novel understanding of the non-zero one-loop beta function of the half-twisted B-model solely in terms of holomorphic data. Furthermore, for the half-twisted B-model on a non-K\"ahler, parallelised, smooth manifold with torsion such as $\bf{S}^3 \times \bf{S}^1$, a study of the corresponding sheaf of mirror CDR reveals a direct relationship between the modulus of sheaves and the level of the underlying $SU(2)$ WZW theory. 
   
In Section 8, we will show, using the geometrical properties of $\bf{S}^3 \times \bf{S}^1$ elaborated in Section 3 and the concept of fibrewise duality, how the relationship between the modulus of sheaves and the level of the underlying $SU(2)$ WZW theory (found in both the half-twisted B-model on $\bf{S}^3 \times \bf{S}^1$ $\it{and}$ the half-twisted A-model on a $\it{mirror}$ $\bf{S}^3 \times \bf{S}^1$ in \cite{MC}), is consistent with the mathematics of mirror symmetry on generalised complex manifolds established by Ben-Bassat in \cite{ben}.

\bigskip\noindent{\it Beyond Perturbation Theory} 
 
As pointed out in \cite{CDO}, instanton effects can change the picture radically, triggering a spontaneous breaking of supersymmetry, hence making the chiral algebra trivial as the elliptic genus vanishes.    Hence, out of perturbation theory, the sigma model may no longer be described by the theory of CDO's. This non-perturbative consideration is beyond the scope of the present paper. However, we do hope to address it in a future publication.       

\newpage      
 
\newsection{A Twisted Heterotic Sigma Model}

\vspace{-0.5cm}

\newsubsection{The Heterotic Sigma Model with $(0,2)$ Supersymmetry}

To begin, let us first recall the two-dimensional heterotic non-linear sigma model with $(0,2)$ supersymmetry on a rank-$r$ holomorphic gauge bundle     $\mathcal E$ over a  K\"ahler manifold $X$. It governs maps $\Phi : \Sigma \to X$, with $\Sigma$ being the worldsheet Riemann surface. By picking local coordinates $z$, $\bar z$ on $\Sigma$, and $\phi^{i}$, $\phi^{\bar i}$ on $X$, the map $\Phi$ can then be described locally via the functions $\phi^{i}(z, \bar z)$ and $\phi^{\bar i}(z, \bar z)$. Let $K$ and ${\overline K}$ be the canonical and anti-canonical bundles of $\Sigma$ (the bundles of one-forms of types $(1,0)$ and $(0,1)$ respectively), whereby the spinor bundles of $\Sigma$ with opposite chiralities are given by $K^{1/2}$ and ${\overline K}^{1/2}$. Let $TX$ and $\overline {TX}$ be the holomorphic and anti-holomorphic tangent bundle of $X$. 
The left-moving fermi fields of the model consist of $\lambda^a$ and $\lambda_a$, which are smooth sections of the bundles ${ K}^{1/2} \otimes \Phi^*{\mathcal E}$ and ${ K}^{1/2} \otimes \Phi^*{\mathcal E^*}$, respectively. On the other hand, the right-moving fermi fields consist of $\psi^i$ and $\psi^{\bar i}$, which are smooth sections of the bundles ${\overline K}^{1/2} \otimes \Phi^*{TX}$ and ${\overline K}^{1/2} \otimes \Phi^*{\overline {TX}}$, respectively. Here, $\psi^i$ and $\psi^{\bar i}$ are superpartners of the scalar fields $\phi^i$ and $\phi^{\bar i}$, while $\lambda^a$ and $\lambda_a$ are superpartners to a set of $\it{auxiliary}$ scalar fields $l^a$ and $l_a$, which are in turn smooth sections of the bundles $K^{1/2} \otimes {\overline K}^{1/2} \otimes \Phi^*{\mathcal E}$ and $K^{1/2} \otimes {\overline K}^{1/2} \otimes \Phi^*{\mathcal E^*}$. Let $g$ be the hermitian metric on $X$. The action is then given by
\begin{eqnarray}
S& = & \int_{\Sigma} |d^2z| \left( {1\over 2}g_{i{\bar j}} (\partial_z \phi^i \partial_{\bar z}\phi^{\bar j} + \partial_{\bar z} \phi^i \partial_z\phi^{\bar j} ) + g_{i{\bar j}} \psi^i D_z \psi^{\bar j} + {\lambda}_a D_{\bar z} {\lambda}^a  \right.  \nonumber \\
&&   \qquad \qquad  \qquad  \left. + F^a{}_{ b i {\bar j}}(\phi) {\lambda}_a {\lambda}^b \psi^i \psi^{\bar j}  - l_a l^a  \right), 
\label{action}
\end{eqnarray} 
whereby $i, {\bar i} = 1 \dots, n={\textrm {dim}}_{\mathbb C}X$, $a= 1 \dots, r$,\footnote{As we will be studying the sigma model in the peturbative limit,  worldsheet instantons are absent, and one considers only (degree zero) constant maps $\Phi$, such that  $\int_{\Sigma}{\Phi^*c_1(\cal E)} = 0$. Since the selection rule from the requirement of anomaly cancellation states that the number of $\lambda^a$s must be given by $\int_{\Sigma}{\Phi^* c_1(\cal E)} + r(1-g)$, where $g$ is the genus of $\Sigma$, we find that at string tree level, the number of $\lambda^a$s must be given by $r$.} $|d^2 z| = i dz \wedge d{\bar z}$, and $F^a{}_{ b i {\bar j}}(\phi) = A^a{}_{b i, {\bar j}}(\phi)$ is the curvature 2-form of the holomorphic gauge bundle $\mathcal E$ with connection $A$. In addition,  $D_z$ is the $\partial$ operator on ${\overline K}^{1/2} \otimes \phi^*{\overline {TX}}$ using the pull-back of the Levi-Civita connection on $TX$, while $D_{\bar z}$ is the $\bar {\partial}$ operator on ${K}^{1/2} \otimes \Phi^*{\mathcal E}$ using the pull-back of the connection $A$ on $\mathcal E$. In formulas (using a local trivialisation of ${\overline K}^{1/2}$ and ${K}^{1/2}$ respectively), we have\footnote{Note that we have used a flat metric and hence vanishing spin connection on the Riemann surface $\Sigma$ in writing these formulas.} 
\be
D_z \psi^{\bar j} = \partial_z \psi ^{\bar j} + \Gamma^{\bar j}_{\bar l \bar k} \partial _z \phi^{\bar l} \psi^{\bar k},
\ee   
and  
\be
D_{\bar z} \lambda^a = \partial_{\bar z} \lambda^a + A^a{}_{b i} (\phi) \partial_{\bar z} \phi^i \lambda^b.
\ee 
Here,  $\Gamma^{\bar j}_{\bar l \bar k}$ is the affine connection of $X$, while $A^a{}_{b i} (\phi)$ is the connection on $\mathcal E$  in component form.

The infinitesimal transformation  of the fields generated by the supercharge $\overline Q_+$ under the first right-moving supersymmetry,  is given by
\begin{eqnarray}
\label{tx1}
\delta \phi^{i} = 0, & \quad & \delta \phi^{\bar i} = {{\bar \epsilon}_-} \psi ^{\bar i}, \nonumber \\
\delta \psi ^{\bar i} = 0, & \quad & \delta \psi^i = - {{\bar \epsilon}_-}\partial_{\bar z}\phi^i, \nonumber \\
\delta \lambda^a = 0, & \quad &\delta \lambda_a = {{\bar \epsilon}_-} l_a, \\
\delta l_a = 0, & \quad & \delta l^a = {{\bar \epsilon}_-} \left ( D_{\bar z} \lambda^a + F^a{}_{b i {\bar j}}(\phi) \lambda^b \psi^i \psi^{\bar j} \right),\nonumber
\end{eqnarray}
while the infinitesimal transformation  of the fields generated by the supercharge $Q_+$ under the second right-moving supersymmetry,  is given by
\begin{eqnarray}
\label{tx2}
\delta \phi^{i} = {{\epsilon}_-}\psi ^{i}, & \quad & \delta \lambda^a = {{\epsilon}_-} \left (l^a + A^a{}_{b i}(\phi) \lambda^b \psi^i \right), \nonumber \\
\delta \psi ^{i} = 0, & \quad & \delta l^a = - {{\epsilon}_-} A^a{}_{bi} (\phi) l^b \psi^i, \nonumber \\
\delta \phi^{\bar i} = 0 , & \quad & \delta \psi^{\bar i} = -{{\epsilon}_-} \partial_{\bar z} \phi^{\bar i}, \\
\delta \lambda_a = 0, & \quad & \delta l_a = {{\epsilon}_-} \partial_{\bar z} \lambda_a,\nonumber
\end{eqnarray}
where ${{\epsilon}_-}$ and ${{\bar \epsilon}_-}$ are anti-holomorphic sections of ${\overline K}^{-1/2}$. Since we are considering a holomorphic vector bundle $\mathcal E$, the supersymmetry algebra is trivially satisfied.\footnote{The supersymmetry algebra is satisfied provided the $(2,0)$ part of the curvature vanishes i.e.,  $A^{a}{}_{b[i,j]}-A^{a}{}_{c[i}A^{c}{}_{bj]}=0$. For a real gauge field $A$ of a unitary gauge group whereby $A^\dag_i = A_{\bar i}$, this just means that $\mathcal E$ must be a holomorphic vector bundle \cite{GSW2}.}

\newsubsection{Twisting the Model}

Classically, the action (\ref{action}) and therefore the model that it describes, possesses a left-moving flavour symmetry and a right-moving R-symmetry, giving rise to a $U(1)_L \times U(1)_R$ global symmetry group. Denoting $(q_L, q_R)$ to be the left- and right-moving charges of the fields under this symmetry group, we find that $\lambda_a$ and $\lambda^a$ have charges $(\pm 1, 0)$, $\psi^{\bar i}$ and $\psi^i$ have charges $(0, \pm 1)$, and $l_a$ and $l^a$ have charges $(\pm 1, \pm 1)$ respectively. Quantum mechanically however, these symmetries are anomalous because of non-perturbative worldsheet instantons; the charge violations for the left- and right-moving global symmetries are given by $\Delta {q_L} = \int_{\Sigma} \Phi^* c_1(\mathcal E)$ and $\Delta {q_R} = \int_{\Sigma} \Phi^* c_1(TX)$, respectively.      

In order to define a twisted variant of the model, the spins of the various fields need to be shifted by a linear combination of their corresponding left- and right-moving charges $(q_L, q_R)$ under the global $U(1)_L \times U(1)_R$ symmetry group; by considering a shift in the spin $S$     via $S \to S + {1\over 2} \left [(1-2s)q_L + (2{\bar s} -1)q_R \right ]$ (where $s$ and $\bar s$ are real numbers), the various fields of the twisted model will transform as smooth sections of the following bundles:  
\begin{eqnarray}
\lambda_a  \in  \Gamma \left( K^{(1-s)} \otimes \Phi^*{\mathcal E^*}  \right), & \qquad & \lambda^a  \in  \Gamma \left(K^{s} \otimes \Phi^*{\mathcal E} \right), \nonumber \\
\psi^i \in  \Gamma \left({\overline K}^{(1- \bar s)} \otimes \Phi^*{TX} \right), & \qquad &  \psi^{\bar i} \in  \Gamma \left({\overline K}^{\bar s} \otimes \Phi^*{\overline{TX}}\right), \\ 
l_a \in \Gamma \left(K^{(1-s)} \otimes {\overline K}^{\bar s} \otimes \Phi^*{\mathcal E^*} \right), & \qquad & l^a \in \Gamma \left(K^{s} \otimes {\overline K}^{(1- \bar s)} \otimes \Phi^*{\mathcal E} \right). \nonumber 
\end{eqnarray}    
Notice that for $s = \bar s = {1\over 2}$, the fields transform as smooth sections of the same tensored bundles defining the original heterotic sigma  model, i.e., we get back the untwisted model.  

In order for a twisted model to be physically consistent, one must ensure that the $\it{new}$ Lorentz symmetry (which has been modified from the original due to the twist) continues to be non-anomalous quantum mechanically. Note that similar to the untwisted case, the $U(1)_L$ and $U(1)_R$ symmetries are anomalous in the quantum theory. The charge violations on a genus-$g$ Riemann surface $\Sigma$ are given by
\be
\Delta {q_L}  =   r(1-2s)(1-g) + \int_{\Sigma} \Phi^* c_1(\mathcal E),   
\label{anomaly1}
\ee
\be
\Delta {q_R}  =  n(2{\bar s} -1)(g-1) + \int_{\Sigma} \Phi^* c_1(TX).
\label{anomaly2}     
\ee        

If one has the condition $c_1(\mathcal E) = c_1 (TX)$, one can see from (\ref{anomaly1}) and (\ref{anomaly2}) that an example of a non-anomalous combination of global currents that one can use to twist the model with is ${1\over 2} (J_L - J_R)$, where $s = \bar s =0$. If one has the additional condition that $c_1(\mathcal E) = c_1(TX) =0$, i.e., $X$ is a Calabi-Yau, one can also consider the non-anomalous current combination $-{1\over 2} (J_L + J_R)$, where $s = 1$ and $\bar s = 0$. 
  
Note that the former twist was  considered in \cite{MC}, where it was shown that the model one will get at the $(2,2)$ locus is a half-twisted variant of the topological A-model. Now, recall that we would like to study a twisted heterotic model which can be related to a half-twisted variant of the topological B-model when $\mathcal E =TX$ at the $(2,2)$ locus.   Since the former twist  maps to the latter twist when we make the replacement $J_L \to -J_L$, one should study the twisted variant of the heterotic sigma model defined by $s= 1$ and $\bar s =0$, i.e., we should consider the twisted model associated with the current combination $-{1\over 2}(J_L + J_R)$. As required, the various fields in this twisted model of interest will thus transform as smooth sections of the following bundles:
\begin{eqnarray}    
\lambda_a  \in  \Gamma \left(\Phi^*{\mathcal E^{\vee}}  \right), & \qquad & \lambda^a_{z}  \in  \Gamma \left( K \otimes \Phi^*{\mathcal E}\right), \nonumber \\
\psi^i_{\bar z} \in  \Gamma \left({\overline K} \otimes \Phi^*{TX} \right), & \qquad &  \psi^{\bar i} \in  \Gamma \left(\Phi^*{\overline{TX}}\right), \\ 
l_{z\bar z}^a \in \Gamma \left( K \otimes {\overline K} \otimes \Phi^*{\mathcal E} \right), & \qquad & l_{a} \in \Gamma \left(\Phi^*{\mathcal E^{\vee}} \right), \nonumber 
\end{eqnarray}
where ${\cal E}^{\vee}$ is the bundle dual to $\cal E$. Notice that we have included additional indices in the above fields so as to reflect their new geometrical characteristics on $\Sigma$; fields without a $z$ or $\bar z$ index transform as worldsheet scalars, while fields with a $z$ or $\bar z$ index transform as $(1,0)$ or $(0,1)$ forms on the worldsheet. In addition, as reflected by the $a$, $i$, and $\bar i$ indices, all fields continue to be valued in the pull-back of the corresponding bundles on $X$. Thus, the action of the twisted variant of the two-dimensional heterotic sigma model will be given by

\begin{eqnarray}
S_{\mathrm twist}& = & \int_{\Sigma} |d^2z| \left( {1\over 2}g_{i{\bar j}} (\partial_z \phi^i \partial_{\bar z}\phi^{\bar j} + \partial_{\bar z} \phi^i \partial_z\phi^{\bar j} ) + g_{i{\bar j}} \psi_{\bar z}^i D_z \psi^{\bar j} + {\lambda}_{a} D_{\bar z} {\lambda}^a_z  \right.  \nonumber \\
&&   \qquad \qquad  \qquad  \left. + F^a{}_{ b i {\bar j}}(\phi) {\lambda}_{a} {\lambda}^b_z \psi_{\bar z}^i \psi^{\bar j}  - l_{a} l_{z\bar z}^a  \right). 
\label{actiontwist}
\end{eqnarray} 
 
A twisted theory is the same as an untwisted one when defined on a $\Sigma$ which is flat. Hence, locally (where one has the liberty to select a flat metric),  the twisting does nothing at all. However, what happens non-locally may be non-trivial. In particular, note that globally, the supersymmetry parameters $\epsilon_-$ and ${\bar \epsilon}_-$ must now be interpreted as sections of different line bundles; in the twisted model, the transformation laws given by (\ref{tx1}) and (\ref{tx2}) are still valid, and because of the shift in the spins of the various fields, we find that for the laws to remain physically consistent, ${\bar \epsilon}_-$ must now be a function on $\Sigma$ while $\epsilon_-$ must be a section of the non-trivial bundle ${\overline K}^{-1}$. One can therefore canonically pick ${\bar \epsilon}_-$ to be a constant and $\epsilon_-$ to vanish, i.e., the twisted variant of the two-dimensional heterotic sigma model has just $\it{one}$ canonical global fermionic symmetry generated by the supercharge ${\overline Q}_+$. Hence, the infinitesimal transformation of the (twisted) fields under this single canonical symmetry must read (after setting ${\bar \epsilon}_-$ to 1)
\begin{eqnarray}
\label{txtwist}
\delta \phi^{i} = 0, & \quad & \delta \phi^{\bar i} = \psi^{\bar i}, \nonumber \\
\delta \psi ^{\bar i} = 0, & \quad & \delta \psi_{\bar z}^i = - \partial_{\bar z}\phi^i, \nonumber \\
\delta \lambda^a_z = 0, & \quad &\delta \lambda_{a} =  l_{a}, \\
\delta l_{a} = 0, & \quad & \delta l_{z \bar z}^a = \left ( D_{\bar z} \lambda^a_z + F^a{}_{b i {\bar j}}(\phi) \lambda^b_z  \psi_{\bar z}^i \psi^{\bar j} \right). \nonumber
\end{eqnarray}
From the $(0,2)$ supersymmetry algebra, we have ${{\overline Q}^2_+} = 0$. In addition, (after twisting) ${{\overline Q}_+} $ transforms as a scalar. Consequently, we find that the symmetry is nilpotent i.e., $\delta^2 = 0$ (off-shell), and behaves as a BRST-like symmetry.     

Note at this point that the transformation laws of (\ref{txtwist}) can be expressed in terms of the BRST operator ${\overline Q}_+$, whereby $\delta W = \{{\overline Q}_+, W\}$ for any field $W$. One can then show that the action  (\ref{actiontwist}) can be written as
\be
S_{\mathrm twist} = \int_{\Sigma} |d^2z| \{{\overline Q}_+, V\} + S_{\mathrm top} 
\label{Stwist}
\ee
where
\be
V = - g_{i \bar j}  \psi_{\bar z}^i \partial_z \phi^{\bar j} - \lambda_{a} l_{z\bar z}^a, 
\label{chi}
\ee
while
\be
S_{\mathrm top} = {1\over 2}\int_{\Sigma} g_{i \bar j} \left( \partial_z \phi^i \partial_{\bar z} \phi^{\bar j} - \partial_{\bar z} \phi^i \partial_z \phi^{\bar j} \right)
\ee
is $\int_{\Sigma} \Phi^*(K)$, the integral of the pull-back to $\Sigma$ of the $(1,1)$ K\"ahler form $K= {i \over 2} g_{i \bar j} d\phi^i \wedge d\phi^{\bar j}$. 

Notice that since ${{\overline Q}^2_+} = 0$, the first term on the RHS of (\ref{Stwist}) is invariant under the transformation generated by ${\overline Q}_+$. In addition, because $dK =0$  on a K\"ahler manifold, $\int_{\Sigma} \Phi^* (K)$ depends only on the cohomology class of $K$ and the homotopy class of $\Phi_* (\Sigma)$, i.e., the class of maps $\Phi$. Consequently, $S_{\mathrm top}$ is a topological term, invariant under local field deformations and the transformation $\delta$. Thus, the action given in (\ref{Stwist}) is invariant under the BRST symmetry as required. Moreover, for the transformation laws of (\ref{txtwist}) to be physically consistent, ${\overline Q}_+$ must have charge $(0,+1)$ under the global $U(1)_L \times U(1)_R$ gauge group. Since $V$ has a corresponding charge of $(0, -1)$, while $K$ has zero charge, $S_{\mathrm twist}$ in (\ref{Stwist}) continues to be invariant under the $U(1)_L \times U(1)_R$ symmetry group at the classical level.         

As mentioned in the introduction, we will be studying the twisted model in $\it{perturbation}$ theory, where one does an expansion in the inverse of the large-radius limit. Hence, only the degree-zero maps of the term $\int_{\Sigma} \Phi^*(K)$ contribute to the path integral factor $e^{-S_{\mathrm twist}}$. Therefore, in the perturbative limit, one can set $\int_{\Sigma} \Phi^*(K) = 0$ since $dK=0$, and the model will be independent of the K\"ahler structure of $X$. This also means that one is free to study an $\it{equivalent}$ action obtained by setting $S_{\mathrm top}$ in (\ref{Stwist}) to zero. After eliminating the $l_{a} l_{z\bar z}^a$ term via its own equation of motion $l_{z \bar z}^a = 0$, the equivalent action in perturbation theory reads
\begin{eqnarray}             
S_{\mathrm pert}& = & \int_{\Sigma} |d^2z| \left( g_{i{\bar j}} \partial_z \phi^{\bar j} \partial_{\bar z}\phi^i + g_{i{\bar j}} \psi_{\bar z}^i D_z \psi^{\bar j} + {\lambda}_{a} D_{\bar z} {\lambda}^a_z  + F^a{}_{ b i {\bar j}}{\lambda}_{a} {\lambda}^b_z \psi_{\bar z}^i \psi^{\bar j}  \right), 
\label{actionpert}
\end{eqnarray} 
where it can also written as 
\be
S_{\mathrm pert} = \int_{\Sigma} |d^2z| \{{\overline Q}_+, V\}.  
\label{Spert}
\ee

Note that the original symmetries of the theory persist despite limiting ourselves to perturbation theory; even though $S_{\mathrm top} = 0$,  from (\ref{Spert}),  one finds that $S_{\mathrm pert}$ is invariant under the nilpotent BRST symmetry generated by ${\overline Q}_+$. It is also invariant under the $U(1)_L \times U(1)_R$ global symmetry. $S_{\mathrm pert}$ shall henceforth be the action of interest in all our subsequent discussions.

\newsection{Chiral Algebras from the Twisted Heterotic Sigma Model}

\vspace {-0.5cm}

\newsubsection{The Chiral Algebra}

Classically, the model is conformally invariant. The trace of the stress tensor from $S_{\mathrm pert}$ vanishes, i.e., $T_{z \bar z} = 0$. The other non-zero components of the stress tensor, at the classical level, are given by
\be
T_{zz} = g_{i \bar j} \partial_z \phi^i \partial_{z} \phi^{\bar j} + \lambda^a_z D_z \lambda_a,
\label{Tzz}
\ee
and
\be
T_{\bar z \bar z} =g_{i \bar j}  \partial_{\bar z} \phi^i \partial_{\bar z} \phi^{\bar j} + g_{i \bar j}  \psi_{\bar z}^i \left ( \partial_{\bar z} \psi^{\bar j} + \Gamma^{\bar j}_{\bar l \bar k}\partial_{\bar z} \phi^{\bar l} \psi^{\bar k} \right). 
\ee 
Furthermore, one can go on to show that 
\be
T_{\bar z \bar z} = \{ {\overline Q}_+ , - g_{i \bar j} \psi_{\bar z}^i \partial_{\bar z} \phi^{\bar j} \},
\label{tZZ}
\ee   
and (since $l_a =0$ from its equation of motion)     
\begin{eqnarray} 
\label{tzz}
[{\overline Q}_+ , T_{zz} ] & = &  \left( g_{i \bar j} D_z \psi^{\bar j} + F^a{}_{b i \bar j} (\phi) \lambda_{a} \lambda^b_z \psi^{\bar j} \right) \partial_z \phi^i \nonumber \\
& = & 0 \hspace{0.2cm} (\textrm {on-shell}).
\end{eqnarray}
From (\ref{tzz}) and (\ref{tZZ}), we see that all components of the stress tensor are ${\overline Q}_+$-invariant; $T_{zz}$ is an operator in the ${\overline Q}_+$-cohomology while $T_{\bar z \bar z}$ is ${\overline Q}_+$-exact and thus trivial in ${\overline Q}_+$-cohomology. The fact that $T_{zz}$ is not ${\overline Q}_+$-exact even at the classical level implies that the twisted model is not a 2D $\it{topological}$ field theory; rather, it is a 2D $\it{conformal}$ field theory. This because the original model has $(0,2)$  and not $(2,2)$ supersymmetry. On the other hand, the fact that $T_{\bar z \bar z}$ is ${\overline Q}_+$-exact has some non-trivial consequences on the nature of the local operators in the ${\overline Q}_+$-cohomology. Let us discuss this further.

We say that a local operator $\cal O$ inserted at the origin has
dimension $(n,m)$ if under a rescaling $z\to \lambda z$, $\bar
z\to \bar\lambda z$ (which is a conformal symmetry of the classical theory),
it transforms as $\partial^{n+m}/\partial z^n\partial\bar z^m$,
that is, as $\lambda^{-n}\bar\lambda{}^{-m}$. Classical local
operators have dimensions $(n,m)$ where $n$ and $m$ are
non-negative integers.\footnote{Anomalous
dimensions under RG flow may shift the values of $n$ and $m$ quantum mechanically, but the spin given by
$(n-m)$, being an intrinsic property, remains unchanged.} However, only local operators with $m = 0$ survive in ${\overline Q}_+$-cohomology. The reason for the last statement is that the rescaling of $\bar z$ is generated by $\bar L_0=\oint d\bar z\, \bar z T_{\bar
z\bar z}$.  As we noted in the previous paragraph, $T_{\bar z\,\bar z}$
is of the form $\{{\overline Q}_+,\dots\}$, so $\bar L_0=\{{\overline Q}_+,V_0\} $ for some $V_0$. If $\cal O$ is to be admissible as a local physical operator, it must at least be true that $\{{\overline Q}_+, {\cal O}\}=0$. Consequently, $[\bar L_0,{\cal
O}]=\{{\overline Q}_+,[V_0,{\cal O}]\}$.  Since the eigenvalue of $\bar L_0$ on $\cal O$ is $m$, we have $[\bar L_0,{\cal O}]=m{\cal O}$. Therefore, if $m\not= 0$, it follows that ${\cal O}$ is $\overline Q_+$-exact and thus trivial in $\overline Q_+$-cohomology.
    
By a similar argument, we can show that $\cal O$,  as an element of the $\overline Q_+$-cohomology, varies homolomorphically with $z$. Indeed, since the momentum operator (which acts on $\cal O$ as $\partial_{\bar z}$) is given by $\bar L_{-1}$, the term $\partial_{\bar z} \cal O$ will be given by the commutator $[ \bar L_{-1}, \cal O]$. Since $\bar L_{-1} = \oint d\bar z\,T_{\bar z\bar z}$, we will  have $\bar L_{-1}=\{{\overline Q_+},V_{-1}\}$ for some $V_{-1}$. Hence, because $\cal O$ is physical such that ${\{\overline Q_+, \cal O\}} =  0$, it will be true that $\partial_{\bar z}{\cal O}=\{{\overline Q_+},[V_{-1},{\cal O}]\}$ and thus vanishes in $\overline Q_+$-cohomology.          

The observations that we have made so far are based solely on classical grounds. The question that one might then ask is whether these observations will continue to hold when we eventually consider the quantum theory. The key point to note is that if it is true classically that a cohomology vanishes, it should continue to do so in perturbation theory, when quantum effects are small enough. Since the above observations were made based on the classical fact that $T_{\bar z \bar z}$ vanishes in $\overline Q_+$-cohomology, they will continue to hold at the quantum level. Let us look at the quantum theory more closely.  

\vspace{0.4cm}{\noindent{\it The Quantum Theory}}

Quantum mechanically, the conformal structure of the theory is violated by a non-zero one-loop $\beta$-function; renormalisation adds to the classical action $S_{\mathrm pert}$ a term of the form:
\be
 \Delta_{\textrm 1-loop}\ = \ c_{1} \ R_{ i \bar j} \partial_{z}\phi^{\bar j}\psi_{\bar z}^{i}+ c_{2} \ g^{ i \bar j}F^{a}{}_{b  i \bar j}\lambda_{ a}l_{z\bar z}^{b}
\label{1-loop}
\ee
for some divergent constants $c_{{1,2}}$, where $R_{ i \bar j}$ is the Ricci tensor of $X$. In the Calabi-Yau case, one can choose a Ricci-flat metric and a solution to the Uhlenbeck-Yau equation, $g^{ i \bar j}F^{a}{}_{b  i \bar j}=0$, such that $\Delta_{\textrm 1-loop}$ vanishes and the original action is restored. In this case, the classical observations made above continue to hold true. On the other hand, in the ``massive models'' where $c_1(X) \neq 0$, there is no way to set $\Delta_{\textrm 1-loop}$ to zero. Conformal invariance is necessarily lost, and there {\it is} nontrivial RG running. However, one can continue to express $T_{\bar z \bar z}$ as $\{\overline Q_+, \dots \}$, i.e., it remains $\overline Q_+$-exact, and thus continues to vanish in $\overline Q_+$-cohomology. Hence, the above observations about the holomorphic nature of the local operators having dimension $(n,0)$ $\it{continue}$ to hold in the quantum theory.

We would also like to bring to the reader's attention another important feature of the $\overline Q_+$-cohomology at the quantum level. Recall that classically, we had $[\overline Q_+, T_{zz}] = 0$ via the classical equations of motion. Notice that the classical expression for $T_{zz}$ is not modified at the quantum level (at least up to one-loop), since even in the non-Calabi-Yau case, the additional term of $\Delta_{\mathrm 1-loop}$ in the quantum  action does not contribute to $T_{zz}$. However, due to one-loop corrections to the action of $\overline Q_+$, we have, at the quantum level   
\be
[\overline Q_+, T_{zz}] = \partial_z (R_{ i \bar j}\partial_z \phi^i    \psi^{\bar j}) + \dots
\label{tzzanomaly}
\ee    
(where `$\dots$' is also a partial derivative of some terms with respect to $z$). Note that the term on the RHS of (\ref{tzzanomaly}) $\it{cannot}$ be eliminated through the equations of motion in the quantum theory. Neither can we modify $T_{zz}$ (by subtracting a total derivative term) such that it continues to be $\overline Q_+$-invariant. This implies that in a `massive'  model, operators do not remain in the $\overline Q_+$-cohomology after general holomorphic coordinate transformations on the worldsheet, i.e., the model is $\it{not}$ conformal at the level of the    $\overline Q_+$-cohomology.\footnote{In section 5.7, we will examine more closely, from a different point of view, the one-loop correction to the action of $\overline Q_+$ associated with the beta-function, where (\ref{tzzanomaly}) will appear   in a different guise.} However, $T_{zz}$ continues to be holomorphic in $z$ up to $\overline Q_+$-trivial terms; from the conservation of the stress tensor, we have $\partial_{\bar z}T_{zz} = - \partial_z T_{z \bar z}$, and $T_{z \bar z}$, while no longer zero, is now given by $T_{z \bar z} = \{\overline Q_+, G_{z \bar z}\}$ for some $G_{z \bar z}$, i.e., $\partial_z T_{z \bar z}$ continues to be $\overline Q_+$-exact,  and  $\partial_{\bar z}T_{zz} \sim 0$ in $\overline Q_+$-cohomology. The holomorphy of $T_{zz}$, together with the relation (\ref{tzzanomaly}), has further implications for the $\overline Q_+$-cohomology of local operators; by a Laurent expansion of $T_{zz}$,\footnote{Since we are working modulo $\overline Q_+$-trivial operators, it suffices for $T_{zz}$ to be holomorphic up to $\overline Q_+$-trival terms before an expansion in terms Laurent coefficients is permitted.} one can use (\ref{tzzanomaly}) to show that $[\overline Q_+, L_{-1} ] = 0$. This means that operators remain in the $\overline Q_+$-cohomology after global translations on the worldsheet. In addition, recall that $\overline Q_+$ is a scalar with spin zero in the twisted model. As shown few paragraphs before, we have the condition $\bar L_0 = 0$. Let the spin be $S$, where $S= L_0 - \bar L_0$. Therefore, $[\overline Q_+, S]= 0$ implies that $[\overline Q_+, L_0 ] = 0$. In other words, operators remain in the $\overline Q_+$-cohomology after global dilatations of the worldsheet coordinates.    

One can also make the following observations about the correlation functions of these local operators. Firstly, note that $\left < \{ \overline Q_+, W \} \right> = 0$ for any $W$, and recall      that for any local physical operator ${\cal O_{\alpha}}$, we have ${\{\overline Q_+, {\cal O_{\alpha}}\}} = 0$. Since the $\partial_{\bar z}$ operator on $\Sigma$ is given by ${\bar L}_{-1} = \oint d{\bar z}\ T_{\bar z \bar z}$, where  $T_{\bar z \bar z} = \{ \overline Q_+, \dots \}$, we find that ${\partial_{\bar z}\left <{\cal O}_1(z_1) {\cal O}_2(z_2) \dots {\cal O}_s(z_s)  \right >}$ is given by $ \oint d{\bar z} \left <\{ \overline Q_+, \dots \} \ {\cal O}_1(z_1) {\cal O}_2(z_2) \dots {\cal O}_s(z_s) \right > = \oint d{\bar z} \left < \{\overline Q_+, \dots \prod_{i} {{\cal O}_i(z_i)} \} \right> = 0$. Thus, the correlation functions are always holomorphic in $z$. Secondly, $T_{z \bar z} = \{\overline Q_+, G_{z \bar z}\}$ for some $G_{z \bar z}$ in the `massive' models. Hence, the variation of the correlation functions due to a change in the scale of $\Sigma$ will be given by $\left <{{\cal O}_1(z_1)} {{\cal O}_2(z_2)} \dots {{\cal O}_s(z_s)} {\{\overline Q_+, G_{z \bar z} \}} \right >= \left < \{\overline Q_+, \prod_{i} {{\cal O}_i(z_i)} \cdot G_{z \bar z} \} \right> = 0$.   In other words, the correlation functions of local physical operators will continue to be invariant under arbitrary scalings of $\Sigma$. Thus, the correlation functions are always independent of the K\"ahler structure on $\Sigma$ and depend only on its complex structure.

\vspace{0.4cm}{\noindent{\it  A Holomorphic Chiral Algebra $\cal A$}}

Let ${\cal O} (z)$ and $\widetilde {\cal O} (z')$ be two $\overline Q_+$-closed operators such that their product is $\overline Q_+$-closed as well. Now, consider their operator product expansion or OPE:
\be
{\cal O}(z)  {\widetilde {\cal O}}(z') \sim \sum_k f_k (z-z') {\cal O}_k (z'),
\label{OPE}
\ee 
in which the explicit form of the coefficients $f_k$ must be such that the scaling dimensions and $U(1)_L  \times U(1)_R$ charges of the operators agree on both sides of the OPE. In general, $f_k$ is not holomorphic in $z$. However, if we work modulo $\overline Q_+$-exact operators in passing to the $\overline Q_+$-cohomology, the $f_k$'s which are non-holomorphic and are thus not annihilated by $\partial / \partial {\bar z}$, drop out from the OPE because they multiply operators ${\cal O}_k$ which are $\overline Q_+$-exact. This is true because $\partial / \partial{\bar z}$ acts on the LHS of (\ref{OPE}) to give terms which are cohomologically trivial.\footnote{Since $\{\overline Q_+,{\cal O}\}=0$, we have $\partial_{\bar z}{\cal O}=\{\overline Q_+, V(z)\}$ for some $V(z)$, as argued before. Hence $\partial_{\bar z}{\cal O}(z)\cdot {\widetilde {\cal O}}(z')=\{\overline Q_+,V(z){\widetilde {\cal O}}(z')\}$.} In other words, we can take the $f_k$'s to be holomorphic coefficients in studying the $\overline Q_+$-cohomology. Thus, the OPE of (\ref{OPE}) has a holomorphic structure.

In summary, we have established that the $\overline Q_+$-cohomology of holomorphic local operators has a natural structure of a holomorphic chiral algebra (as defined in the mathematical literature) which we shall henceforth call $\cal A$; it is always preserved under global translations and dilatations, though (unlike the usual physical notion of a chiral algebra) it may not be preserved under general holomorphic coordinate transformations on the Riemann surface $\Sigma$. Likewise, the OPE's of the chiral algebra of local operators obey the usual relations of holomorphy, associativity, and invariance under translations and scalings of $z$, but not necessarily invariance under arbitrary holomorphic reparameterisations of $z$. The local operators are of dimension (n,0) for $n \geq 0$, and the chiral algebra of such operators requires a flat metric up to scaling on $\Sigma$ to be defined.\footnote{Notice that we have implicitly assumed the flat metric on $\Sigma$ in all of our analysis thus far.} Therefore, the chiral algebra that we have obtained can only be globally-defined on a Riemann surface of genus one, or be locally-defined on an arbitrary but curved $\Sigma$. To define the chiral algebra globally on a surface of higher genus requires more in-depth analysis, and is potentially obstructed by an anomaly involving $c_1(\Sigma)$ and $({c_1(\cal E)} + c_1(X))$ that we will discuss in sections 4 and 5.6. Last but not least, as is familiar for chiral algebras, the correlation functions of these operators depend on $\Sigma$ only via its complex structure. The correlation functions are holomorphic in the parameters of the theory and are therefore protected from perturbative corrections.

\newsubsection{The Moduli of the Chiral Algebra}

We shall now discuss the moduli of the chiral algebra $\cal A$. Note that the chiral algebra does depend on the complex structure of $X$ because it enters in the definition of the fields and the fermionic symmetry transformation generated by ${\overline Q}_+$. In addition, the moduli also depends on a certain type of cohomology class. We shall now determine what this cohomology class is. To this end, we shall consider adding to $S_{\mathrm pert}$, a term which will represent a modulus of $\cal A$.  

To proceed, let $T = {1\over 2}T_{ij} d\phi^i \wedge d\phi^j$ be any two-form on $X$ that is of type $(2,0).$\footnote{As noted in \cite{MC}, the restriction of $T$ to be a gauge field of type $(2,0)$, will enable us to associate the moduli of the chiral algebra with the moduli of sheaves of vertex superalgebras.} The term  that deforms $S_{\mathrm pert}$ will then be given by
\be
{S_T=\int_{\Sigma} |d^2z| \{{\overline Q}_+, T_{ij}\psi_{\bar z}^i\partial_z\phi^j\}}.
\label{ST}
\ee       
By construction, $S_T$ is ${\overline Q}_+$-invariant. Moreover, since it has vanishing charge, it is also invariant under the global $U(1)_L \times U(1)_R$ symmetry. Hence,  as required, the addition of $S_T$ preserves  the classical symmetries of the theory. Explicitly, we then have 
\be
S_T = \int_{\Sigma} |d^2 z| \left( T_{ij, {\bar k}}\psi^{\bar k} \psi_{\bar z}^i \partial_z \phi^j - T_{ij} \partial_{\bar z} \phi^i \partial_z \phi^j  \right),
\label{STex}
\ee
where $T_{ij,{\bar k}} = \partial T_{ij}/ \partial \phi^{\bar k}$. Note that since $|d^2 z| = i dz \wedge d{\bar z}$, we can write the second term on the RHS of (\ref{STex}) as 
\be
S^{(2)}_T = {{i \over 2} \int_{\Sigma} T_{ij} d\phi^i \wedge d\phi^j }= {i \int_{\Sigma} \Phi^*(T)}. 
\label{T}
\ee  
Recall that in perturbation theory, we are considering degree-zero maps $\Phi$ with no multiplicity. Hence, for $S^{(2)}_T$ to be non-vanishing, $T$ must $\it{not}$ be closed, i.e. $dT \neq 0$. In other words, one must have a non-zero flux ${\cal H} = {dT}$.  As $T$ is of type $(2,0)$, $\cal H$ will be a three-form of type $(3,0) \oplus (2,1)$. 

Notice here that the first term on the RHS of (\ref{STex}) is expressed in terms of $\cal H$, since $T_{ij,{\bar k}}$  is simply the $(2,1)$ part of $\cal H$. In fact, $S^{(2)}_T$ can also be written in terms of $\cal H$ as follows. Suppose that $C$ is a three-manifold whose boundary is $\Sigma$ and over which the map $\Phi : \Sigma \to X$ extends. Then, if $T$ is globally-defined as a $(2,0)$-form, the relation ${\cal H} =dT$ implies, via Stoke's theorem, that 
\be
S^{(2)}_T = i \int_{C} \Phi^*(\cal H).
\label{H}
\ee                   
Hence, we see that $S_T$ can be expressed solely in terms of the three-form flux $\cal H$ (modulo terms that do not affect perturbation theory). A relevant fact for the present paper is that $\cal H$ represents a class in the Cech cohomology group  $H^1(X, \Omega^{2,cl}_X)$, where $\Omega^{2,cl}_X$ is the sheaf of $\partial$-closed $(2,0)$-forms on $X$. This has been shown in \cite{CDO} and reviewed in \cite{MC}. Thus, the modulus of the chiral algebra is represented by a class in $H^1(X, \Omega^{2,cl}_X)$.          

One last thing to note is that we do not actually want to limit ourselves to the case where $T$ is globally-defined; as is clear from (\ref{ST}), if $T$ were to be globally-defined, $S_T$ and therefore the modulus of the chiral algebra would vanish in ${\overline Q}_+$-cohomology. Fortunately, the RHS of (\ref{H}) makes sense as long as $\cal H$ is globally-defined, with the extra condition that $\cal H$ be closed, since $C$ cannot be the boundary of a four-manifold.\footnote{From homology theory, the boundary of a boundary is zero. Hence, since $\Sigma$ exists as the boundary of $C$, the three-manifold $C$ itself cannot be a boundary of a higher-dimensional four-manifold.} Therefore, it suffices for $T$ to be locally-defined such that ${\cal H} =dT$ is true only $\it{locally}$. Hence, $T$ must be interpreted a a two-form gauge field in string theory (or a $\it{non}$-$\it{trivial}$ connection on gerbes in mathematical theories). This has been emphasised in a similar context in \cite{CDO, MC}.

\newsubsection{The Moduli as a Non-K\"ahler Deformation of $X$}           

As shown above, in order to incorporate the moduli so that we can obtain a family of chiral algebras, we need to turn on the three-form $\cal H$-flux. As was shown in \cite{CDO, MC}, this term results in a non-K\"ahler deformation of the target space $X$. Thus, $X$ will be a complex, hermitian manifold in all our following discussions. For brevity, we shall simply state the results derived in \cite{MC} that are essential to the present paper.         

Firstly, for a complex, hermitian, non-K\"ahler manifold, one can define a $(1,1)$-form $\omega_T$, which is an analog of a K\"ahler $(1,1)$-form $\omega$ on a K\"ahler manifold. In contrast to $\omega$, which obeys $\partial \omega = \bar \partial \omega = 0$, $\omega_T$ obeys the weaker condition $\bar \partial \partial {\omega_T} = 0$ instead.      
         
Secondly, if we are to consider an example of a unitary sigma model (as we will do so in section 7.2), $\cal H$ must be restricted to just $(2,1)$-forms. In addition, it must also be expressible as $2i \partial {\omega_T}$, i.e., $\omega_T$ defines the torsion $\cal H$ of $X$. 

Consequently, we see that a non-vanishing $\cal H$ will mean that $\partial\omega_T \neq 0$. Thus, by turning on the moduli of the chiral algebra via a deformation $S_T$ of the action $S_{pert}$ by the three-form flux $\cal H$, one will effectively induce a non-K\"ahler deformation of the target space $X$ as claimed.

\newsubsection{An Example of a Non-K\"ahler Complex Manifold with Torsion}

We shall now describe an example of a non-K\"ahler complex manifold with torsion which will play a central role in sections 7.2 and 8 as the target space of a supersymmetric sigma model. To this end, we shall summarise some of the relevant results derived in \cite{CDO} (where the geometrical properties of the manifold have been elucidated in some detail).
   
The example that we will be considering is the group manifold $X={\bf S}^1\times {\bf S}^3$. Despite the geometrical simplicity of this manifold, its relevance to WZW models makes it rather interesting from the viewpoint of conformal field theory \cite{rocek, spindel}. $X$ is also interesting from the mathematical perspective - it serves as the simplest non-trivial example of what is known in the mathematical literature as a twisted generalised complex manifold, and has been considered in the recent study of generalised complex geometry \cite{Gua}. In fact,  $X$ will also play a role in the mirror symmetry of (compact) twisted generalised complex manifolds as we will show in section 8.                                    
           
First, note that the complex structure of $X$ can be constructed as follows. By
composing the (trivial) projection onto the second factor $X\to {\bf S}^3$
with the (non-trivial) Hopf fibration $\pi:{\bf S}^3\to {\bf S}^2\cong{\Bbb{
CP}}^1$, whose fibres are copies of ${\bf S}^1$, $X$ can viewed as a (non-trivial) fibration
of ${\Bbb{CP}}^1$ with fibres $E={\bf S}^1\times {\bf
S}^1$. Giving $E={\bf{T}}^2$ the structure of a complex Riemann surface of           
genus one, $X$ becomes a complex manifold.     

Alternatively, $X$ can be constructed as $\Bbb{C}^2/\Bbb{Z}$,
where $\Bbb{Z}$ acts on coordinates $z^i,$ $i=1,2$ of $\Bbb{C}^2$
by $z^i\to \lambda^nz^i$, with $\lambda$ a nonzero complex number    
of modulus less than 1.  The choice of $\lambda$ determines the
complex structure of $E$ in the former description and therefore that of $X$. The two
descriptions are related by simply regarding the $z^ i$'s as
homogeneous coordinates of $\Bbb{CP}^1$.

Second, it has been shown and explained in \cite{CDO} that as required, one can find a hermitian $(1,1)$-form $\omega_T$ on $X$ that obeys
$\partial\bar\partial\omega_T=0$ (and corresponds to real $\lambda$). It is given by
\be
{\omega_T}= dt\wedge \zeta + \pi^*(\omega_0),    
\label{omegaT}   
\ee
where $dt$ is a one-form on $X$ that is invariant under a $U(1)$ symmetry which acts by rotation of $\bf{S}^1$, $\zeta$ is a unique $U(2)$-invariant one-form where $U(2)$ is a symmetry which acts on $\bf{S}^3$, and $\omega_0$ is an   $SO(3)$-invariant form on $\bf{S}^2$ that integrates to 1.  One way to prove that $\partial\bar\partial\omega_T=0$, is to note that
this is the same as $d((\bar\partial-\partial)\omega/2)$, and so on a
four-manifold without boundary such as ${\bf S}^1\times {\bf S}^3$, it will integrate to zero. Since $\omega_T$ and therefore $\partial\bar\partial\omega_T$ are $U(1)\times U(2)$-invariant by construction, $\partial\bar\partial\omega_T$ can only integrate to zero if it vanishes pointwise. 

Third, notice that the full symmetry group of $X$ is actually $U(1) \times SO(4)$, where $SO(4)$ is the full rotation symmetry of $\bf{S}^3$. One can in fact define a $U(1) \times SO(4)$-invariant metric on $\bf{S}^1 \times \bf{S}^3$. Such a metric will be determined by two positive numbers, namely the radii of the two factors $\bf{S}^1$ and $\bf{S}^3$. More can be said about these two parameters as follows. The ratio of radii of ${\bf S}^1$ and ${\bf S}^3$ is determined by the choice of $dt$ (since $dt$ determines the radius $r=\int_{{\bf S}^1}dt$ of $\bf{S}^1$ for a particular $\bf{S}^3$).  The choice of $dt$ is also correlated with the choice of complex structure, since $\omega_T$ must be of type $(1,1)$. Hence, when the complex structure of ${\bf S}^1\times {\bf S}^3$ is chosen, the ratio of radii is fixed.  Note that one is free to rescale the ${\bf
S}^1$ and ${\bf S}^3$ radii by a common positive constant by
multiplying the sigma model action by this constant, since this will leave the complex structure invariant. In short, the complex
structure determines the ratio of radii, and there is one overall free parameter determined by this common positive constant. As we will see momentarily, this free parameter must be related to the level $k$ of an underlying WZW model associated with the supersymmetric sigma model under study.    
         
Now let us write down $H={\rm Re}\,{\cal H}$, the curvature of the $B$-field. As shown in \cite{CDO}, we have $H=\zeta\wedge\pi^*(\omega_0)$. It follows that  $\int_{{\bf S}^3}H=1$ and in particular $H$ is topologically non-trivial.  Therefore, to obtain a consistent quantum theory,  we must
multiply the sigma model action by a constant chosen so that $\int_{{\bf S}^3}H=2\pi k$ for some integer $k$,\footnote{This is to ensure that the theory which the action represents does not depend on the way it is being parameterised.} which must be positive so   that the hermitian metric of ${\bf S}^1\times {\bf S}^3$ is positive. This constant is simply the common positive constant which determines the overall free parameter mentioned in the previous paragraph.     

How then is this common positive constant which determines the overall free parameter, related to the level $k$ of an underlying WZW model? As explained in \cite{rocek, spindel}, the $U(1)\times SO(4)$-invariant supersymmetric sigma model of ${\bf S}^1\times{\bf S}^3$ is simply a product of a WZW model of  the group $SU(2)\cong {\bf S}^3$ with a free field theory,\footnote{The free field theory will be described in greater detail in section 7.2.} and the level of the WZW model is $k$.  In fact, as we will show from the point of
view of the perturbative theory of CDO's in an example in section 7.2, the parameter $k$ is a complex parameter
associated with $H^1({\bf S}^1\times {\bf S}^3,\Omega^{2,cl})\cong
\Bbb{C}$, and thus from the discussion in section 3.2 of \cite{MC}, we learn that $k$ must actually be an integer in order for the model to be well-defined
non-perturbatively. This is consistent with the observation made in the preceding paragraph. Another important point to note is that since the overall free parameter determines the scale of the radii of $\bf{S}^1$ and $\bf{S}^3$, we see from the representation of $X$ as a ${\bf{T}}^2$ fibration of $\mathbb {CP}^1$, that the level $k$ will also determine the K\"ahler moduli of the fibre $E = {\bf{T}}^2$ in $X$.           
     
The above facts about $X= \bf{S}^1 \times \bf{S}^3$ will be essential to our analysis in section 8, where we make first-contact with the mirror symmetry of twisted generalised complex manifolds, and show that our results are consistent with the recent mathematical observations made by Ben-Bassat in \cite{ben}.

\newsection{Anomalies of the Twisted Heterotic Sigma Model}

In this section, we will study the anomalies of the twisted heterotic sigma model. In essence, the model will fail to exist in the quantum theory if the anomaly  conditions are not satisfied. We aim to determine what these conditions are. In this discussion, we shall omit the additional term $S_T$ as anomalies do not depend on continuously varying couplings such as this one. 

To begin, let us first note from the action $S_{\mathrm pert}$ in (\ref{actionpert}), that the kinetic energy term quadratic in the fermi fields  $\psi^i$ and $\psi^{\bar i}$ is given by $(\psi, D \psi )= \int |d^2 z| g_{i \bar j} \psi^i D \psi^{\bar j}$, where $D$ is the $\partial$ operator on $\Sigma$ acting on sections $\Phi^*(\overline{TX})$, constructed using a pull-back of the Levi-Civita connection on ${TX}$. The other kinetic energy term quadratic in the fermi fields $\lambda_a$ and $\lambda^a$ is given by  $(\lambda, {\overline D} \lambda) = \int |d^2z| \lambda_{a} {\overline D} \lambda^a$, where $\overline D$ is the $\bar \partial$ operator on $\Sigma$ acting on sections $K \otimes \Phi^*(\mathcal E)$, constructed using a pull-back of the gauge connection $A$ on $\mathcal E$. (Notice that we have omitted the $z$ and $\bar z$ indices of the fields as they are irrelevant in the present discussion.)  By picking a spin structure on $\Sigma$, one can equivalently interpret $D$ and $\overline D$ as the Dirac operator and its complex conjugate on $\Sigma$, acting on  sections of ${\cal V} ={\overline K}^{-1/2} \otimes \Phi^*(\overline {TX})$ and ${{\cal W}} = K^{1/2} \otimes \Phi^*(\mathcal E)$ respectively, where $K$  is the canonical bundle of $\Sigma$ and $\overline K$ its complex conjugate.

Next, note that the anomaly arises as an obstruction to defining the functional Grassmann integral of the action quadratic in the Fermi fields $\lambda_{a}$,  $\lambda^a$, and $\psi^{i}$, $\psi^{\bar i}$, as a general function on the configuration space $\cal C$ of inequivalent connections \cite{moore}. Via the last paragraph, the Grassmann integral is given by the product of the determinant of $D$ with the determinant of $\overline D$. This can also be expressed as the determinant of $D + \overline D$. As argued in \cite{moore}, one must think of the functional integral as a section of a complex determinant line bundle $\cal L$ over $\cal C$. Only if $\cal L$ is trivial   would the integral be a global section and therefore a function on $\cal C$. Hence, the anomaly is due to the non-triviality of $\cal L$. The bundle $\cal L$ can be characterised completely by its restriction to a non-trivial two-cycle in $\cal C$ such as a two-sphere \cite{Bott}.       

To be more precise, let us consider a family of maps $\Phi : \Sigma \to X$, parameterised by a two-sphere base which we will denote as $B$. In computing the path integral, we actually want to consider the universal family of $\it{all}$ maps from $\Sigma$ to $X$. This can be represented by a $\it{single}$ map $\hat{\Phi} : {\Sigma \times B} \to X$. The quantum path integral is anomaly-free if $\cal L$, as a complex line bundle over $B$, is trivial. Conversely, if $\cal L$ is trivial, it can be trivialised by a local Green-Schwarz anomaly-cancellation mechanism and the quantum theory will exist. 

From the theory of determinant line bundles, we find that the basic obstruction to triviality of  $\cal L$ is its first Chern class. By an application of the family index theorem to anomalies \cite{CDO20, CDO21}, the first Chern class of $\cal L$ is given by ${\pi} ({ch_2({\cal W})} - {ch_2 ( {\cal V})} )$,  whereby $\pi: H^4(\Sigma \times B) \to H^2(B)$. Note that the anomaly lives in $H^4(\Sigma \times B)$ and not  $H^2(B)$;  ${\pi}({ch_2({\cal W})} -{ch_2({\cal V})} )$ vanishes if $({ch_2({\cal W})} -{ch_2({\cal V})} )$ in $H^4(\Sigma \times B)$ vanishes $\it{before}$ it is being mapped to $H^2(B)$. However, if $({ch_2({\cal W})} -{ch_2({\cal V})} ) \neq 0$ but ${\pi}({ch_2({\cal W})} -{ch_2({\cal V})} ) = 0$, then even though $\cal L$ is trivial, it $\it{cannot}$ be trivialised by a Green-Schwarz mechanism. 

To evaluate the anomaly, first note that we have a Chern character identity $ch(E \otimes F) = ch(E)ch(F)$, where $E$ and $F$ are any two bundles. Hence, by tensoring $\Phi^*(\cal E)$ with $K^{1/2}$ to obtain ${\cal W}$, we get an additional term $-{1\over 2} c_1(\Sigma) c_1(\cal E)$. Next, note that $ch_2(\overline {E}) = ch_2(E)$, and by tensoring $\Phi^*(\overline {TX})$ with ${\overline K}^{-1/2} $ to obtain $\cal V$, we get an additional term ${1\over 2} c_1(\Sigma) c_1 (TX)$.  Therefore, the condition for vanishing anomaly will be given by 
\be
 0= {-{1\over 2} c_1(\Sigma)  (  {c_1 (\cal E)} +{c_1 (TX)}  )}= {{ch_2(\cal E)} - {ch_2(TX)}}.
\label{an}
\ee
The first condition means that we can either restrict ourselves to Riemann surfaces $\Sigma$ with $c_1(\Sigma) = 0$ and ${(  {c_1 (\cal E)} + c_1 (TX) )} \neq 0$, or allow $\Sigma$ to be arbitrary while ${(  {c_1 (\cal E)} + c_1 (TX) )} =  0$. Notice also that both the anomalies automatically vanish if the bundles $TX$ and $\cal E$ are trivial such that $c_n(TX) = {c_n(\cal E)} = 0$ for any $n \geq 1$, while the second anomaly vanishes if ${\cal E} = TX$. The latter condition will be important when we discuss what happens at the $(2,2)$ locus in section 6.  

The characteristic class $({{ch_2(\cal E)} - {ch_2(TX)}})$ corresponds to an element of the Cech cohomology group $H^2(X, \Omega^{2, cl}_X)$.\footnote{As had been shown in \cite{CDO}, $ch_2(TX)$ can be interpreted as an element of $H^2(X, \Omega^{2,cl}_X)$, while $c_k (TX)$ can be interpreted as an element of $H^1(X, \Omega^{1,cl}_X)$. Using similiar arguments, we can also show that $ch_2(\cal E)$ corresponds to an element of $H^2(X, \Omega^{2,cl}_X)$  as follows. On any complex hermitian manifold, $ch_2(\cal E)$ can be represented by a closed form of type $(2,2)$. This can be seen by picking any connection on the holomorphic vector bundle $\cal E$ over $X$, whose $(0,1)$ part is the natural $\bar\partial$ operator of this bundle. Since $\bar\partial^2=0$, the curvature of such a connection is of type $(2,0)\oplus (1,1)$. However, as discussed in footnote 3, the $(2,0)$ part of the curvature must vanish. Hence, the curvature is of type $(1,1)$. Therefore, for every $k \geq 0$, $c_k(\cal E)$ is described by a closed form of type $(k,k)$. Thus, via the Cech-Dolbeault isomorphism, $c_k (\cal E)$ represents an element of $H^k(X,\Omega^{k,cl}_X)$. In particular, $c_1(\cal E)$ represents an element of $H^1(X,\Omega^{1,cl}_X)$, and ${ch_2 (\cal E)} = {1\over 2}({c^2_1(\cal E)}- {2c_2(\cal E)})$ represents an element of $H^2(X,\Omega^{2,cl}_X)$.} We will encounter it in this representation in sections 5.5 and 5.6. Similiarly, as explained in the footnote, $({c_1(\cal E)} + c_1(X))$ corresponds to an element of $H^1(X, \Omega^{1,cl}_X)$, while $c_1(\Sigma)$ corresponds to a class in $H^1(\Sigma, \Omega^{1, cl}_X)$. These will make a later appearance as well. 

Note that the $({{ch_2(\cal E)} - {ch_2(TX)}})$ anomaly appears in a heterotic sigma model with $(0,2)$ supersymmetry regardless of any topological twisting.  The ${{1\over 2} c_1(\Sigma)  (  {c_1 (\cal E)} + c_1 (TX) )}$ anomaly however,  $\it{only}$ occurs in a heterotic $(0,2)$ theory that has been twisted.

\bigskip\noindent{\it Additional Observations}

Recall from section 3.1 that the chiral algebra of local holomorphic operators, requires a flat metric up to scaling on $\Sigma$ to be globally-defined. Therefore, it can be defined over all of $\Sigma$ for genus one. The obstruction to its global definition on $\Sigma$ of higher genera is captured by the ${1\over 2}c_1(\Sigma) (c_1({\cal E)} + c_1(TX))$ anomaly. This can be seen as follows.

Note that at this stage, we are  considering the case where ${\cal E} \neq TX$. So in general, ${c_1(\cal E)} \neq - c_1(TX)$. In such an event,  the anomaly depends solely on $c_1(\Sigma)$. If $c_1(\Sigma) \neq 0$, such as when $\Sigma$ is curved or of higher genera, the Ricci scalar $R$ of $\Sigma$ is non-vanishing. Thus, the expression of $T_{z \bar z}$ will be modified,    such that
\be
{T_{z \bar z}} =  {\{ \overline Q_+, G_{z \bar z}\}+ {c \over {2 \pi}} R },
\label{tzbarzq}
\ee
where $c$ is a non-zero constant related to the central charge of the sigma model. The additional term on the RHS of (\ref{tzbarzq}), given by a multiple of $R$, represents a soft conformal anomaly on the worldsheet due to a curved $\Sigma$.  $R$ scales as a $(1,1)$ operator as required. 

There are consequences on the original nature of the $\overline Q_+$-cohomology of operators due to this additional term. Recall from section 3.1 that the holomorphy of $T_{zz}$ holds so long as $\partial_z T_{z \bar z} \sim 0$. However, from the modified expression of $T_{z \bar z}$ in (\ref{tzbarzq}),  we now find that $\partial_z T_{z \bar z} \nsim 0$. Hence, the invariance of the $\overline Q_+$-cohomology of operators under translations on the worldsheet, which requires $T_{zz}$ to be holomorphic in $z$, no longer holds. Therefore, the local holomorphic operators fail to define a chiral algebra that is globally   valid  over $\Sigma$, since one of the axioms of a chiral algebra is invariance under translations on the worldsheet. 

On the other hand, the second term on the RHS of (\ref{tzbarzq}), being a $c$-number anomaly, will affect only the partition function and not the normalised correlation functions. Thus, as argued in section 3.1, the correlation functions of local holomorphic operators will continue to depend on $\Sigma$ only via its complex structure (as is familiar for chiral algebras).

\newsection{Sheaf of Perturbative Observables} 

\vspace{-0.5cm}

\newsubsection{General Considerations}

In general, a local operator is an operator $\cal F$ that is a function of the $\it{physical}$ fields $\phi^i$, $\phi^{\bar i}$, $\psi^i_{\bar z}$, $\psi^{\bar i}$, $\lambda_{a}$, $\lambda_z^a$, and their derivatives with respect to $z$ and $\bar z$.\footnote{Notice that we have excluded the auxiliary fields $l_{a}$ and $l_{z\bar z}^a$ as they do not contribute to the correlation functions since their propagators are trivial.}$^,$\footnote{Note here that since we are interested in local operators which define a holomorphic chiral algebra on the Riemann surface $\Sigma$, we will work locally on a flat $\Sigma$ with local parameter $z$. Hence, we need not include in our operators the dependence on the scalar curvature of $\Sigma$.} However, as we saw in section 3.1, the $\overline Q_+$-cohomology vanishes for operators of dimension $(n,m)$ with $m \neq 0$. Since $\psi^i_{\bar z}$ and the derivative $\partial_{\bar z}$ both have $m=1$ (and recall from section 3.1 that a physical operator cannot have negative $m$ or $n$), $\overline Q_+$-cohomology classes can be constructed from just $\phi^i$, $\phi^{\bar i}$, $\psi^{\bar i}$, $\lambda_{a}$, $\lambda^a_z$ and their derivatives with respect to $z$. Note that the equation of motion for $\psi^{\bar i}$ is   $D_z \psi^{\bar i}= {-F^a{}_b{}^{\bar i}{}_{\bar j}(\phi) \lambda_{a} \lambda^b_z \psi^{\bar j}}$. Thus, we can ignore the $z$-derivatives of $\psi^{\bar i}$, since it can be expressed in terms of the other fields and their corresponding derivatives. Therefore, a chiral (i.e. $\overline Q_+$-invariant) operator which represents a $\overline Q_+$-cohomology class is given by
\be
{\cal F}(\phi^i,\partial_z\phi^i,\partial_z^2\phi^i,\dots;
\phi^{\bar i},\partial_z\phi^{\bar i},\partial_z^2\phi^{\bar i},\dots; \lambda_{a}, \partial_z\lambda_{a}, \partial_z^2\lambda_{a} \dots; \lambda^a_z, \partial_z\lambda^a_z, \partial_z^2\lambda^a_z \dots ; \psi^{\bar i}),
\ee
where we have tried to indicate that $\cal F$ might depend on $z$ derivatives of $\phi^{i}$, $\phi^{\bar i}$, $\lambda_{a}$ and $\lambda ^a_z$ of arbitrarily high order, though not on derivatives of $\psi^{\bar i}$. If the scaling dimension of $\cal F$ is bounded, it will mean that $\cal F$ depends only on the derivatives of fields up to some finite order, is a polynomial of bounded degree in those, and/or is a bounded polynomial in $\lambda_z^a$. Notice that $\cal F$ will always be a polynomial of finite degree in $\lambda^a_z$, $\lambda_{a}$ and $\psi ^{\bar i}$, simply because $\lambda^a_z$, $\lambda_{a}$ and $\psi^{\bar i}$ are fermionic and can only have a finite number of components before they vanish due to their anticommutativity. However, the dependence of $\cal F$ on $\phi^i$, $\phi^{\bar i}$ (as opposed to their derivatives) need not have any simple form. Nevertheless, we can make the following observation - from the $U(1)_L \times U(1)_R$ charges of the fields listed in section 2.2, we see that if $\cal F$ is homogeneous
of degree $k$ in $\psi^{\bar i}$, then it has $U(1)_L \times U(1)_R$-charge $(q_L, q_R) =( p, k)$, where $p$ is determined by the net number of $\lambda_{a}$ over $\lambda^a_z$ fields (and/or of their corresponding derivatives) in $\cal F$. 
     
\hspace {-0.2cm}A general $q_R= k$ operator ${\cal F} (\phi^i,\partial_z\phi^i,\dots; \phi^{\bar i},\partial_z\phi^{\bar i},\dots; \lambda_{a}, \partial_z \lambda_{a}, \dots; \lambda^a_z, \partial_z \lambda^a_z, \dots ; \psi^{\bar i})$ can be interpreted as a $(0,k)$-form on $X$ with values in a
certain tensor product bundle. In order to illustrate the general idea behind this interpretation, we will make things explicit for      
operators of dimension $(0,0)$ and $(1,0)$. Similiar arguments will likewise apply for operators of higher dimension. For dimension $(0,0)$, the most general operator takes the form ${\cal F}(\phi^i,\phi^{\bar i}; \lambda_a ; \psi^{\bar j})= f_{\bar j_1,\dots,\bar j_k}^{a_1, \dots, a_q}(\phi^i, \phi^{\bar i}) \psi^{\bar j_i}\dots \psi^{\bar j_k} \lambda_{a_1} \dots \lambda_{a_q}$; thus, $\cal F$ may depend on $\phi^i$, $\phi^{\bar i}$ and $\lambda_a$, but not on their
derivatives, and is $k^{th}$ order in $\psi^{\bar j}$. Mapping $\psi^{\bar j}$ to $d\phi^{\bar j}$, such an operator corresponds to an ordinary $(0,k)$-form $f_{\bar j_1,\dots,\bar j_k}(\phi^i, \phi^{\bar i})d\phi^{\bar j_1}\dots d\phi^{\bar j_k}$ on $X$ with values in the bundle $\Lambda^q {\cal E}$.\footnote{Note that $q \leq \textrm{rank}({\cal E})$ due to the anticommutativity of $\lambda_a$.} For dimension $(1,0)$, there are four general cases. In the first case, we have an operator ${\cal F}(\phi^l,\partial_z\phi^i, \phi^{\bar l};\lambda_a; \psi^{\bar j})=f_{i,\bar j_1,\dots,\bar j_k}^{a_1, \dots, a_q}(\phi^l,\phi^{\bar l}) \partial_z\phi^i \psi^{\bar j_1}\dots\psi^{\bar j_k}\lambda_{a_1} \dots \lambda_{a_q}$ that is linear in $\partial_z\phi^i$ and does not depend on any other derivatives. It is a $(0,k)$-form on $X$ with values in the tensor product bundle of $T^*X$ with $\Lambda^q {\cal E}$; alternatively, it is a $(1,k)$-form on $X$ with values in the bundle $\Lambda^q {\cal E}$. Similarly, in the second case, we have an operator ${\cal F}(\phi^l, \phi^{\bar l}, \partial_z \phi^{\bar s};\lambda_a ; \psi^{\bar j})=f_{\bar j_1,\dots,\bar j_k}^{i; a_1, \dots, a_q}(\phi^l, \phi^{\bar l}) g_{i \bar s}\partial_z \phi^{\bar s}\psi^{\bar j_i}\dots \psi^{\bar j_k} \lambda_{a_1} \dots \lambda_{a_q}$ that is linear in $\partial_z \phi^{\bar s}$ and does not depend on any other derivatives. It is a $(0,k)$-form on $X$ with values in the tensor product bundle of $TX$ with $\Lambda^q {\cal E}$. In the third case, we have an operator ${\cal F}(\phi^l, \phi^{\bar l};\lambda_a, \partial_z \lambda_a; \psi^{\bar j})=  f_{\bar b; \bar j_1,\dots,\bar j_k }^{a_1, \dots, a_q}(\phi^l,\phi^{\bar l}) h^{\bar b a}\partial_z\lambda_{a} \psi^{\bar j_1}\dots\psi^{\bar j_k} \lambda_{a_1} \dots \lambda_{a_q}$ that is linear in $\partial_z\lambda_{a}$ and does not depend on any other derivatives. Such an operator corresponds to a $(0,k)$-form on $X$ with values in the  (antisymmetric)   tensor product bundle of ${\overline {E}}$ with $\Lambda^q {\cal E}$, where the local holomorphic sections of the bundle $E$   are spanned by $\partial_z \lambda_a$. In the last case, we have an operator ${\cal F} (\phi^l, \phi^{\bar l}; \lambda_{a}, \lambda^a_z ; \psi^{\bar j})=f_{a; \bar j_1,\dots,\bar j_k}^{a_1, \dots, a_q}(\phi^l, \phi^{\bar l}) \lambda^a_{z}\psi^{\bar j_i}\dots \psi^{\bar j_k} \lambda_{a_1} \dots \lambda_{a_q}$; here, $\cal F$ may depend on $\phi^i$, $\phi^{\bar i}$, $\lambda_{a}$ and $\lambda^a_z$, but not on their derivatives. Such an operator corresponds to a $(0,k)$-form on $X$ with values in the (antisymmetric) tensor product bundle of $\cal E^{\vee}$ with $\Lambda^q{\cal E}$. In a similiar fashion, for any integer $n>0$, the operators of dimension $(n,0)$ and charge $q_R = k$ can be interpreted as $(0,k$)-forms with values in a certain tensor product bundle over $X$. This structure persists in quantum perturbation theory, but there  may be  perturbative corrections to the complex structure of the bundle.

The action of $\overline Q_+$ on such operators can be easily described at the classical level. If we interpret $\psi^{\bar i}$ as $d\phi^{\bar i}$, then $\overline Q_+$ acts on functions of $\phi^i$ and $\phi^{\bar i}$, and  is simply the $\bar\partial $ operator on $X$. This follows from the transformation laws $\delta\phi^{\bar i}=\psi^{\bar i}$, ${\delta\phi^i} = 0$, ${\delta \psi^{\bar i}}=0$, and (on-shell) $\delta \lambda_{a} = \delta \lambda^a_z = 0$. Note that if the holomorphic vector bundle $\cal E$ has vanishing curvature, the interpretation of $\overline Q_+$ as the $\bar\partial$ operator will remain valid when $\overline Q_+$ acts on a more general operator ${\cal F}(\phi^i,\partial_z\phi^i,\dots;\phi^{\bar i},\partial_z \phi^{\bar i},\dots; \lambda_{a}, \dots ; \lambda^a_z, \dots; \psi^{\bar i})$ that does depend on the derivatives of $\phi^i$ and $\phi^{\bar i}$. The reason for this is that if $\cal E$ is a trivial bundle with zero curvature, we will have the equation of motion $D_z \psi^{\bar i}=0$.  This means that one can neglect the action of $\overline Q_+$ on derivatives $\partial_z^m\phi^{\bar i}$ with $m>0$. On the other hand, if $\cal E$ is a non-trivial     holomorphic vector bundle, $\overline Q_+$ will only act as the $\bar \partial$ operator on  physical  operators that $\it{do}$ $\it{not}$ contain the derivatives $\partial_z^m\phi^{\bar i}$ with $m>0$.      

Perturbatively, there will be corrections to the action of $\overline Q_+$. In fact, as briefly mentioned in section 3.1 earlier, (\ref{tzzanomaly}) provides such an example - the holomorphic stress tensor $T_{zz}$, though not corrected at 1-loop, is no longer $\overline Q_+$-closed because the action of $\overline Q_+$ has received perturbative corrections. Let us now attempt to better understand the nature of such perturbative corrections. To this end, let $Q_{cl}$  denote the classical approximation to $\overline Q_+$. The perturbative corrections in $\overline Q_+$ will then modify the classical expression $Q_{cl}$. Note that since sigma model perturbation theory is local on $X$, and it depends on an expansion of fields such as the metric tensor of $X$ in a Taylor series up to some given order, the perturbative corrections to $Q_{cl}$ will also be local on $X$, where order by order, they consist  of differential operators whose possible degree grows with the order of perturbation theory.   

Let us now perturb the classical expression $Q_{cl}$ so that $\overline Q_+ = Q_{cl} + \epsilon Q' + O(\epsilon^2) $, where $\epsilon$ is a parameter that controls the magnitude of the perturbative quantum corrections at each order of the expansion. To ensure that we continue to have ${\overline Q}_+^2 = 0$, we require that $\{Q_{cl}, Q' \} = 0$. In addition, if $Q'=\{Q_{cl},\Lambda\}$ for some $\Lambda$, then via the conjugation of $\overline Q_+$ with $\exp(-\epsilon\Lambda)$ (which results in a trivial change of basis in the space of $\overline Q_+$-closed local operators), the correction by $Q'$ can be removed. Hence, $Q'$ represents a $Q_{cl}$-cohomology class. Since $Q'$ is to be generated in sigma model perturbation theory, it must be constructed locally from the fields appearing in the sigma model action. 

It will be useful for later if we discuss the case when $\cal E$ is a trivial bundle now. In such a case, $Q_{cl}$ will always act as the $\bar \partial$ operator as argued above. In other words, perturbative corrections to $\overline Q_+$ will come from representatives of $\bar \partial$-cohomology classes on $X$. An example would be the Ricci tensor in (\ref{tzzanomaly}) which represents a $\bar\partial$-cohomology class in $H^1(X,T^*X)$. It  is also constructed locally from the metric of $X$, which appears in the action. Hence, it satisfies the conditions required of a perturbative correction $Q'$. Another representative of a $\bar\partial$-cohomology class on $X$ which may contribute as a perturbative correction to the classical expression $\overline Q_+ = Q_{cl}$, would be an element of $H^1(X, \Omega^{2, cl}_X)$. It is also constructed locally from fields appearing in the action $S_{\mathrm pert}$, and is used to deform the action. In fact, its interpretation as a perturbative correction $Q'$ is consistent with its interpretation as the moduli of the chiral algebra. To see this, notice that its interpretation as $Q'$ means that it will parameterise a family of $\overline Q_+ = Q_{cl} + \epsilon Q'$ operators at the quantum level. Since the chiral algebra of local operators is defined to be closed with respect to the $\overline Q_+$ operator, it will vary with the $\overline Q_+$ operator  and consequently with $H^1(X, \Omega^{2,cl}_X)$, i.e., one can associate the moduli of the chiral algebra with $H^1(X, \Omega^{2,cl}_X)$. Apparently, these classes are the only one-dimensional $\bar \partial$-cohomology classes on $X$ that can be constructed locally from fields appearing in the action, and it may be that they completely determine
the perturbative corrections to $\overline Q_+ = Q_{cl}$.\footnote{Since we are considering a  holomorphic vector bundle $\cal E$ whose curvature two-form vanishes in this case, the second term of $\Delta_{1-loop}$ in (\ref{1-loop}) will be zero. Consequently, only the first term on the RHS of (\ref{1-loop}) remains. In other words, only $R_{i \bar j}$ will contribute to the correction of $Q_{cl}$ from $\Delta_{1-loop}$. Since an element of $H^1(X, \Omega^{2,cl}_X)$ is the only other $\bar \partial$-cohomology class which can appear in the quantum action, it would contribute as the only other perturbative correction to $Q_{cl}$.} The observations  in this paragraph will be important in section 5.4, when we discuss the $\overline Q_+$-cohomology of local operators  (on a small open set $U \subset X$) furnished by a sheaf of vertex superalgebras associated with a free $bc$-$\beta\gamma$ system.
 
The fact that $\overline Q_+$ does not always act as the $\bar \partial$ operator even at the classical level, seems to suggest that one needs a more general framework than just ordinary Dolbeault or $\bar \partial$-cohomology to describe the $\overline Q_+$-cohomology of the twisted heterotic sigma model. Indeed, as we will show shortly in section 5.3, the appropriate description of the $\overline Q_+$-cohomology of local operators spanning the chiral algebra will be given in terms of the more abstract notion of Cech cohomology.

\newsubsection{ \it A Topological Chiral Ring}

Next, let us make an interesting and relevant observation about the ground operators in the $\overline Q_+$-cohomology. Note that we had already shown in section 3.1, that the $\overline Q_+$-cohomology of operators has the structure of a chiral algebra with holomorphic operator product expansions. Let the local operators of the $\overline Q_+$-cohomology  be given by ${\cal F}_a$, ${\cal F}_b$, $\dots$ with scaling dimensions $(h_a, 0)$, $(h_b, 0)$, $\dots$. By holomorphy, and the conservation of scaling dimensions and $U(1)_L \times U(1)_R$ charges, the OPE of  these  operators take the form 
\be
{{\cal F}_a (z) {\cal F}_b (z')} = {\sum_{q_c =  q_a + q_b} { {C_{abc}\ {\cal F}_c(z')} \over {(z- z')^{h_a + h_b -h_c}} } },
\label{OPEab}
\ee 
where we have represented the $U(1)_L \times U(1)_R$  charges $(q_L, q_R)$ of the operators ${\cal F}_a$, ${\cal F}_b$ and ${\cal F}_c$ by $q_a$, $q_b$ and $q_c$ for brevity of notation. Here, $C_{abc}$ is a structure constant that is (anti)symmetric in the indices. If ${\widetilde {\cal F}}_a$ and ${\widetilde {\cal F}}_b$ are ground operators of dimension $(0,0)$, i.e., $h_a = h_b =0$, the OPE will then be given by
\be
{{\widetilde {\cal F}}_a (z) {\widetilde {\cal F}}_b (z')} = {\sum_{q_c =  q_a + q_b} { {C_{abc}\ {{\cal F}}_c(z')} \over {(z- z')^{-h_c}} } }.
\label{OPEc}
\ee   
Notice that the RHS of (\ref{OPEc}) is only singular if $h_c < 0$. Also recall that all physical operators in the $\overline Q_+$-cohomology cannot have negative scaling dimension, i.e., $h_c \geq 0$.\footnote{As mentioned in the footnote 4, for an operator of classical dimension $(n, m)$, anomalous dimensions due to RG flow may shift the values of $n$ and $m$ in the quantum theory. However, the spin $n-m$ remains unchanged. Hence, since the operators in the $\overline Q_+$-cohomology of the quantum theory will continue to have $m =0$ (due to a $\overline Q_+$-trivial anti-holomorphic stress tensor $T_{\bar z \bar z}$ at the quantum level),  the value of $n$ is unchanged as we go from the classical to the quantum theory, i.e., $n\geq 0$ holds even at the quantum level.} Hence, the RHS of (\ref{OPEc}), given by $(z-z')^{h_c} {\cal F}_c (z')$, is non-singular as $z \to z'$, since a pole does not exist. Note that $(z-z')^{h_c}  {\cal F}_c (z')$ must also be annihilated by $\overline Q_+$ and be in its cohomology, since ${\widetilde {\cal F}}_a$ and ${\widetilde {\cal F}}_b$ are. In other words, we can write ${\widetilde {\cal F}}_c (z, z') = (z-z')^{h_c} {\cal F}_c (z')$, where ${\widetilde {\cal F}}_c (z, z')$ is a dimension $(0,0)$ operator that represents a $\overline Q_+$-cohomology class.  Thus, we can express the OPE of the ground operators as

\be
{{\widetilde {\cal F}}_a (z) {\widetilde {\cal F}}_b (z')} = {\sum_{q_c =  q_a + q_b}  C_{abc} \ {\widetilde {\cal F}}_c(z , z')}.
\label{OPEgnd}
\ee
Since the only holomorphic functions without a pole on a Riemann surface are constants, it will mean that the operators $\widetilde {\cal F}$ are independent of the coordinate `$z$' on $\Sigma$. Hence, they are completely independent of their insertion points and the metric on $\Sigma$. Therefore, we conclude that the ground operators of the $\overline Q_+$-cohomology define a $\it{topological}$ chiral ring via their OPE  
\be
{{\widetilde {\cal F}}_a  {\widetilde {\cal F}}_b } = {\sum_{q_c =  q_a + q_b}  C_{abc} \ {\widetilde {\cal F}}_c}.
\label{OPEgnd1}
\ee

In perturbation theory, the chiral ring will have a ${\mathbb Z} \times {\mathbb Z}$ grading by the $U(1)_L \times U(1)_R$ charges of the operators. However, since each charged, anti-commuting, fermionic field cannot appear twice in the same operator,  each operator will consist of only a finite number of them. Consequently, the individual ${\mathbb Z}$ grading will be reduced mod 2 to $\mathbb Z_2$, such that the ring is effectively ${\mathbb Z_2} \times {\mathbb Z_2}$ graded. Non-perturbatively, due to worldsheet instantons, the continuous $U(1)_L \times U(1)_R$ symmetry is reduced to a discrete subgroup. In order for this discrete symmetry to be non-anomalous, the values of the corresponding $U(1)_L \times U(1)_R$ charges can only be fractional multiples of $\pi$. More precisely, from the relevant index theorems, we find that the initial ${\mathbb Z} \times {\mathbb Z}$ grading by the $U(1)_L \times U(1)_R$ charges will be reduced to $\mathbb Z_{2p} \times \mathbb Z_{2k}$ by worldsheet instantons, where $2p$ and $2k$ are the greatest divisors of $c_1(\cal E)$ and $c_1(TX)$ respectively. 

At the classical level (i.e. in the absence of perturbative corrections),   $\overline Q_+ = Q_{cl}$ will act on a dimension $(0,0)$ operator (i.e., one that does not contain the derivatives $\partial_z^m\phi^{\bar i}$ with $m>0$) as the $\bar \partial$ operator. Moreover, recall that any dimension $(0,0)$ operator $\widetilde {\cal F}_d$ with $(q_L, q_R) = (q,k)$, will correspond to an ordinary $(0,k)$-form $f_{\bar j_1,\dots,\bar j_k}(\phi^i, \phi^{\bar i})d\phi^{\bar j_1}\wedge \dots \wedge d\phi^{\bar j_k}$ on $X$ with values in the bundle $\Lambda^q {\cal E}$.  Hence, via the Cech-Dolbeault isomorphism in ordinary differential geometry, the classical ring is just the graded Cech cohomology ring $H^{*}(X, \Lambda^{*}{\cal E})$. In any case, the operators will either be non-Grassmannian or Grassmannian, obeying either commutators or anti-commutators, depending on whether they contain an even or odd number of fermionic fields.

\newsubsection{A Sheaf of Chiral Algebras}

We shall now explain the idea of a ``sheaf of chiral algebras'' on $X$. To this end, note that both the $\overline Q_+$-cohomology of local operators (i.e., operators that are local on the Riemann surface $\Sigma$), and the fermionic symmetry generator $\overline Q_+$, can be described locally on $X$. Hence, one is free to restrict the local operators to be well-defined not throughout $X$, but only on a given open set $U \subset X$. Since in perturbation theory, we are considering trivial maps $\Phi :\Sigma \to X$ with no multiplicities, an operator defined in an open set $U$ will have a sensible operator product expansion with another operator defined  in $U$. From here, one can naturally proceed to restrict the definition of the operators to smaller open sets, such that  a global definition of the operators can be obtained by gluing together the open sets on their unions and intersections. From this description, in which one associates a chiral algebra, its OPE's, and chiral ring to every open set $U \subset X$, we get what is known mathematically as a ``sheaf of chiral algebras''. We shall call this sheaf $\widehat {\cal A}$.

\bigskip \noindent{\it Description of $\cal A$ via Cech Cohomology}

In perturbation theory, one can also describe    the  $\overline Q_+$-cohomology   classes by a form of Cech cohomology. This alternative description will take us to the mathematical point of view on the subject \cite{MSV1, GMS1, GMS3}. In essence, we will show that the chiral algebra $\cal A$ of the $\overline Q_+$-cohomology classses of the twisted heterotic sigma model on a  holomorphic vector bundle $\cal E$ over $X$, can be represented, in perturbation theory, by the classes of the Cech cohomology     of the sheaf $\widehat {\cal A}$ of locally-defined chiral operators. To this end, we shall generalise the argument in section 3.2 which provides a Cech cohomological description of a $\bar \partial$-cohomology, to demonstrate an isomorphism  between the $\overline Q_+$-cohomology classes and the classes of the Cech cohomology of $\widehat {\cal A}$.

Let us start by considering an open set $U \subset X$ that is isomorphic to a contractible space such as an open ball in $\mathbb C^n$, where $n = {\textrm {dim}}_{\mathbb C} (X)$. Because $U$ is a contractible space, any bundle over $U$ will be trivial. By applying this statement on the holomorphic vector bundle $\cal E$ over $U$, we find that the curvature of $\cal E$ vanishes. From the discussion in section 5.1, we find that $\overline Q_+$ will then act as the $\bar \partial$ operator on any local operator $\cal F$ in $U$. In other words, $\cal F$ can be interpreted as a $\bar\partial$-closed $(0,k)$-form with values in a certain  tensor product bundle $\widehat F$ over $U$. Thus, in the absence of perturbative corrections at the classical level, any operator ${\cal F}$ in the $\overline Q_+$-cohomology will be classes of $H^{0,k}_{\bar \partial}(U, \widehat {F})$ on $U$. As explained, $\widehat F$ will also be a trivial bundle over $U$, which means that $\widehat {F}$ will always possess a global section, i.e., it corresponds to a soft sheaf.  Since the higher Cech cohomologies of a soft sheaf are trivial \cite{Wells}, we will have $ {H_{\textrm{Cech}}^{k} (U, {\widehat {F}} )} = 0$ for $k > 0$. Mapping this back to Dolbeault cohomology via the Cech-Dolbeault isomorphism, we find that $H^{0,k}_{\bar \partial}(U, \widehat {F}) = 0$ for $k > 0$.  Note that small quantum corrections in the perturbative limit can only annihilate cohomology classes and not create them. Hence, in perturbation theory, it follows that the local operators ${\cal F}$  with positive values of $q_R$,  must vanish in    $\overline Q_+$-cohomology on $U$.     

Now consider a good cover of  $X$ by open sets $\{U_a \}$.  Since the intersection of open sets $\{U_a \}$ also give open sets (isomorphic to open balls in $\mathbb C^n$),  $\{U_a \}$ and all of their intersections have the same property as $U$    described above: $\bar\partial$-cohomology and hence $\overline Q_+$-cohomology vanishes for positive values of $q_R$ on $\{U_a \}$ and their intersections.  

Let the operator ${\cal F}_1$ on $X$ be a $\overline Q_+$-cohomology class with $q_R = 1$. It is here that we shall import the usual arguments relating a $\bar\partial$ and Cech cohomology, to demonstrate an isomorphism between the $\overline Q_+$-cohomology and a Cech cohomology.  When restricted to an open set        $U_a$, the operator ${\cal F}_1$ must be trivial in $\overline Q_+$-cohomology, i.e., ${\cal F}_1 =\{\overline Q_+,{\cal C}_a\}$, where ${\cal C}_a$ is an operator of $q_R=0$ that is well-defined in $U_a$. 

Now, since $\overline Q_+$-cohomology classes such as ${\cal F}_1$ can be globally-defined on $X$, we have ${\cal F}_1 =\{\overline Q_+ ,{\cal C}_a\}=\{\overline Q_+,{{\cal C}_b}\}$ over the intersection  $U_a\cap U_b$, so $\{\overline Q_+,{{\cal C}_a}- {{\cal C}_b}\}=0$. Let
${\cal C}_{ab}= {{\cal C}_a}- {{\cal C}_b}$.  For each $a$ and $b$, ${\cal C}_{ab}$ is defined in
$U_a\cap U_b$. Therefore, for all $a,b,c$, we have
\be
{\cal C}_{ab}=     -{\cal C}_{ba}, \quad {{\cal C}_{ab}}+ {{\cal C}_{bc}} + {{\cal C}_{ca}} =0.
\label{cab}
\ee
Moreover, for ($q_R =0$) operators ${\cal K}_a$ and ${\cal K}_b$, whereby $\{ \overline Q_+, {{\cal K}_a} \} = \{ \overline Q_+, {{\cal K}_b} \}= 0$, we have an equivalence relation 
\be
{\cal C}_{ab} \sim  {{\cal C'}_{ab} = {{\cal C}_{ab} + {\cal K}_a - {\cal K}_b}}.
\label{cab1}
\ee
Note that the collection $\{{\cal C}_{ab} \}$ are operators in the $\overline Q_+$-cohomology with well-defined operator product expansions, and whose dimension $(0,0)$ subset furnishes a topological chiral ring with $q_R =0$.  
 
Since the local operators with positive values of $q_R$ vanish in $\overline Q_+$-cohomology on an arbitrary open set $U$, the sheaf $\widehat {\cal A}$ of the chiral algebra of operators has for its local sections  the $\psi^{\bar i}$-independent (i.e. $q_R =0$) operators  ${\widehat {\cal F}} (\phi^i,\partial_z\phi^i,\dots; \partial_z\phi^{\bar i},\dots; \lambda_{a}, \partial_z \lambda_{a}, \dots; \lambda^a_z, \partial_z \lambda^a_z, \dots)$ 
that are annihilated by $\overline Q_+$. Each ${\cal C}_{ab}$ with $q_R =0$ is thus a section of
$\widehat{\cal A}$ over the intersection $U_a\cap U_b$. From (\ref{cab}) and (\ref{cab1}), we find that the collection $\{{\cal C}_{ab} \}$ defines the elements  of the first Cech cohomology group $H_{{\rm Cech}}^1(X, \widehat{\cal A})$.

Next, note that the $\overline Q_+$-cohomology classes are defined as those operators which are $\overline Q_+$-closed, modulo those which can be globally written as $\{ \overline Q_+, \dots \}$ on $X$. In other words, ${\cal F}_1$ vanishes in $\overline Q_+$-cohomology if we can write it as ${\cal F}_1 = \{\overline Q_+ ,{\cal C}_a\}=\{\overline Q_+,{{\cal C}_b}\} =  \{\overline Q_+ ,{\cal C}\}$, i.e., ${\cal C}_a = {\cal C}_b$ and hence ${\cal C}_{ab} = 0$. Therefore, a vanishing $\overline Q_+$-cohomology with $q_R =1$ corresponds to a vanishing first Cech cohomology. Thus, we have obtained a map between the $\overline Q_+$-cohomology with $q_R =1$ and a first Cech cohomology.      

Similar to the case of relating a $\bar \partial$ and Cech cohomology,  one can also run everything backwards and construct an inverse of this map. Suppose we are given a family $\{ {\cal C}_{ab} \}$ of sections of $\widehat {\cal A}$ over the corresponding intersections $\{U_a \cap U_b \}$, and they obey (\ref{cab}) and (\ref{cab1}) so that they define the elements of $H^1(X , \widehat{\cal A})$. We can then proceed as follows.  Let the set $\{f_a \}$ be partition of unity subordinates to the open cover of $X$ provided by $\{U_a\}$. This means that the elements of $\{f_a \}$ are continuous functions on $X$, and they  vanish outside the corresponding elements in  $\{U_a \}$ whilst obeying $\sum_a f_a=1$. Let ${\cal F}_{1,a}$ be a chiral operator defined in $U_a$ by ${\cal F}_{1,a}= \sum_c[ \overline Q_+, f_c]  {\cal C}_{ac}$.\footnote{Normal ordering of the operator product
of $[\overline Q_+, f_c(\phi^i,\phi^{\bar i})]$ with ${\cal C}_{ac}$ is needed for regularisation purposes.} ${\cal F}_{1,a}$ is well-defined throughout $U_a$, since in $U_a$, $[\overline Q_+, f_c]$ vanishes wherever ${\cal C}_{ac}$ is not defined. Clearly, ${\cal F}_{1,a}$ has $q_R =1$, since ${\cal C}_{ac}$ has $q_R =0$ and $\overline Q_+$ has $q_R=1$. Moreover, since ${\cal F}_{1,a}$ is a chiral operator defined in $U_a$, it will mean that $\{ \overline Q_+, {\cal F}_{1,a}\} = 0$  over $U_a$. For any $a$ and $b$, we have ${\cal F}_{1,a}- {\cal F}_{1,b} = \sum_c [\overline Q_+ , f_c]  ({\cal C}_{ac}- {\cal C}_{bc})$. Using (\ref{cab}), this is $\sum_c [\overline Q_+ , f_c] {\cal C}_{ab} = [\overline Q_+, \sum_c f_c ] {\cal C}_{ab}$. This vanishes since $\sum_c f_c = 1$. Hence, ${\cal F}_{1,a} = {\cal F}_{1,b}$ on $U_a \cap U_b$, for $\it{all}$ $a$ and $b$. In other words, we have found a globally-defined $q_R =1$ operator ${\cal F}_1$ that obeys $\{\overline Q_+, {\cal F}_1 \} = 0$ on $X$. Notice that ${\cal F}_{1,a}$ and thus ${\cal F}_1$ is not defined to be of the form $\{\overline Q_+, \dots \}$. Therefore, we have obtained a map from the Cech cohomology group $H^1(X, \widehat {\cal A})$ to the $\overline Q_+$-cohomology group with $q_R =1$, i.e., $\overline Q_+$-closed $q_R =1 $ operators modulo those that can be globally written as $\{\overline Q_+, \dots \}$. The fact that this map is an inverse of the first map can indeed be verified.      

Since there is nothing unique about the $q_R =1$ case, we can repeat the above procedure for operators with $q_R > 1$. In doing so, we find that  the $\overline Q_+$-cohomology coincides with the Cech cohomology of $\widehat {\cal A}$ for all $q_R$. Hence, the chiral algebra $\cal A$ of the twisted heterotic sigma model will be given by  $\bigoplus_{q_R} H^{q_R}_{\textrm {Cech}} (X, {\widehat{\cal A}})$ as a vector space. As there will be no ambiguity, we shall henceforth omit the label ``Cech'' when referring to the cohomology of $\widehat {\cal A}$.

Note that in the mathematical literature, the sheaf $\widehat {\cal A}$, also known as a sheaf of vertex superalgebras, is studied purely from the Cech viewpoint; the field $\psi^{\bar i}$ is omitted and locally on $X$, one considers operators constructed only from $\phi^i$, $\phi^{\bar i}$, $\lambda_{a}$, $\lambda^a_z$ and their  $z$-derivatives. The chiral algebra $\cal A$ of $\overline Q_+$-cohomology classes with positive $q_R$ are correspondingly constructed as Cech $q_R$-cocycles. However, in the physical description via a Lagrangian and $\overline Q_+$ operator, the sheaf  $\widehat {\cal A}$ and its cohomology are given a $\bar \partial$-like description, where Cech $q_R$-cycles are represented by operators that are $q^{th}_R$ order in the field $\psi^{\bar i}$. Notice that the mathematical description does not involve any form of perturbation theory at all. Instead, it utilises the abstraction of Cech cohomology to define the spectrum of operators in the quantum sigma model. It is in this sense that the study of the sigma model is given a rigorous foundation in the mathematical literature.       

\bigskip \noindent{\it The Constraint $\Lambda^r {\cal E} \cong K_X$}

In a physical heterotic string compactification on a gauge bundle $\cal E$ over a space $X$, the  charged massless RR states are represented (in the perturbative limit, ignoring worldsheet instantons) by classes in the Cech cohomology group \cite{Greene}
\be
H^q (X, \Lambda^p {\cal E}), 
\label{RR}
\ee
and  the corresponding vertex operators representing these states contain $p$ left-moving and $q$ right-moving fermi fields. Notice that the classes of (\ref{RR}) can be represented by the dimension (0,0) local operators of the $\overline Q_+$-cohomology in the twisted heterotic sigma model with $U(1)_L \times U(1)_R$ charge $(p, q)$. It is here that the physical relevance of  the sigma model is readily manifest.    

In the context of the physical heterotic string with $(0,2)$ worldsheet supersymmetry, one can sometimes speak sensibly of a heterotic chiral ring. This ring is described additively by the sum of Cech cohomology groups of the form in (\ref{RR}) above, i.e., 
\be
 H^{*,*}_{\textrm{het}} = \sum_{p,q} H^q (X, \Lambda^p {\cal E}).
\label{RR ring}  
\ee
Note that Serre duality in $(0,2)$ theories require that states in $H^{*,*}_{\textrm{het}}$ be dual to other states in $H^{*,*}_{\textrm{het}}$ \cite{Greene}. Serre duality acts as 
\begin{eqnarray}
\label{Serre}
H^i (X, \Lambda^j {\cal E}) & \cong & H^{n-i} (X, \Lambda^j{\cal E}^{\vee} \otimes K_X)^* \nonumber \\
& \cong & H^{n-i} (X, \Lambda^{r-j} {\cal E} \otimes \Lambda^r {\cal E}^{\vee} \otimes K_X)^*,
\end{eqnarray}
where $n = {\textrm {dim}}_{\mathbb C} X$ and $r$ is the rank of $\cal E$. $K_X$ is simply the canonical bundle of $X$ (i.e. the bundle over $X$ whose holomorphic sections are $(n,0)$-forms on $X$). Hence, from (\ref{Serre}), the states of $H^{*,*}_{\textrm{het}}$ only close back onto themselves under a duality relation if and only if the line bundle $\Lambda^r {\cal E}^{\vee} \otimes K_X$ on $X$ is trivial, i.e., $\Lambda^r {\cal E} \cong K_X$. Thus, if the twisted heterotic sigma model is to be physically relevant such as to have a geometrical background that is consistent with one that will be considered in the actual, physical heterotic string theory, this constraint needs to be imposed. In fact, $\Lambda^r {\cal E} \cong K_X$ implies $c_1(TX)=-c_1({\cal E})$. This condition on the first Chern class of the bundles is simply the first anomaly cancellation condition in (\ref{an}) for a general worldsheet $\Sigma$.

\newsubsection{Relation to a Free $bc$-$\beta\gamma$ System}

Now, we shall express in a physical language a few key points that are made in the mathematical literature \cite{GMS1, GMS3} starting from a Cech viewpoint.  Let us start by providing a convenient description of the local structure of the sheaf $\widehat {\cal A}$. To this end, we will describe in a new way the $\overline Q_+$-cohomology of operators that are regular in a small open set $U \subset X$. We assume that $U$ is isomorphic to an open ball in $\mathbb C^n$ and is thus contractible. 

Notice from $S_{\mathrm pert}$ in (\ref{Spert}) and $V$ in (\ref{chi}), that the hermitian metric on $X$ and the fibre metric of $\cal E$ (implicit in the second term $\lambda_{a} l^a_{z\bar z}$ of $V$), only appear inside a term of the form $\{\overline Q_+, \dots \}$  in the action. Thus, any shift in the metrics will also appear inside $\overline Q_+$-exact (i.e. $\overline Q_+$-trivial) terms. Consequently, for our present purposes, we can arbitrarily redefine the values of the hermitian metric on $X$ and the fibre metric of $\cal E$, since they do not affect the analysis of the $\overline Q_+$-cohomology. Therefore, to describe the local structure, we can pick a hermitian metric that is flat when restricted to $U$. Similarly, we can pick a fibre metric of $\cal E$  that is flat over $U$ as well. In fact, this latter choice is automatically satisfied in $U$ - the bundle $\cal E$ over a contractible space $U$ is trivial. The action, in general, also contains terms derived from an element of $H^1(X,\Omega^{2,cl}_X)$, as we explained in section 3.2.  From (\ref{ST}), we see that these terms are also $\overline Q_+$-exact locally, and therefore can be discarded in analysing the local structure in $U$. Thus, the local action (derived from the flat fibre and hermitian metric) of the twisted heterotic sigma model on ${\cal E}_f \times U$ (where ${\cal E}_f$ denotes the fibre space of $\cal E$) is 
\be
I = {1 \over 2 \pi} \int_{\Sigma} |d^2 z| \sum_{i, \bar j} \delta_{i \bar j} \left ( \partial_z \phi^{\bar j} \partial_{\bar z}\phi^i +   \psi ^i_{\bar z} \partial_z \psi^{\bar j} \right ) + \sum_{a,\bar b}  \delta_{a \bar b}\lambda^{\bar b} \partial_{\bar z} \lambda^a_z,
\label{Su}
\ee 
where $\lambda^{\bar b}$ is a scalar on $\Sigma$ with values in the pull-back bundle $\Phi^*(\overline {\cal E})$, such that for an arbitrary fibre metric $h_{a\bar b}$, we have $\lambda_{a} = h_{a \bar b} \lambda^{\bar b}$.   
       
Now let us describe the $\overline Q_+$-cohomology classes of operators regular in $U$.  As explained earlier, these are operators of dimension $(n,0)$ that are independent of $\psi^{\bar i}$. In general, such operators  are of the form ${\widehat {\cal F}} (\phi^i,\partial_z\phi^i,\dots; \phi^{\bar i},\partial_z \phi^{\bar i},\dots; \lambda_{a}, \partial_z \lambda_{a}, \dots ; \lambda^a_z, \partial_z \lambda^a_z, \dots)$. Note that since $\cal E$ has vanishing curvature over $U$, from the discussion in section 5.1, we see that $\overline Q_+$ will act  as the $\bar \partial$ operator at the classical level. In this case, the $\overline Q_+$ operator can receive perturbative corrections from $\bar\partial$-cohomology classes such as the Ricci tensor and classes in $H^1(X,\Omega^{2,cl}_X)$. However, note that since we have picked a flat hermitian metric on $U$, the corresponding Ricci tensor on $U$ is zero. Moreover, as explained above, classes from $H^1(X, \Omega^{2,cl}_X)$ do not contribute  when analysing the $\overline Q_+$-cohomology on $U$. Hence, we can ignore the perturbative corrections to $\overline Q_+$ for our present purposes. Therefore, on the classes of operators in $U$, $\overline Q_+$ acts as $\bar \partial =\psi^{\bar i}\partial/\partial\phi^{\bar i}$, and the condition that $\widehat {\cal F}$ is annihilated by $\overline Q_+$ is precisely that, as a function of $\phi^i$, $\phi^{\bar i}$, $\lambda_{a}$, $\lambda^a_z$ and their $z$-derivatives, it is independent of $\phi^{\bar i}$ (as opposed to its derivatives), and depends only on the other variables, namely $\phi^i$, $\lambda_{a}$, $\lambda^a_z$ and the derivatives of $\phi^i$, $\phi^{\bar i}$, $\lambda_{a}$ and $\lambda^a_z$.\footnote{We can again ignore the action of $\overline Q_+$ on $z$-derivatives of $\phi^{\bar i}$ because of the equation of motion $\partial_z\psi^{\bar i}=0$ and the symmetry transformation law $\delta \phi^{\bar i} = \psi^{\bar i}$.} Hence, the $\overline Q_+$-invariant operators are of the form ${\widehat {\cal F}}(\phi^i,\partial_z\phi^i,\dots;\partial_z\phi^{\bar i},\partial_z^2\phi^{\bar i},\dots ; \lambda_{a}, \partial_z \lambda_{a}, \partial_z^2 \lambda_{a}, \dots; \lambda^a_z, \partial_z \lambda^a_z, \partial_z^2 \lambda^a_z, \dots )$. In other words, the operators, in their dependence on the center of mass coordinate of the string whose worldsheet theory is the twisted heterotic sigma model, is holomorphic. The local sections of $\widehat {\cal A}$ are just given by the operators in the $\overline Q_+$-cohomology  of the local, twisted heterotic sigma model with action (\ref{Su}).

Let us set $\beta_i  =  \delta_{i \bar j} \partial_z \phi^{\bar j}$ and $\gamma^i = \phi^i$, whereby $\beta_i$ and $\gamma^i$ are bosonic operators of dimension $(1,0)$ and $(0,0)$ respectively. Next, let us set $\delta_{a\bar b} \lambda^{\bar b} = c_a$ and $\lambda^a_z = b^a$, whereby $b^a$ and $c_a$ are fermionic operators of dimension $(1,0)$ and $(0,0)$ accordingly. Then, the $\overline Q_+$-cohomology of operators regular in $U$ can be represented by arbitrary local functions of $\beta$, $\gamma$, $b$ and $c$, of the form ${\widehat {\cal F}} (\gamma, \partial_z \gamma, \partial_z^2 \gamma, \dots, \beta, \partial_z \beta, \partial_z^2 \beta, \dots, c, \partial_z c, \partial_z^2 c, \dots, b, \partial_z b, \partial_z^2 b, \dots)$. The operators $\beta$ and $\gamma$ have the operator products of a standard $\beta\gamma$ system.  The products $\beta\cdot\beta$ and
$\gamma\cdot\gamma$ are non-singular, while
\be
\beta_i(z)\gamma^j(z')=-{\delta^j_i \over z-z'}+{\rm regular}.
\ee

Similarly, the operators $b$ and $c$ have the operator products of a standard $bc$ system. The products $b \cdot b$ and $c \cdot c$ are non-singular, while   
\be
b^a(z) c_b(z')={\delta^a_{b}\over z-z'}+{\rm regular}.
\ee
These statements can be deduced from the flat action (\ref{Su}) by standard field theory methods. We can write down an action for the fields $\beta$, $\gamma$, $b$ and $c$, regarded as free elementary fields, which reproduces these OPE's.  It is simply the following action of a $bc$-$\beta\gamma$ system: 
\be
I_{bc \textrm{-} \beta\gamma}= {1\over
2\pi} \int_{\Sigma} |d^2z| \left ( \sum_i\beta_i \partial_{\bar z}\gamma^i + \sum_a b^a \partial_{\bar z} c_a \right).
\label{bcaction}
\ee
Hence, we find that the local $bc$-$\beta\gamma$ system above reproduces the $\overline Q_+$-cohomology of $\psi^{\bar i}$-independent operators of the sigma model on $U$ and their appropriate OPE's, i.e., the local sections of the sheaf $\widehat{\cal A}$.

At this juncture, one can make another important observation concerning the relationship between the local twisted heterotic sigma model with action (\ref{Su}) and the local version of the $bc$-$\beta\gamma$ system of (\ref{bcaction}). To begin with, note that the holomorphic stress tensor ${\widehat T}(z) = -2 \pi T_{zz}$ of the local, free field sigma model is given by
\be
{\widehat T}(z) =  - \delta_{i \bar j} \partial_z \phi^{\bar j}\partial_z \phi^i -  \lambda_z^{a} \partial_z \lambda_a
\label{T(z)}
\ee
(Here and below, normal ordering is understood for ${\widehat T}(z)$). Via the respective identification of the fields $\beta$ and $\gamma$ with $\partial_z \phi$ and $\phi$, $\lambda_{a}$ and $\lambda^a_z$ with $c_a$ and $b^a$, we find that ${\widehat T}(z)$ can be written in terms of the $b$ and $c$ fields as
\be
{\widehat T}(z) = - \beta_i \partial_z \gamma^i - b^a \partial_z c_a.
\label{Tz}
\ee 
${\widehat T}(z)$, as given by (\ref{Tz}), coincides with the holomorphic stress tensor of the local $bc$-$\beta \gamma$ system. Simply put, the twisted heterotic sigma model and the $bc$-$\beta\gamma$ system have the same $\it{local}$ holomorphic stress tensor. This means that locally on $X$ (and hence ${\cal E} \to X$), the sigma model and the $bc$-$\beta\gamma$ system have the same generators of general holomorphic coordinate transformations on the worldsheet.

One may now ask the following question: does the $bc$-$\beta\gamma$ system reproduce the $\overline Q_+$-cohomology of $\psi^{\bar i}$-independent operators and their respective OPE's globally on $X$, or only in a small open set $U$? Well, the $bc$-$\beta\gamma$ system will certainly reproduce the $\overline Q_+$-cohomology of $\psi^{\bar i}$-independent operators and their OPE's globally on $X$ if there is no obstruction to defining the system globally on $X$, i.e., one finds, after making global sense of the action (\ref{bcaction}), that the corresponding theory remains anomaly-free. Let's look at this more closely.  

First and foremost, the classical action (\ref{bcaction}) makes sense globally if we interpret the bosonic fields $\beta$, $\gamma$, and the fermionic fields $b$, $c$, correctly.  $\gamma$ defines a map $\gamma:\Sigma\to X$, and $\beta$ is a $(1,0)$-form on $\Sigma$ with values in the pull-back $\gamma^*(T^*X)$. The field $c$ is a scalar on $\Sigma$ with values in the pull-back $\gamma^*({\cal E}^{\vee})$, while the field $b$ is a $(1,0)$-form on $\Sigma$ with values in the pull-back $\gamma^* (\cal E)$. With this interpretation, (\ref{bcaction}) becomes the action of what one might call a non-linear $bc$-$\beta\gamma$ system. However, by choosing $\gamma^i$ to be local coordinates on a small open set $U\subset X$, and $c_a$ to be local sections of the pull-back $\gamma^* ({\cal E}^{\vee})$ over $U$, one can make the action linear. In other words, a local version of (\ref{bcaction}) represents the action of a linear $bc$-$\beta \gamma$ system. To the best of the author's knowledge, the non-linear $bc$-$\beta\gamma$ system with action (\ref{bcaction}) does not seem to have been studied anywhere in the physics literature. Nevertheless, the results derived in this paper will definitely serve to provide additional insights into future  problems involving the application of this non-linear $bc$-$\beta\gamma$ system.

Now that we have made global sense of the action of the $bc$-$\beta\gamma$ system at the classical level, let us move on to discuss what happens at the quantum level. The anomalies that enter in the twisted heterotic sigma model also appear in the non-linear $bc$-$\beta\gamma$ system.    Expand around a classical solution of the non-linear $bc$-$\beta\gamma$ system, represented by a holomorphic map $\gamma_0:\Sigma \to X$, and a section $c_0$ of the pull-back $\gamma_0^*({\cal E}^{\vee})$. Setting ${\gamma} =\gamma_0 +\gamma'$, and $c = c_0 + c'$, the action, expanded to quadratic order about this solution, is $(1/2\pi) \left [ (\beta , {\overline D \gamma'}) + (b, \overline D c') \right]$. $\gamma'$, being a deformation of the coordinate $\gamma_0$ on $X$, is a section of the pull-back $\gamma_0^* (TX)$. Thus, the kinetic operator of the $\beta$ and $\gamma$ fields is the $\overline D$ operator on sections of $ \gamma_0^*(TX)$; it is the complex conjugate of the $D$ operator whose anomalies we encountered in section 4.  Complex conjugation reverses the sign of the anomalies, but here the fields are bosonic, while in section 4, they were fermionic; this gives a second sign change.  
(Notice that the $D$ operator in section 4 acts on sections of the pull-back of the anti-holomorphic bundle $\overline {TX}$ instead of the holomorphic bundle $TX$. However, this difference is irrelevant with regard to anomalies since $ch_2(\overline E) = ch_2(E)$ for any holomorphic vector bundle $E$.) Next, since $c'$ is a deformation of $c_0$, it will be a section of the pull-back $\gamma_0^*({\cal E}^{\vee})$. The kinetic operator of the $b$ and $c$ fields is therefore the $\overline D$ operator on sections of $\gamma_0^*({\cal E}^{\vee})$. Now, introduce a spin structure on $\Sigma$, so that we can equivalently interpret $\overline D$ as the complex conjugate of the Dirac operator acting on sections $K^{-1/2} \otimes \gamma_0^*({\cal E}^{\vee})$. Using the same argument found in section 4, we find that by tensoring $K^{-1/2}$ with $\gamma_0^*({\cal E}^{\vee})$, one will get an additional term ${1\over 2}c_1(\Sigma) c_1({\cal E}^{\vee})$. However, since $\cal E$ is a complex vector bundle, we will have ${\cal E}^{\vee} = {\overline {\cal E}}$, and because $c_1(\overline {\cal E}) = - c_1(\cal E)$, the additional term can actually be written as $-{1\over 2} c_1(\Sigma)c_1(\cal E)$. Moreover, we also have $ch_2({\cal E}^{\vee}) = ch_2(\overline {\cal E}) = ch_2(\cal E)$. Thus, the anomalies due to the kinetic operator of the $b$ and $c$ fields, are the same as those due to the $\overline D$ operator of section 4. Hence, the non-linear $bc$-$\beta\gamma$ system has exactly the same anomalies as the underlying twisted heterotic sigma model. And if the anomalies vanish, the $bc$-$\beta\gamma$ system will reproduce the $\overline Q_+$-cohomology of $\psi^{\bar i}$-independent operators and their OPE's globally on $X$, i.e., one can find a global section of $\widehat {\cal A}$.            

Via the identification of the various fields mentioned above, the left-moving fields $b^a$ and $c_a$ will have $U(1)_L$ charges $q_L = -1$ and $q_L = 1$ respectively. Notice that this $U(1)_L$ symmetry is nothing but the usual $U(1)$ R ghost number symmetry of the action (\ref{bcaction}) with the correct charges. However, note that the $bc$-$\beta\gamma$ system lacks the presence of right-moving fermions and thus the $U(1)_R$ charge $q_R$ carried by the fields $\psi^i_{\bar z}$ and $\psi^{\bar i}$ of the underlying twisted heterotic sigma model. Locally, the $\overline Q_+$-cohomology of the sigma model is non-vanishing only for $q_R =0$. Globally however, there can generically be cohomology in higher degrees. Since the chiral algebra of operators furnished by the linear $bc$-$\beta\gamma$ system gives the correct description of the   $\overline Q_+$-cohomology of $\psi^{\bar i}$-independent operators on $U$, one can then expect the globally-defined chiral algebra of operators furnished by the non-linear $bc$-$\beta\gamma$ system to correctly describe the $\overline Q_+$-cohomology classes of zero degree (i.e. $q_R =0$) on $X$. How then can one use the non-linear $bc$-$\beta\gamma$ system to describe the higher cohomology? The answer lies in the analysis carried out in section 5.3. In the $bc$-$\beta\gamma$ description, we do not have a close analog of $\bar \partial$ cohomology at our convenience. Nevertheless, we can use the more abstract notion  of Cech cohomology. As before, we begin with a good cover of $X$ by small open sets $\{U_a \}$, and, as explained in section 5.3, we can then describe the         $\overline Q_+$-cohomology classes of positive degree (i.e. $q_R > 0$) by Cech $q_R$-cocycles, i.e., they can be described by the $q^{th}_R$ Cech cohomology of the sheaf $\widehat {\cal A}$ of the chiral algebra of the linear $bc$-$\beta\gamma$ system with action being a linearised version of (\ref{bcaction}). Although unusual from a physicist's perspective, this Cech cohomology approach has been taken as a starting point for the present subject in the mathematical literature \cite{MSV1, MSV2, GMS1, GMS3}. Other more algebraic approaches to the subject have also been taken in \cite{BD}.     

Another issue that remains to be elucidated is the appearance of the respective moduli of the sigma model in the non-linear $bc$-$\beta\gamma$ system. Recall from section 3.2  that the moduli of the chiral algebra of the sigma model consists of the complex and holomorphic structure of $X$ and $\cal E$ respectively, as well as a class in $H^1(X, \Omega^{2, cl}_X)$. The complex and holomorphic structures are built into the the classical action (\ref{bcaction}) via the definition of the fields themselves. However, one cannot incorporate a class from $H^1(X, \Omega^{2,cl}_X)$ within the action in this framework. Nevertheless, as we will explain in section 5.6, the modulus represented by a class in $H^1(X, \Omega^{2, cl}_X)$ can be built into the definition of specific Cech cocycles through which one can define a family of sheaves of chiral algebras. This approach has also been taken in the mathematical literature \cite{GMS1, GMS3}.                      

A final remark to be made is that in the study of quantum field theory, one would like to be able to do more than just define the $\overline Q_+$-cohomology classes or a sheaf of chiral algebras. One would also like to be able to compute physically meaningful quantities such as the correlation functions of these cohomology classes of local operators. In the sigma model, the correlation functions can be computed from standard methods in quantum field theory. But at first sight, there seems to be an obstacle in doing likewise for the non-linear  $bc$-$\beta\gamma$ system. This can be seen as follows. Let the correlation function of $s$ local operators ${\cal O}_1$, ${\cal O}_2$, $\dots$, ${\cal O}_s$ on a genus $g$ Riemann surface $\Sigma$ be given by $\left < {\cal O}_1(z_1) \dots {\cal O}_s(z_s) \right>_g$, where ${\cal O}_i(z_i)$ has $U(1)_R$ charge $q_R = q_i$. Note that the $U(1)_R$ anomaly computation of (\ref{anomaly2}) in section 2.2 means that for the correlation functions of our model to be non-vanishing, they must satisfy $\sum_i q_i = n (1-g)$ in perturbation theory (in the absence of worldsheet instantons). Thus, generic non-zero correlation functions require that not all the $q_i$'s be zero. In particular, correlation functions at string tree level vanish unless $\sum_i q_i = n$, where $n ={\textrm dim}_{\mathbb C}X$. However, the operators of $q_i \neq 0$ cannot be represented in a standard way in the non-linear $bc$-$\beta\gamma$ system. They are instead described by Cech $q_i$-cocycles. This means that in order for one to compute the corresponding correlation functions using the non-linear $bc$-$\beta\gamma$ system, one must translate the usual quantum field theory recipe employed in the sigma model into a Cech language. The computation in the Cech language will involve cup products of Cech cohomology groups and their maps into complex numbers. An illuminating example would be to consider a computation of the  correlation function of dimension $(0,0)$ operators  on the sphere. To this end, first recall from section 5.1 that a generic dimension $(0,0)$ operator ${\cal O}_i$ with $U(1)_L \times U(1)_R$ charge $(p_i, q_i)$ can be interpreted as a $(0, q_i)$-form with values in the bundle $\Lambda^{p_i} {\cal E}$. Thus, from section 5.3, we find that it represents a class in the Cech cohomology group $H^{q_i}( X, \Lambda^{p_i}{\cal E})$. Secondly, note that the additional $U(1)_L$ anomaly computation of (\ref{anomaly1}) means that for the correlation functions of our model to be non-vanishing on the sphere, they must also satisfy $\sum_i p_i = r$  in perturbation theory. Thirdly, via the fixed-point theorem \cite{mirror manifolds} and the BRST transformation laws in (\ref{txtwist}), we find that the path integral reduces to an integral over the moduli space of holomorphic maps. Since we are considering degree-zero maps in perturbation theory, the moduli space of holomorphic maps is $X$ itself, i.e., the path integral reduces to an integral over the target space $X$. In summary, we find that a non-vanishing $\it{perturbative}$ correlation function involving $s$ dimension $(0,0)$ operators ${\cal O}_1$, ${\cal O}_2$, $\dots$, ${\cal O}_s$ on the sphere, can be computed as 
\be
\left < {\cal O}_1(z_1) \dots {\cal O}_s(z_s) \right>_0 = \int_X W^{n,n}, 
\label{corr1}
\ee                             
where $W^{n,n}$ is a top-degree form on $X$ which represents a class in the Cech cohomology group $H^n(X, K_X)$. This $(n,n)$-form  is obtained via the sequence of  maps
\be
{ H^{q_1} (X, \Lambda^{p_1} {\cal E}) \otimes \dots \otimes H^{q_s} (X, \Lambda^{p_s} {\cal E}) } \to H^n(X, \otimes_{i=1}^s \Lambda^{p_i} { \cal E}) \to H^n(X, \Lambda^{r} { \cal E})\cong H^n(X, K_X),
\ee          
where $\sum_{i=1}^s q_i = n$ and $\sum_{i=1}^s p_i = r$. The first map is given by the cup product of Cech cohomology classes      which represent  the corresponding dimension $(0,0)$ operators. The second map is given by a wedge product of holomorphic bundles. The last isomorphism follows from the required constraint $\Lambda^r {\cal E} \cong K_X$. Therefore, (\ref{corr1}) just defines a map $H^n(X, K_X) \to \mathbb C$. Although this procedure is unusual for a physicist, it has been utilised in \cite{other} as a powerful means to compute certain quantum (i.e. non-perturbative) correlation functions in heterotic string theory. Analogous procedures follow for the computation of correlation functions involving higher dimension operators.  

Note that in the computation of a $\it{non}$-$\it{perturbative}$ correlation function of the above dimension $(0,0)$ operators, the operators will  be represented by Cech cohomology classes in the moduli space of worldsheet instantons (See \cite{other}). An extension of this recipe to compute the non-perturbative correlation functions of local operators of higher dimension, will therefore serve as  the basis of a chiral version of $(0,2)$ quantum cohomology.

\newsubsection{Local Symmetries}

So far, we have obtained an understanding of the local structure of the $\overline Q_+$-cohomology. We shall now proceed towards our real objective of obtaining an understanding of its global structure. In order to do, we will need to glue the local descriptions that we have studied above together.

To our end, let us first note that the $bc$-$\beta\gamma$ action given by (\ref{bcaction}) can also be written as 
\be     
I_{bc \textrm{-} \beta\gamma}= {1\over
2\pi} \int_{\Sigma} |d^2z| \left ( \sum_i\beta_i \partial_{\bar z}\gamma^i + \sum_m b_m \partial_{\bar z} c^m \right),
\label{bcactionalternative}       
\ee
where $b_m$ is a $(1,0)$-form on $\Sigma$ with values in $\gamma^*({\widetilde {\cal E}}^{\vee})$ and ${\widetilde {\cal E}} = {\cal E}^{\vee}$, while $c^m$ is a scalar on $\Sigma$ with values in $\gamma^*(\widetilde{\cal E})$. Notice that the action of (\ref{bcactionalternative}) just represents a conventional\footnote{Conventional in the sense that as commonly defined in the physics and math literature, the $b$ and $c$ fields have lower and upper target-space indices respectively.} $bc$-$\beta\gamma$ system on the bundle $\widetilde{\cal E} \to X$. In other words, $\it{locally}$ on $X$, the underlying twisted heterotic sigma model on the bundle ${\cal E} \to X$ is $\it{equivalent}$ to the above $bc$-$\beta\gamma$ system on the bundle ${\widetilde{\cal E}} \to X$, where ${\widetilde{\cal E}}$ is just the $\it{dual}$ bundle of ${\cal E}$.  We shall refer to this equivalent $bc$-$\beta\gamma$ system henceforth in all our discussions.

Next, let us cover $X$ by small open sets $\{U_a\}$.  Recall here that in each $U_a$, the $\overline Q_+$-cohomology is described by the chiral algebra of local operators of the above free $bc$-$\beta\gamma$ system on ${\widetilde{\cal E}}_f \times U_a$ (with action a linearised version of (\ref{bcactionalternative})). Next, we will need to glue these local descriptions together over the intersections $\{ U_a \cap U_b \}$, so as to describe the global structure of the $\overline Q_+$-cohomology  in terms of a globally-defined sheaf of chiral algebras over the entire manifold $X$.

Note that the gluing has to be carried out using the automorphisms of the free $bc$-$\beta\gamma$ system. Thus, one must first ascertain the underlying symmetries of the system, which are in turn divided  into  geometrical and non-geometrical symmetries. The geometrical symmetries are used in gluing together the local sets $\{ {\widetilde{\cal E}}_f \times {U_a} \}$ into the entire holomorphic bundle ${\widetilde{\cal E}} \to X$. The non-geometrical symmetries on the other hand, are used in gluing the local descriptions at the algebraic level.

As usual, the generators of these symmetries will be given by the charges of the conserved currents of the free $bc$-$\beta\gamma$ system. In turn, these generators will furnish the Lie algebra $\mathfrak g$ of the symmetry group. Let the elements of $\mathfrak g$ which generate the non-geometrical and geometrical symmetries be   written as $\mathfrak c$ and ${\mathfrak h} = ({\mathfrak v} , \mathfrak f)$   respectively, where  $\mathfrak v$ generates the geometrical symmetries of $U$, while $\mathfrak f$ generates the fibre space symmetries of the bundle ${\widetilde{\cal E}} \to U$. Since the conserved charges must also be conformally-invariant, it will mean that an element of $\mathfrak g$ must be given by an integral of a dimension one current, modulo total derivatives. In addition, the currents must also be invariant under the $U(1)$ R-symmetry of the action (\ref{bcaction}),  under which the $b$ and $c$ fields have charges $-1$ and $1$ respectively. With these considerations in mind, the dimension one currents of the free $bc$-$\beta \gamma$ system can be constructed as follows.

Let us start by describing the currents which are associated with the geometrical symmetries first. Firstly, if we have a holomorphic vector field $V$ on $X$ where $V = V^i (\gamma) {\partial \over {\partial \gamma^i}}$, we can construct a $U(1)$ R-invariant

dimension one current $J_V=-V^i \beta_i$. The corresponding
conserved charge is then given by $K_V=\oint J_V dz  $. A computation of the operator product expansion with the elementary fields $\gamma$ gives 
\be
J_V(z)\gamma^k(z')\sim {V^k(z')\over z-z'}.
\label{jv}
\ee 
Under the symmetry transformation generated by $K_V$, we have $\delta \gamma^k = i \epsilon [ K_V, \gamma^k ]$, where $\epsilon$ is a infinitesinal transformation parameter. Thus, we see from (\ref{jv}) that $K_V$ generates the infinitesimal diffeomorphism $\delta\gamma^k=i \epsilon V^k$ of $U$. In other words, $K_V$ generates the holomorphic diffeomorphisms of the target space $X$. Therefore, $K_V$ spans the $\mathfrak v$ subset
of $\mathfrak g$. For finite diffeomorphisms, we will have a coordinate transformation ${\tilde \gamma}^k = g^k (\gamma)$, where each $g^k (\gamma)$ is a holomorphic function in the $\gamma^k$s. Since we are using the symmetries of the $bc$-$\beta \gamma$ system to glue the local descriptions over the intersections $\{U_a \cap U_b\}$, on an arbitrary intersection $U_a \cap U_b$, $\gamma^k$ and ${\tilde \gamma}^k$ must be defined in $U_a$ and $U_b$ respectively.      

Next, let $[t(\gamma)]$ be an arbitrary $r \times r$ matrix over $X$ whose components are holomorphic functions in $\gamma$. One can then construct a $U(1)$ R-invariant dimension one current involving the fermionic fields $b$ and $c$ as $J_F = c^m [t(\gamma)] _m{}^n b_n$, where the indices $m$ and $n$ on the matrix $[t(\gamma)]$ denote its $(m,n)$ component, and $m,n = 1, 2, \dots, r$. The corresponding conserved charge is thus given by $K_F = \oint J_F dz$. A computation of the operator product expansion with the elementary fields $c$ gives 
\be
J_F(z) c^n(z') \sim {  {c^m (z') t_m{}^n  }  \over z-z'}, 
\label{jf1}
\ee 
while a computation of the operator product expansion with the elementary fields $b$ gives 
\be
J_F(z) b_n(z') \sim - {  { t_n{}^m b_m (z')  }  \over z-z'}.
\label{jf2}
\ee 
Under the symmetry transformation generated by $K_F$, we have $\delta c^n = i \epsilon [ K_F, c^n]$ and $\delta b_n = i \epsilon [ K_F, b_n]$. Hence, we see from (\ref{jf1}) and (\ref{jf2}) that $K_F$ generates the infinitesimal transformations $\delta c^n=i \epsilon c^m t_m{}^n$ and $\delta b_n= - i \epsilon t_n{}^m b_m$. For finite transformations, we will have ${\tilde c}^n = c^m A_m{}^n$ and ${\tilde b}_n = (A^{-1})_n{}^mb_m$, where $A$ is an $r \times r$ matrix holomorphic in $\gamma$ and given by $[A(\gamma)] = e^{i \alpha [t(\gamma)]}$, where $\alpha$ is a finite transformation parameter.  As before, since we are using the symmetries of the $bc$-$\beta \gamma$ system to glue the local descriptions over the intersections $\{U_a \cap U_b\}$, on an arbitrary intersection $U_a \cap U_b$, $(c^n, b_n)$ and $({\tilde c}^n, {\tilde b}_n)$ must be defined in $U_a$ and $U_b$ respectively. Recall at this point that the $c^n$'s transform as holomorphic sections of the pull-back $\gamma^*({\widetilde{\cal E}})$, while the $b_n$'s transform as holomorphic sections of the pull-back $\gamma^* ({\widetilde{\cal E}}^{\vee})$. Moreover, note that the transition function matrix of a dual bundle is simply the inverse of the transition function matrix of the original bundle. This means that we can consistently identify $[A(\gamma)]$ as the holomorphic transition matrix of the gauge bundle ${\widetilde{\cal E}}$, and that $K_F$ spans the $\mathfrak f$ subset   of $\mathfrak g$. It is thus clear from the discussion so far how one can use the geometrical symmetries generated by $K_V$ and $K_F$ to glue the local sets $\{ {\widetilde{\cal E}}_f \times U_a \}$ together on  intersections of small open sets to form the entire bundle $ {\widetilde{\cal E}} \to X$. Note however, that ${\mathfrak h} = {\mathfrak v} \oplus {\mathfrak f}$ is {\it not} a Lie subalgebra of $\mathfrak g$, but only a linear subspace. This is because $\mathfrak h$ does not close upon itself as a Lie algebra.  This leads to non-trivial consequences for $\mathfrak g$. In fact,  this property of $\mathfrak h$ is related to the physical anomalies of the underlying sigma model.  We will explain this as we go along. For the convenience of our later discussion, let us denote the current and charge associated with the geometrical symmetries by $J_H = J_V + J_F$ and $K_H = K_V + K_F$ respectively.    

Before we proceed any further, note that one can also interpret the results of the last paragraph in terms of a spacetime gauge symmetry as follows. Recall that the fermionic fields $c^n$ ($b_n$) are identified with the matter fields $\lambda_a$ ($\lambda^a_z$) of the underlying twisted heterotic sigma model, thus leading to their interpretation as sections of the pull-back $\gamma^*({\widetilde{\cal E}})$ ($\gamma^*({\widetilde{\cal E}}^{\vee})$). This in turn allows us to interpret the relation  ${\tilde c}^n = c^m A_m{}^n$ as a local gauge transformation, where $[A(\gamma)]$ is the holomorphic gauge transformation matrix in the $r$-dimensional representation of the corresponding gauge group (associated with the gauge bundle ${\widetilde{\cal E}}$). One should then be able to find a basis of matrices such that $[t(\gamma)] = \sum_{r=1}^{\textrm{dim}\mathfrak s} \theta^r (\gamma)  t_r$, where $\mathfrak s$ is the Lie algebra of the corresponding spacetime gauge group linearly realised by the $r$ left-moving fermi fields $\lambda_a$ of the sigma model, $\theta^r (\gamma)$ is a spacetime-dependent gauge transformation parameter, and the $t_r$'s are the constant generator matrices of the Lie algebra $\mathfrak s$. $[A(\gamma)]$ will then take the correct form of a gauge transformation matrix, i.e., $[A(\gamma)] = e^{ i \theta^r(\gamma)  t_r}$.

We shall now determine the current associated with the non-geometrical symmetries. Let $B = \sum_i B_i (\gamma) d{\gamma^i}$ be a holomorphic $(1,0)$-form on $X$. We can then construct a $U(1)$ R-invariant dimension one current $J_B=B_i\partial_z\gamma^i$. The conserved charge is then given by $\oint J_B dz$. Let's assume that $B$ is an exact form on $X$, so that $B = \partial H = \partial_i H d{\gamma^i}$, where $H$ is some local function on $X$ that is holomorphic in $\gamma$. This in turn means that  $B_i = \partial_i  H$. In such a case, $\oint J_B dz  = \oint \partial_i H \partial_z\gamma^i dz$. From the action (\ref{bcaction}), we have the equation of motion  $\partial_{\bar z} \gamma^i = 0$. Hence, $\oint J_B dz  = \oint \partial_i H d\gamma^i  = \oint dH  = 0$ by Stoke's theorem. In other words, the conserved charge constructed from $B$ vanishes if $B$ is exact and vice-versa. Let us now ascertain the conditions under which $B$ will be exact. To this end, note that it suffices to work locally on $X$, since non-local instanton effects do not contribute in perturbation theory. Via Poincare's lemma, $B$ is locally exact if and only if $B$ is a closed form on $X$, i.e., $\partial B = \partial_i B_j - \partial_j B_i = 0$. Thus, for every non-vanishing holomorphic $(2,0)$-form $C = \partial B$, we will have a non-vanishing conserved charge $K_C = \oint J_B dz $. Notice that $C$ is annihilated by $\partial$ since $\partial^2 = 0$, i.e., $C$ must be a local holomorphic section of the sheaf $\Omega^{2,cl}$. Notice also that the current $J_B$ is constructed from $\gamma$ and its derivatives only. Consequently, the $\gamma^i$, $b_n$ and $c^n$ fields are invariant under the symmetry transformations generated by $K_C$. This means that $K_C$ generates non-geometrical symmetries only. Hence, $K_C$ spans the $\mathfrak c$ subset  of $\mathfrak g$.

\bigskip\noindent{\it Local Field Transformations} 

Let us now describe how the different fields of the      free $bc$-$\beta \gamma$ system on ${\widetilde{\cal E}}_f \times U$ transform under the geometrical and  non-geometrical symmetries generated by $K_H= K_V + K_F$ and $K_C$ of $\mathfrak g$ respectively. Firstly, note that the symmetries generated by $K_F$ and $K_C$ act trivially on the $\gamma$ fields, i.e., the $\gamma$ fields have non-singular OPE's with $J_F$ and $J_B$. Secondly, note that the symmetries generated by $K_V$ and $K_C$ act trivially on both the $b$ and $c$ fields, i.e., the $b$ and $c$ fields have non-singular OPE's with $J_V$ and $J_B$. As for the $\beta$ fields,    they transform non-trivially under $\it{all}$ the symmetries, i.e., the OPE's of the $\beta$ fields with $J_V$, $J_F$ and $J_C$ all contain simple poles. In summary, via a computation of the relevant OPE's, we find that the fields transform under the symmetries of the free $bc$-$\beta\gamma$ system on ${\widetilde{\cal E}}_f \times U$ as follows:      
\begin{eqnarray}
\label{auto1}
{\tilde \gamma}^i & = & g^i (\gamma) ,\\
\label{autobeta}
{\tilde \beta}_i  & = &  \beta_k   D^k{}_i + b_m c^n A_n{}^l D^k{}_i (\partial_k A^{-1})_l{}^m + \partial_z \gamma^j E_{i j} ,  \\
{\tilde c}^n & = & c^m A_m{}^n, \\
\label{auto2}
{\tilde b}_n & = & (A^{-1})_n{}^m b_m ,
\end{eqnarray} 
where $i,j,k = 1, 2, \dots, N={\textrm{dim}_{\mathbb C} X}$, and $l,m,n = 1, 2, \dots, r$. Here, $D$ and $E$ are $N \times N$ matrices such that $[D]^T = [\partial g]^{-1}$ and $[E] = [\partial B]$, that is, $[(D^T)^{-1}]_i{}^k = \partial_i g^k$ and $[E]_{ij} = \partial_i B_j$. It can be verified that $\tilde \beta$, $\tilde \gamma$, $\tilde b$ and $\tilde c$ obey the correct OPE's amongst themselves. We thus conclude that the fields must undergo the above transformations (\ref{auto1})-(\ref{auto2}) when we glue a local description (in a small open set) to another local description (in another small open set) on the mutual intersection of open sets using the automorphism of the free $bc$-$\beta\gamma$ system. Note that the last term in $\tilde \beta$ is due to the non-geometrical symmetry transformation generated by $K_C$, while the first and second term in $\tilde \beta$ is due to the geometrical symmetry transformation generated by $K_V$ and $K_F$ respectively. This observation will be important when we discuss what happens at the $(2,2)$ locus later.     

Another important comment to be made is that in computing (\ref{auto1})- (\ref{auto2}), we have just rederived, from a purely physical perspective, the set of field transformations (7.2a)-(7.2d) in \cite{GMS1}, which defines the valid automorphisms of the sheaf of vertex superalgebras obtained from a mathematical model that is equivalent to a free $bc$-$\beta\gamma$ system with action (\ref{bcactionalternative})! Hence, since the actions given by (\ref{bcactionalternative}) and (\ref{bcaction}) are equivalent, we learn that  the sheaf $\widehat {\cal A}$ is mathematically known as a sheaf of vertex superalgebras spanned by chiral differential operators on the exterior algebra $\Lambda {\widetilde{\cal E}} = \oplus_{i=1}^{rk({\widetilde{\cal E}})} \Lambda^i {\widetilde{\cal E}}$ of the holomorphic vector bundle ${\widetilde{\cal E}}$ over $X$ \cite{GMS1, GMS3}. 

\bigskip\noindent{\it A Non-Trivial Extension of Lie Algebras and Groups} 

We shall now study the properties of the symmetry algebra $\mathfrak g$ of the free $bc$-$\beta\gamma$ system on ${\widetilde{\cal E}}_f \times U$. From the analysis thus far, we find that we can write $\mathfrak g = {\mathfrak c} + {\mathfrak h}$ as a linear space, where $\mathfrak h = {\mathfrak v} + {\mathfrak f}$. Note that $\mathfrak c$ is a trivial abelian subalgebra of $\mathfrak g$. This because the commutator of $K_C$ with itself vanishes - the OPE of $J_B$ with itself is non-singular since the current is constructed from $\gamma$ and its derivatives only. Hence, $\mathfrak g$ can be expressed in an extension of Lie algebras as follows: 
\be
0 \to {\mathfrak c} \to {\mathfrak g} \to {\mathfrak h} \to 0.
\label{extension}
\ee   
In fact, (\ref{extension}) is an exact sequence of Lie algebras as we will show shortly that $[{\mathfrak h}, {\mathfrak c}] \subset {\mathfrak c}$. This means that $\mathfrak c$ is `forgotten' when we project $\mathfrak g$ onto $\mathfrak h$. 

The action of $\mathfrak h$ on $\mathfrak c$ can be found from the $J_H (z) J_C(z')$ OPE
\begin{eqnarray}
\label{OPE1}
\left[-V^i\beta_i(z) + c^m t_m{}^n b_n(z)\right] \cdot B_j\partial_{z'}\gamma^j(z') & \sim & {1\over
z-z'} \left[V^i(\partial_i B_k-\partial_k B_i) + \partial_k(V^iB_i) \right] \partial_{z'}\gamma^k \nonumber \\
&&    + {1\over
(z-z')^2}V^iB_i(z'). 
\end{eqnarray}

The commutator of $K_H$ with $K_C$, and thus $[{\mathfrak h}, {\mathfrak c}]$, is simply the residue of the simple pole on the RHS of (\ref{OPE1}). The numerator of the first term on the RHS of (\ref{OPE1}), given by $V^i(\partial_iB_k-\partial_k B_i)+\partial_k(V^iB_i)$, is the same
as $({\cal L}_V(B))_k$, the $k^{th}$ component of the one-form that results from the action of a Lie derivative of the vector field $V$ on the one-form $B$.  This observation should not come as a surprise since the charges of $J_V$ generate diffeomorphisms of $U$, and only the $J_V \cdot J_C$ part of the OPE in (\ref{OPE1}) is non-trivial (since $J_F$ has non-singular OPE's with $J_C$). Hence, $[{\mathfrak h}, {\mathfrak c}] \subset {\mathfrak c}$ as claimed.     

Let us now compute the commutator of two elements of $\mathfrak h$. To this end, let $V$ and $W$ be two vector fields on $U$ that are holomorphic in $\gamma$. Let $t(\gamma)$ and $\tilde t(\gamma)$ be $r \times r$ matrices holomorphic in $\gamma$. Let $V$ and $W$ be associated with the currents $J_V(z) \subset  J_H(z)$ and $J_W (z') \subset J_H(z')$ respectively. Likewise, let $t$ and $\tilde t$ be associated with the currents $J_F(z) \subset J_H(z)$ and $J_{\widetilde F}(z') \subset J_H(z')$ respectively.     The $J_H (z)  J_H (z')$ OPE is then computed to be 
\begin{eqnarray}
\label{OPE2}
J_H (z) J_H (z')  & \sim & -{(V^i\partial_iW^j-W^i\partial_iV^j)\beta_j \over z-z'}
-{(\partial_k\partial_jV^i)(\partial_iW^j\partial_{z'}\gamma^k)\over
z-z'} + { { c^m {\{t, {\tilde t} \}}_m{}^c b_c } \over z-z'} \nonumber \\
&&  + {{Tr[ \tilde t \partial_i t]  \partial_{z'}} \gamma^i \over z-z'}   -{\partial_jV^i\partial_iW^j(z')\over
(z-z')^2}+ {{Tr [\tilde t  t ] (z')} \over (z-z')^2}.
\end{eqnarray} 
The last two terms on the RHS of (\ref{OPE2}), being double poles, do not contribute to the commutator. From the mathematical relation $[V,W]^j= ({\cal L}_V(W))^j = V^i\partial_iW^k-W^i\partial_iV^j$, we see that the first term takes values in $\mathfrak v  \subset \mathfrak h$, the second term takes values in $\mathfrak c$, the third term takes values in   $\mathfrak f \subset \mathfrak h$, and the fourth term takes values in $\mathfrak c$. The first and third terms which come from a single contraction of elementary fields in evaluating the OPE, arise from the expected results     $J_V(z) J_W(z') \sim J_{[V,W]}/ (z-z')$ and $J_F(z) J_{\widetilde F}(z') \sim J_{\{t,\tilde t \}} / (z-z')$ respectively. We would have obtained the same results by computing the commutator of $J_V$ and $J_W$, and that of $J_F$ and $J_{\widetilde F}$, via Poisson brackets in the classical $bc$-$\beta \gamma$ theory. The second and fourth terms are the reason why  $[\mathfrak h, \mathfrak h ] \nsubseteq {\mathfrak h}$. Note that these two terms result from multiple contractions of elementary fields, just like the anomalies of conformal field theory. Hence, since $\mathfrak h$ does not closed upon itself as a Lie algebra, $\mathfrak g$ is not a semi-direct product of $\mathfrak h$ and $\mathfrak c$. Consequently, the extension of Lie algebras in (\ref{extension}) is non-trivial. Is the non-triviality of the extension of Lie algebras of the symmetries of the $bc$-$\beta\gamma$ system on ${\widetilde{\cal E}}_f \times U$ then related to the physical anomalies of the underlying sigma model? Let us study this further.               

The exact sequence of Lie algebras in (\ref{extension}) will result in the following group extension when we exponentiate the elements of $\mathfrak g$:
\be
 1 \to \widetilde C \to \widetilde G \to \widetilde H \to 1.
\label{group extension}
\ee  
Here, $\widetilde G$ is the symmetry group of all admissible automorphisms of the $bc$-$\beta\gamma$ system, $\widetilde C$ is the symmetry group of the non-geometrical automorphisms, and $\widetilde H$ is the symmetry group of the geometrical automorphisms. Just as in (\ref{extension}), (\ref{group extension}) is an exact sequence of groups, i.e., the kernel of the map $\widetilde G \to \widetilde H$ is given by $\widetilde C$. This means that the non-geometrical symmetries are `forgotten' when we project the full symmetries onto the geometrical symmetries. Since (\ref{group extension}) is derived from a non-trivial extension of Lie algebras in (\ref{extension}), it will be a non-trivial group extension. In fact, the cohomology class of  the group extension that captures its non-triviality is given by \cite{GMS1} 
\be
c_1^2 - 2c_2 - ({c'_1}^2 - 2c'_2) \in H^2({\widetilde H}, \Omega_{\widetilde H}^{2,cl}), 
\label{group obstruction}
\ee
where $\Omega_{\widetilde H}^{2,cl}$ is a sheaf of an $\widetilde H$-module of closed two-forms, and $ c_i, c'_i \in  H^i({\widetilde H}, \Omega_{\widetilde H}^{2,cl})$ are the $\it{universal}$ $\it{Chern}$ $\it{classes}$. The cohomology class $ H^2({\widetilde H}, \Omega_{\widetilde H}^{2,cl})$ vanishes if and only if the kernel of the map $\widetilde G \to \widetilde H$ is empty, i.e., $\widetilde G = \widetilde H$. Thus, the group extension is trivial if the admissible automorphisms of the $bc$-$\beta\gamma$ system are solely of a geometrical kind. This observation will be essential to our discussion of the sigma model at the $(2,2)$ locus later. Let us return back to the issue of (\ref{group obstruction})'s relevance to the physical anomalies of the underlying sigma model. Note that the mathematical arguments in \cite{GMS1} and a detailed computation in \cite{GMS3}, show that (\ref{group extension}), together with its cohomology class (\ref{group obstruction}), imply that the obstruction to a globally-defined  sheaf of chiral algebras $\widehat {\cal A}$ of chiral differential operators on the exterior algebra $\Lambda {\widetilde{\cal E}}$, must be captured by the cohomology class
\be
2ch_2(TX) - 2ch_2({\widetilde{\cal E}}).
\label{group anomaly}   
\ee 
Since ${\widetilde{\cal E}} = {\cal E}^{\vee}$, and $ch_2({\cal E}^{\vee}) = ch_2(\overline{\cal E}) = ch_2(\cal E)$, the cohomology class capturing the obstruction can actually be written as
\be
2ch_2(TX) - 2ch_2(\cal E),    
\label{group anomaly actual}   
\ee      
which in turn represents an element of $ {H^2(X, \Omega^{2,cl}_X)}$ (as explained in footnote 14). Notice that the vanishing of (\ref{group anomaly}) coincides with one of the anomaly-cancellation conditions of the underlying twisted heterotic sigma model in (\ref{an})! In hindsight, this `coincidence' should not be entirely surprising - note that a physically valid sigma model must be defined over all of ${\cal E} \to X$ (and $\Sigma$). Since (\ref{group anomaly}) captures the obstruction to gluing the local descriptions together to form a global description, this implies that the sigma model, which is described locally by the free $bc$-$\beta\gamma$ system on ${\widetilde{\cal E}}_f \times U$, cannot be globally-defined over all of ${\cal E} \to X$ unless (\ref{group anomaly}) vanishes. Hence, the anomaly which obstructs the physical validity of the underlying sigma model must be given by (\ref{group anomaly}). Thus, the non-triviality of the extension of Lie algebras of the symmetries of the $bc$-$\beta\gamma$ system on ${\widetilde{\cal E}}_f \times U$, is indeed related to the physical anomaly of the underlying sigma model.

\newsubsection{Gluing the Local Descriptions Together}

Now, we will describe explicitly, how one can glue the local descriptions together using the automorphisms of the free $bc$-$\beta\gamma$ system on ${\widetilde{\cal E}}_f \times U$ to obtain a globally-defined sheaf of chiral algebras. In the process, we will see how the cohomology class in (\ref{group anomaly}) emerges as an obstruction to gluing the locally-defined sheaves of  chiral algebras globally on $X$. Moreover, we can also obtain the other anomaly in (\ref{an}) which is not captured in (\ref{group anomaly}) (for reasons we will explain shortly) when we consider gluing the sheaves of chiral algebras globally over $X$ $\it{and}$ $\Sigma$. In addition, we will see that the moduli of the resulting sheaf emerges as a Cech cohomology class generated by a relevant Cech cocycle.

To begin with, let's take a suitable collection of small open sets $U_a\subset \Bbb{C}^n$, where $n=\textrm{dim}_{\mathbb C}X$. Next, consider the corresponding set of product spaces $\{{\widetilde{\cal E}}_f \times U_a \}$. We want to glue these trivial product spaces together to make a good cover of the holomorphic vector bundle ${\widetilde{\cal E}} \to X$. On each $U_a$, the sheaf $\widehat{\cal A}$ of chiral algebras is defined by a free $bc$-$\beta\gamma$ system on $\{ {\widetilde{\cal E}}_f \times U_a \}$ . We need to glue together these free conformal field theories to get a globally-defined sheaf of chiral algebras. 

It will be convenient for us to first describe how we can geometrically glue the set of trivial product spaces $\{ {\widetilde{\cal E}}_f \times U_a\}$ together to form the bundle ${\widetilde{\cal E}} \to X$. For each $a,b$, let us pick a product space ${\widetilde{\cal E}}_f \times U_{ab}\subset {\widetilde{\cal E}}_f \times U_{a}$, and likewise another product space ${\widetilde{\cal E}}_f \times U_{ba}\subset {\widetilde{\cal E}}_f \times U_b$. Let us define a geometrical symmetry ${\hat h}_{ab}$ (given by a product of holomorphic diffeomorphisms on $U$ with holomorphic homeomorphisms of the fibre ${\widetilde{\cal E}}_f$)  between these product spaces as  
\be
{\hat h}_{ab}: {\widetilde{\cal E}}_f \times U_{ab}\cong  {\widetilde{\cal E}}_f \times U_{ba}.
\label{g symmetry}
\ee
Note that ${\hat h}$ can be viewed as a geometrical gluing operator corresponding to an element of the geometrical symmetry group $\widetilde H$. From the above definition, we see that ${\hat h}_{ba}={\hat h}_{ab}^{-1}$. We want to identify an arbitrary point $P \in  {\widetilde{\cal E}}_f \times U_{ab}$ with an arbitrary point $Q \in {\widetilde{\cal E}}_f \times U_{ba}$ if $Q={\hat h}_{ab}(P)$. This identification will be consistent if for any $U_a$, $U_b$, and $U_c$, we have
\be
{\hat h}_{ca} {\hat h}_{bc} {\hat h}_{ab}=1
\label{triple}
\ee
in any triple intersection $U_{abc}$ over which all the maps ${\hat h}_{ca}$,  ${\hat h}_{bc}$ and ${\hat h}_{ab}$ are defined. The relation in (\ref{triple}) tells us that the different pieces ${\widetilde{\cal E}}_f \times U_a$ can be glued together via the set of maps $\{{\hat h}_{ab} \}$ to make a holomorphic vector bundle ${\widetilde{\cal E}} \to X$. The holomorphic structure moduli of the bundle (or that of its dual $\cal E$), and the complex structure moduli  of its base, will then manifest as parameters in the ${\hat h}_{ab}$'s.

Suppose we now have a sheaf of chiral algebras on each $U_a$, and we want to glue them together on overlaps to get a sheaf of chiral
algebras on $X$. The gluing must be done using the automorphisms of the conformal field theories. Thus, for each pair $U_a$ and $U_b$, we select a conformal field theory symmetry $\hat g_{ab}$ that maps the free $bc$-$\beta\gamma$ system on ${\widetilde{\cal E}}_f \times U_a$, restricted to ${\widetilde{\cal E}}_f \times U_{ab}$, to the free $bc$-$\beta\gamma$ system on ${\widetilde{\cal E}}_f \times U_b$, similarly restricted to ${\widetilde{\cal E}}_f \times U_{ba}$. We get a globally-defined sheaf of chiral algebras if the gluing
is consistent: 
\be
\hat g_{ca}\hat g_{bc}\hat g_{ab}=1.
\label{consistent}
\ee
Note that $\hat g$ can be viewed as a gluing operator corresponding to an element of the full symmetry group $\widetilde G$. As usual, we have ${\hat g}_{ba} = {\hat g}^{-1}_{ab}$. Moreover, recall at this point that from the exact sequence of groups in (\ref{group extension}), we have a map $\widetilde G \to \widetilde H$ which `forgets' the non-geometrical symmetry group $\widetilde C \subset \widetilde G$. As such, for $\it{any}$ arbitrary set of $\hat g$'s which obey (\ref{consistent}), the geometrical condition (\ref{triple}) will be automatically satisfied, regardless of  what the non-geometrical gluing operator $\hat c$ corresponding to an element of $\widetilde C$ is. Hence, every possible way to glue the conformal field theories together via $\hat g$, determines a way to geometrically glue the set of product spaces $\{{\widetilde{\cal E}}_f \times U_a \}$ together to form a unique holomorphic vector bundle ${\widetilde{\cal E}} \to X$ over which one defines the resulting conformal field theory.      

The above discussion translates to the fact that for a given set of ${\hat h} _{ab}$'s which obey (\ref{triple}), the corresponding set of  $\hat g_{ab}$'s which obey (\ref{consistent}) are not uniquely determined; for each $U_{ab}$, we can still pick an element ${\cal C}_{ab}\in H^0(U_{ab},\Omega^{2,cl})$ which represents an element of $\mathfrak{c}$ (as discussed in section 5.5), so that $\textrm {exp}({\cal C}_{ab})$ represents an element of $\widetilde C$. One can then transform $\hat g_{ab}\to {\hat
g}'_{ab}=  \textrm{exp}({\cal C}_{ab})\hat g_{ab}$, where ${\hat g}'_{ab}$ is another physically valid gluing operator. The condition that the gluing identity (\ref{consistent}) is obeyed by $\hat g'$, i.e., $\hat g'_{ca}\hat g'_{bc}\hat g'_{ab}=1$, is that in each triple
intersection $U_{abc}$, we should have
\be
{\cal C}_{ca}+ {\cal C}_{bc} + {\cal C}_{ab}=0. 
\label{cocycle}
\ee
From ${\hat g}'_{ba} = ({\hat g}'_{ab})^{-1}$, we have ${\cal C}_{ab} = -{\cal C}_{ba} $. Moreover, ${\widetilde {\cal C}}_{ab} \sim {\cal C}_{ab} + {\cal S}_a - {\cal S}_b$ for some $\cal S$, in the sense that the $\widetilde{\cal C}$'s will obey (\ref{cocycle}) as well. In other words, the $\cal C$'s in (\ref{cocycle}) must define an element of the Cech cohomology group $H^1(X,\Omega^{2,cl}_X)$.  As usual, ${\textrm{exp}}({\cal C}_{ab})$  is `forgotten' when we project from ${\hat g}_{ab}'$ to the geometrical gluing operator ${\hat h}_{ab}$. Therefore, in going from $\hat g$ to ${\hat g}'$, the symmetry $\hat h$, and consequently the bundle ${\widetilde{\cal E}} \to X$, remains unchanged.  Now, let us use a specific $\hat g$ operator to define the specific symmetries of a free $bc$-$\beta\gamma$ system, which in turn will define a unique sheaf of chiral algebras. In this sense, given any sheaf and an element ${\cal C} \in
H^1(X,\Omega^{2,cl}_X)$, one can define a new sheaf by going from $\hat g \to
\exp({\cal C}) \hat g$. So, via the action of $H^1(X,\Omega^{2,cl}_X)$, we get a family of sheaves of chiral algebras, with the $\it{same}$ target space ${\widetilde{\cal E}} \to X$. Hence, the moduli of the sheaf of chiral algebras is represented by a class in $H^1(X,\Omega^{2,cl}_X)$. Together with the results of section 3.3, we learn that the analysis of a family of sheaves of chiral algebras on a unique K\"ahler target space $X$, is equivalent to the analysis of a unique sheaf of chiral algebras on a family $\{X'\}$ of non-K\"ahler target spaces.

\bigskip\noindent{\it The Anomaly}

We now move on to discuss the case when there is an obstruction to the gluing. Essentially, the obstruction occurs when (\ref{consistent}) is not satisfied by the $\hat g$'s. In such a case, one generally has, on triple intersections $U_{abc}$, the following relation
\be
\hat g_{ca} \hat g_{bc}\hat g_{ab}= \textrm{exp}({\cal C}_{abc})
\label{obstruction sheaf}
\ee
 for some ${\cal C}_{abc}\in H^0(U_{abc},\Omega^{2,cl})$. The reason for (\ref{obstruction sheaf}) is as follows. First, note that the LHS of (\ref{obstruction sheaf}) projects purely to the group of geometrical symmetries associated with $\hat h$. If the bundle ${\widetilde{\cal E}} \to X$ is to exist mathematically, there will be no obstruction to its construction, i.e., the LHS of (\ref{obstruction sheaf}) will map to the identity under the projection. Hence, the RHS of (\ref{obstruction sheaf}) must represent an element of the abelian group $\widetilde C$ (generated by $\mathfrak c$) that acts trivially on the coordinates $\gamma^i$ of the $U_a$'s and the local sections $c^m$ of the $({\widetilde{\cal E}}_f \times U_a)$'s.     

 Recall that the choice of ${\hat g}_{ab}$ was not unique. If we transform $\hat g_{ab}\to \exp({\cal C}_{ab})\hat g_{ab}$ via a (non-geometrical) symmetry of the system, we get 
\be
{\cal C}_{abc}\to {\cal C}'_{abc} = {\cal C}_{abc}+ {\cal C}_{ca}+ {\cal C}_{bc}+ {\cal C}_{ab}.
\label{cocycle 3}
\ee
 If it is possible to pick the ${\cal C}_{ab}$'s to set $\it{all}$ ${\cal C}_{abc}'=0$, then there is no obstruction to gluing and one can obtain a globally-defined sheaf of chiral algebras. 

In any case, in quadruple overlaps $U_a\cap U_b\cap U_c\cap U_d$,
the $\cal C$'s obey
\be
{\cal C}_{abc}- {\cal C}_{bcd}+ {\cal C}_{cda} - {\cal C}_{dab} =0.
\label{cocycle 4}
\ee
Together with the equivalence relation (\ref{cocycle 3}), this means that the
$\cal C$'s in (\ref{cocycle 4}) must define an element of the Cech cohomology group $H^2(X, \Omega^{2,cl}_X)$. In other words, the obstruction to gluing the locally-defined sheaves of chiral algebras is captured by a non-vanishing cohomology class  $H^2(X, \Omega^{2,cl}_X)$. As discussed in section 4 and the last paragraph of section 5.5, this class can be represented in de Rham cohomology by $2[ch_2(TX) - ch_2({\cal E})]$. Thus, we have obtained an interpretation of the anomaly in the twisted heterotic sigma model in terms of an obstruction to a global definition of the sheaf of chiral algebras derived from a free $bc$-$\beta\gamma$ system that describes the sigma model locally on $X$.

\bigskip\noindent{\it The Other Anomaly}    

In section 4, we showed that the twisted heterotic sigma model had two anomalies, one involving $ch_2({\cal E}) - ch_2(TX)$, and the other involving $-{1\over 2}c_1(\Sigma) \left ( c_1({\cal E}) + c_1(TX) \right)$. We have already seen how the first anomaly arises from the Cech perspective. How then can we see the second anomaly in the present context? 

So far, we have constructed a sheaf of chiral algebras globally on $X$ but only locally on the worldsheet $\Sigma$. This is because the chiral algebra of the twisted heterotic sigma model is not invariant under holomorphic reparameterisations of the worldsheet coordinates at the quantum level,\footnote{To see this, recall from section 3.1 that in the quantum theory, the holomorphic stress tensor $T_{zz}$ is not in the $\overline Q_+$-cohomology (i.e. $\{ \overline Q_+, T_{zz}\} \neq 0$) unless we have a stable bundle $\cal E$ with $c_1(X) = 0$. This prevents the $\overline Q_+$-cohomology and thus the chiral algebra from being invariant under arbitrary reparameterisations of $\Sigma$.}  and as such, can only be given a consistent definition locally on an arbitrary Riemann surface $\Sigma$.  Since $c_1(\Sigma)$ can be taken to be zero when we work locally on $\Sigma$, the second anomaly vanishes and therefore, we did not get to see it.      

Now, note that the free $bc$-$\beta\gamma$ system is conformally invariant; in other words, it can be defined globally on an arbitrary Riemann surface $\Sigma$. But, notice that the anomaly that we are looking for is given by $-{1\over 2}c_1(\Sigma) ( c_1({\cal E}) + c_1(TX))$ or equivalently, ${1\over 2}c_1(\Sigma) ( c_1({\widetilde {\cal E}}) - c_1(TX))$. Hence,  it will vanish even if we use a free $bc$-$\beta\gamma$ system that can be globally-defined on $\Sigma$  if we continue to work locally on the bundle ${\widetilde {\cal E}} \to X$ where $c_1({\widetilde {\cal E}}) = c_1(TX) = 0$. Therefore, the only way to see the second anomaly is to work globally on both $X$ (and hence ${\widetilde {\cal E}} \to X$) $\it{and}$   $\Sigma$. (In fact, recall that the underlying sigma model is physically defined on all of $\Sigma$ and ${\cal E} \to X$.) We shall describe how to do this next.  
   
Let us cover $\Sigma$ and $X$ with small open sets $\{P_\tau \}$ and $\{U_a \}$ respectively.  This will allow us to cover ${\widetilde {\cal E}} \times \Sigma$ with open sets $W_{a\tau}= {\widetilde {\cal E}}_f \times U_a \times P_\tau$. On each $P_\tau$, we can define a free $bc$-$\beta\gamma$ system with target ${\widetilde {\cal E}}_f \times U_a$. In other words, on each open set $W_{a\tau}$, we define a free $bc$-$\beta\gamma$ system and hence a sheaf of chiral algebras. What we want to do is to glue the sheaves of chiral algebras on the $({\widetilde {\cal E}}_f \times U_a \times P_{\tau})$'s together on overlaps, to get a globally-defined sheaf of chiral algebras, with target space ${\widetilde {\cal E}} \to X$, defined on all of $\Sigma$. As before, the gluing must be done using the admissible automorphisms of the free $bc$-$\beta\gamma$ system. 

Recall from section 5.5 that the admissible automorphisms are given by the symmetry group $\widetilde G$. Note that the set of geometrical symmetries $\widetilde H \subset \widetilde G$ considered in section 5.5 can be extended to include holomorphic diffeomorphisms of the worldsheet $\Sigma$ - as mentioned above, the free $bc$-$\beta\gamma$ system is conformally invariant and is therefore invariant under arbitrary holomorphic reparameterisations of the coordinates on $\Sigma$. Previously in section 5.5, there was no requirement to consider and exploit this additional geometrical symmetry in gluing the local descriptions together simply because we were working locally on $\Sigma$. Then, gluing of the local descriptions at the geometrical level was carried out using $\widetilde H$, where $\widetilde H$ consists of the group of holomorphic diffeomorphisms of $X$ and the group of holomorphic homeomorphisms of the fibre ${\widetilde {\cal E}}_f$. Now that we want to work globally on $\Sigma$ as well, one will need to use the symmetry of the free conformal field theory under holomorphic diffeomorphisms of $\Sigma$ to glue the $P_{\tau}$'s together to form $\Sigma$. In other words, gluing of the local descriptions at the geometrical level must now be carried out using the geometrical symmetry group ${\widetilde H}'$, where ${\widetilde H}'$ consists of the group of holomorphic diffeomorphisms on $\Sigma$ $\it{and}$ $X$, and the group of holomorphic homeomorphisms of the fibre ${\widetilde {\cal E}}_f$. Now, let the conformal field theory gluing map from $W_{a\tau}$ to $W_{b\nu}$ be given by    ${\hat g}_{a\tau,b\nu}$.  Let the corresponding geometrical and non-geometrical gluing maps from $W_{a\tau}$ to $W_{b\nu}$ be given by ${\hat h}'_{a\tau,b\nu}$ and ${\hat c}'_{a\tau,b\nu}$ respectively. Since we have a sensible notion of a holomorphic map $\gamma : \Sigma \to X$, and the bundle ${\widetilde {\cal E}}$ and worldsheet $\Sigma$ are defined to exist mathematically, there is no obstruction to gluing at the geometrical level, i.e., 
\be
{\hat h}'_{c\sigma,a\tau}{\hat h}'_{b\nu,c\sigma} {\hat h}'_{a\tau,b\nu} =1
\ee
in triple intersections. There will be no obstruction to gluing at all levels if one has the relation
\be
{\hat g}_{c\sigma,a\tau} {\hat g}_{b\nu,c\sigma} {\hat g}_{a\tau,b\nu} =1. 
\label{consistency}
\ee
However, (\ref{consistency}) may not always be satisfied. Similar to our previous arguments concerning the anomaly $2ch_2(TX) -2ch_2({\cal E}) \in H^2(X, \Omega^{2,cl}_X)$, since one has a map ${\hat g}_{a\tau,b\nu} \to {\hat h}'_{a\tau,b\nu}$ in which ${\hat c}'_{a\tau,b\nu}$ is `forgotten', in general, we will  have
\be
{\hat g}_{c\sigma,a\tau} {\hat g}_{b\nu,c\sigma} {\hat g}_{a\tau,b\nu} = \textrm{exp}({\cal C}_{a \tau b\nu c\sigma}),   
\ee                    
where the ${\cal C}_{a \tau  b\nu  c\sigma}$'s on any triple overlap defines a class in the two-dimensional Cech cohomology group $H^2(X \times {\Sigma}, {\cal G})$. $\cal G$ is a sheaf associated with the non-geometrical symmetries of the free $bc$-$\beta\gamma$ system. Being non-geometrical in nature, these symmetries will act trivially on the $\gamma^i$ coordinates of $X$ and the sections $c^m$ (and $b_m$) of the pull-back $\gamma^*({\widetilde {\cal E}})$ (and $\gamma^*({\widetilde {\cal E}}^{\vee})$).  

Earlier on in our discussion, when we worked locally on $\Sigma$ but globally on $X$, we constructed a $U(1)$ R-invariant dimension one current $J_B$ from a $(1,0)$-form $B$ on $X$, whose conserved charge $K_C$ was shown to generate the non-geometrical symmetries of the free $bc$-$\beta\gamma$ conformal field theory. Therefore, if one works globally on both $\Sigma$ and $X$, one will need to construct an analogous $U(1)$ R-invariant dimension one current $J_{B'}$ from a $(1,0)$-form $B'$ on $X \times \Sigma$, such that the corresponding conformally-invariant conserved charge will generate the non-geometrical symmetries in this extended case. Since the current $J_{B'}$ should have non-singular OPE's with the $\gamma$, $c$ and $b$ fields, it can only depend linearly on $\partial_z \gamma$ and be holomorphic in $\gamma$ and $z$. Thus, the non-geometrical symmetries will be generated by the conserved charge $\oint J_{B'} dz$, with
\be
J_{B'} = B_i(\gamma, z) \partial_z \gamma^i + B_{\Sigma}(\gamma, z).
\ee                       
Here, $B_i$ and $B_{\Sigma}$ are components of a holomorphic $(1,0)$-form $B' = B_i d \gamma^i + B_{\Sigma} dz$ on $X \times \Sigma$, where $B_i$ and $B_{\Sigma}$ have scaling dimension zero and one respectively, i.e., for $z \to \tilde z = \lambda z$, we have $B_i(\gamma, z) \to B_i(\gamma, \tilde z) = B_i(\gamma, z)$, and $B_{\Sigma} (\gamma, z) \to B_{\Sigma} (\gamma, \tilde z) = \lambda^{-1}B_{\Sigma} (\gamma, z)$.   

If $B'$ is exact, i.e, $B'=\partial H'$ for some local function $H' (\gamma,z)$ on $X\times \Sigma$ holomorphic in $\gamma$ and $z$, we will have $B_i = \partial_i H'$ and $B_{\Sigma} = \partial_z H'$. As a result, the conserved charge $\oint J_{B'} dz = \oint (\partial_i H') d\gamma^i + (\partial_z H') dz = \oint dH' = 0$ by Stoke's theorem. Using the same arguments found in section 5.5 (where we discussed the conserved charge $K_C$), we learn that for every non-vanishing holomorphic $(2,0)$-form $C' = \partial B'$ on $X \times \Sigma$, we will have a non-vanishing conserved charge $K_{C'} = \oint J_{B'} dz$. Since $C'$ is $\partial$-closed, it is a local holomorphic section of $\Omega^{2,cl}_{X\times \Sigma}$. Therefore, we find that the sheaf associated with the non-geometrical symmetries that act trivially on $\gamma$, $c$ and $b$, is isomorphic to $\Omega^{2,cl}_{X\times\Sigma}$. Thus, the obstruction to a globally-defined sheaf of chiral algebras, with target space ${\widetilde {\cal E}} \to X$, defined on all of $\Sigma$, will be captured by a class in the Cech cohomology group $H^2(X\times \Sigma, \Omega^{2,cl}_{X\times\Sigma})$. Hence, the physical anomalies of the underlying sigma model ought to be captured by the de Rham cohomology classes which take values in $H^2(X\times \Sigma, \Omega^{2,cl}_{X\times\Sigma})$. 

In fact, since $\Sigma$ is of complex dimension one, its space of $(2,0)$-forms vanishes. Thus, we will have $\Omega^{2,cl}_{X \times\Sigma} = (\Omega^{2,cl}_X \otimes {\cal O}_\Sigma) \oplus (\Omega^{1,cl}_X \otimes \Omega^{1,cl}_\Sigma)$ (where ${\cal O}_{\Sigma}$ is the sheaf of holomorphic functions on $\Sigma$). In other words, on a compact Riemann surface $\Sigma$, where the only holomorphic functions over it are constants, i.e.,  $H^0(\Sigma,{\cal O})\cong \mathbb{C}$,  we have the expansion 
\be
H^2(X\times \Sigma,\Omega^{2,cl}_{X\times
\Sigma})= H^2 (X,\Omega^{2,cl}_X) \oplus
( H^1(X,\Omega^{1,cl}_X) \otimes
H^1(\Sigma,\Omega^{1,cl}_{\Sigma}) )\oplus \dots, 
\label{sheaf anomaly}    
\ee      

Recall that in section 4, we showed that $c_1(\Sigma) \in H^1(\Sigma, \Omega^{1,cl}_{\Sigma})$ and $(c_1({{\widetilde {\cal E}}}) - c_1(TX)) = -( c_1({\cal E}) + c_1(TX)) \in H^1(X, \Omega^{1,cl}_X)$. Hence, the two physical anomalies $ch_2({\cal E}) - ch_2(TX)$ and ${-{1\over 2} c_1(\Sigma) ( c_1({\cal E}) + c_1(TX))}$, take values in the first and second term on the RHS of (\ref{sheaf anomaly}) respectively. Note that the terms on the RHS of (\ref{sheaf anomaly}) must independently vanish for $H^2(X\times \Sigma, \Omega^{2,cl}_{X\times\Sigma})$ to be zero. In other words, we have obtained a consistent, alternative interpretation of the physical anomalies  which arise due to a non-triviality of the determinant line bundles (associated with the Dirac operators of the underlying sigma model) over the space of gauge-inequivalent connections, purely in terms of an obstruction to the gluing of sheaves of chiral algebras. 

By extending the arguments surrounding (\ref{cocycle}) to the present context,  we find that for a vanishing  anomaly, (apart from the geometrical moduli encoded in the holomorphic and complex structures of the bundle ${\widetilde {\cal E}} \to X$ (or that of its dual ${\cal E} \to X$)), the moduli of the globally-defined sheaf of chiral algebras on $\Sigma$, with target space ${\widetilde {\cal E}} \to X$, (or in the context of the underlying twisted heterotic sigma model that the $bc$-$\beta \gamma$ system describes locally, the target space ${\cal E} \to X$), must be parameterised by $H^1(X\times\Sigma,\Omega^{2,cl}_{X\times\Sigma})$.

\newsubsection{The Conformal Anomaly}

In this section, we will demonstrate an application of the rather abstract discussion thus far. In the process, we will be able to provide a physical interpretation of a computed mathematical result and vice-versa.

From eqn.\,(\ref{Tzz}), we see that the holomorphic stress tensor $T(z) \sim T_{zz}$ of the twisted heterotic sigma model lacks the $\psi^{\bar i}$ fields.\footnote{Recall that this is also true in the quantum theory as the classical expression for $T(z)$ does not receive any perturbative corrections up to 1-loop.} In other words, it is an operator with $q_R = 0$. Hence, from the $\overline Q_+$-Cech cohomology dictionary established in section 5.3, if $T(z)$ is to be non-trivial in $\overline Q_+$-cohomology, such that the sigma model and its chiral algebra are conformally-invariant, it will be given by an element of $H^0(X, \widehat{\cal A})$,   that is, a global section of the sheaf of chiral algebras $\widehat{\cal A}$. Recall that the local sections of $\widehat{\cal A}$ are furnished by the physical operators in the chiral algebra of the free (linear) $bc$-$\beta\gamma$ system. Since the free (linear) $bc$-$\beta\gamma$ system  describes a local version of the underlying twisted heterotic sigma model, one can write the local holomorphic stress tensor of the sigma model as the local holomorphic stress tensor of the free (linear) $bc$-$\beta\gamma$ system, which in turn is given by     
\be
{\cal T} (z) =  - : \beta_i \partial_z \gamma^i:  - :b_m \partial_z c^m:.
\ee 
(see section 5.4).  Under an automorphism of the $bc$-$\beta\gamma$ system, ${\cal T}(z)$ will become 
\be
{\widetilde {\cal T}} (z) = - :{\tilde\beta}_i \partial_z {\tilde \gamma}^i:  - :{\tilde b}_m \partial_z {\tilde c}^m:,
\label{tildefraktz}
\ee 
where the fields $\tilde\beta$, $\tilde \gamma$, $\tilde b$ and $\tilde c$ are defined in the automorphism relations of (\ref{auto1})-(\ref{auto2}). It is clear that on an overlap $U_a \cap U_b$ in $X$, ${{\cal T}(z)}$ will be regular in $U_a$ while ${\widetilde{\cal T}(z)}$ will be regular in $U_b$. Note that both ${\cal T}(z)$ and ${\widetilde{\cal T}(z)}$ are at least local sections of $\widehat {\cal A}$. And if there is no obstruction to ${\cal T}(z)$ or ${\widetilde{\cal T}(z)}$ being a global section of $\widehat {\cal A}$, it will mean that $T(z)$ is non-trivial in $\overline Q_+$-cohomology, i.e., $T(z) \neq \{\overline Q_+, \dots \}$ and $[\overline Q_+, T(z) ] =0$, and the sigma model will be conformally-invariant. For ${\cal T}(z)$ or ${\widetilde{\cal T}(z)}$  to be a global section of $\widehat{\cal A}$, it must be true that ${\cal T}(z) = {\widetilde{\cal T}(z)}$ on any overlap $U_a \cap U_b$ in $X$. Let us examine this further by considering an example.

For ease of illustration, we shall consider an example whereby $\textrm{dim}_{\mathbb C} X = \textrm{rank} ({\cal E}) = 1$, say $\cal E$ is a certain $U(1)$ line bundle over $X = \mathbb {CP}^1$. In order for us to consider an underlying sigma model that is physically-consistent (whereby one can at least define a sheaf of chiral algebras globally over $\mathbb {CP}^1$), we require that $\cal E$ be chosen such that $ch_2({\cal E}) = ch_2(T{\mathbb {CP}^1})$. However, we do not necessarily require that $-c_1 ({\cal E}) = c_1(T{\mathbb {CP}^1})$ or equivalently, $c_1 ({\widetilde{\cal E}}) = c_1(T{\mathbb {CP}^1})$  (and why this is so would be clear momentarily). Since $\mathbb{CP}^1$ can be considered as the complex $\gamma$-plane plus a point at infinity, we can cover it with two open sets, $U_1$ and $U_2$, where $U_1$ is the complex $\gamma$-plane, and $U_2$ is the complex $\tilde\gamma$-plane, such that $\tilde \gamma = 1 /\gamma$. And since $\cal E$ and therefore ${\widetilde{\cal E}}$ is a $U(1)$ line bundle, the transition function $A$ in (\ref{auto1})-(\ref{auto2}) will be given by $e^{i \theta(\gamma)}$,    where $\theta(\gamma)$ is some real, holomorphic function of   $\gamma$. By substituting the definitions of $\tilde\beta$, $\tilde \gamma$, $\tilde b$ and $\tilde c$ from (\ref{auto1})-(\ref{auto2}) into $\widetilde{\cal T}(z)$,  we compute that\footnote{Note that in our computation, we have conveniently chosen the arbitrary, local (1,0)-form $B(\gamma)d\gamma$ on $\mathbb {CP}^1$ (associated with the current $J_B$ of section 5.5) to be one with $B(\gamma) = 2 \gamma$.}   
\be
{\widetilde{\cal T}(z)} - {\cal T}(z)  = \partial_z \left ( {{\partial_z \gamma} \over{\gamma}}\right) + \dots, 
\label{fraktzanomaly}         
\ee
where ``$\dots$" are terms involving the fields $b$, $c$ and the function $\theta(\gamma)$. Note that in general, there is no sensible way to remove the terms on the RHS of (\ref{fraktzanomaly}) through a consistent redefinition of ${\cal T}(z)$ and $\widetilde{\cal T}(z)$ (such that ${\cal T}(z)$ and $\widetilde{\cal T}(z)$ continue to be invariant under the symmetries of the terms on the RHS of (\ref{fraktzanomaly}), and have the correct OPE's, as stress tensors, with the elementary fields $\beta$, $\gamma$, $b$ and $c$). Hence, we find that neither ${\cal T}(z)$ nor ${\widetilde{\cal T}(z)}$ can be a global section of $\widehat{\cal A}$, i.e., ${{\cal T}(z), {\widetilde{\cal T}(z)}} \notin H^0( \mathbb {CP}^1, \widehat {\cal A})$. In other words, $T(z)$ is not in the $\overline Q_+$-cohomology of the sigma model; there is a conformal anomaly.      This is consistent with an earlier observation made in section 3.1 via eqn.\,(\ref{tzzanomaly}),    where $[\overline Q_+, T_{zz}] \neq 0$ in general but
\be
[\overline Q_+, T_{zz}] = \partial_z (R_{ i \bar j} \partial_z \phi^i \psi^{\bar j}) + \dots.
\label{tanomaly}
\ee    
Note that since $\overline Q_+$ generates the BRST symmetry (i.e. an automorphism) of the twisted heterotic sigma model via the field transformations (\ref{txtwist}), (\ref{fraktzanomaly}) will be an analog in Cech cohomology of the relation (\ref{tanomaly}) (as briefly mentioned in footnote 5 of section 3.1). In fact, $R_{ i \bar j} \partial_z \phi^i \psi^{\bar j}$ can be interpreted as the counterpart of the term ${\partial_z \gamma} / {\gamma}$ in conventional physics notation as follows. Apart from an obvious comparison of (\ref{tanomaly}) and (\ref{fraktzanomaly}), note that ${\partial_z \gamma} / {\gamma} = - {\partial_z \tilde\gamma} / {\tilde \gamma}$, i.e., ${\partial_z \gamma} /{\gamma}$ is a holomorphic operator over $U_1 \cap U_2$. Moreover, it cannot be expressed as a difference between an operator that is holomorphic in $U_1$ and an operator that is holomorphic in $U_2$. Thus, it is  a dimension one class in the first Cech cohomology group $H^1(\mathbb {CP}^1, \widehat{\cal A})$. Hence, from our $\overline Q_+$-Cech cohomology dictionary,   ${\partial_z \gamma} /{\gamma}$ will correspond to a dimension one operator in the $\overline Q_+$-cohomology of the sigma model with $q_R =1$, namely $R_{ i \bar j} \partial_z \phi^i \psi^{\bar j}$ (which indeed takes the correct form of a $\overline Q_+$-invariant, dimension $(1,0)$ operator with $q_R=1$ as discussed in section 5.1). Since the Ricci tensor $R_{ i \bar j}$ is proportional to the one-loop beta-function of the sigma model,  this correspondence allows one to interpret the one-loop beta-function purely in terms of holomorphic data.

One can certainly consider other higher-dimensional examples in a similar fashion. In fact, it can be shown mathematically that ${\widetilde{\cal T}(z)} \neq {\cal T}(z)$ for any $X$ and ${\widetilde{\cal E}}$ if $[c_1({\widetilde {\cal E}}) - c_1(TX)] \neq 0$ \cite{GMS1, GMS3}. One can indeed see that $[c_1(\widetilde {\cal E}) - c_1(TX)] = - [c_1({\cal E}) + c_1(TX)]$ characterises a conformal anomaly of the twisted heterotic sigma model as follows.  Recall from section 3.1 that the RHS of (\ref{tanomaly}) captures the violation in the conformal structure of the sigma model by the one-loop beta-function. It will vanish if $X$ is a Ricci-flat manifold and if the curvature of the bundle $\cal E$ obeys the Donaldson-Uhlenbeck-Yau equation. Both these conditions can be trivially satisfied if $c_1(TX) = c_1({\cal E}) = 0$, which then implies that $-[c_1({\cal E}) + c_1(TX)] = 0$.         
  
Thus, the obstruction to a globally-defined $T(z)$ operator, characterised by a non-vanishing cohomology class $[c_1({\cal E}) + c_1(TX)]$, translates to a lack of invariance under arbitrary, holomorphic reparameterisations on the worldsheet $\Sigma$
 of the $\overline Q_+$-cohomology of the underlying  twisted heterotic sigma model.

\newsection{The Half-Twisted B-Model and the Mirror Chiral de Rham Complex}

In this section, we shall consider a specific situation in which one of the two anomalies discussed in section 5.6 automatically vanish, thus enabling us to consider, in section 7, other interesting and physically consistent applications of the sheaf of chiral algebras that we have been studying so far. In the process, we will be able to furnish a purely physical interpretation of the sheaf of  CDO's defined by Malikov et al. in \cite{GMS1} known as the $\it{mirror}$   chiral de Rham complex. From the physical definition of the elliptic genus as a specialisation of the genus one partition function, and the CFT state-operator correspondence for a Calabi-Yau target-space, we can express the elliptic genus in terms of the sheaf cohomology of the mirror CDR. In addition, via an equivalence of elliptic genera under mirror symmetry, we can in turn derive a novel, mathematical expression which relates the sheaf cohomology of the CDR on $\widetilde X$, to the sheaf cohomology of the mirror CDR on $X$, where $X$ and $\widetilde X$ are a mirror pair of Calabi-Yau's.

\newsubsection{The $(2,2)$ Locus and the Half-Twisted B-Model}

The $(2,2)$ locus is defined as the set in the moduli space of holomorphic vector bundles $\cal E$ whereby ${\cal E} =TX$. Thus, notice that on the $(2,2)$ locus, the anomaly of the underlying twisted heterotic sigma model quantified by $ch_2({\cal E}) - ch_2(TX)$, vanishes. However, the second anomaly, now quantified by $c_1(\Sigma) c_1(TX)$, will be non-vanishing for a general worldsheet $\Sigma$ unless $c_1(TX) =0$. Incidentally, this is the same as the anomaly cancellation condition for the topological B-model. In addition, $\textrm{rank}({\cal E}) = r = \textrm{dim}_{\mathbb C}X = n$, and the constraint relation discussed at the end of section 5.3 now becomes $\Lambda^nTX \cong K_X$, which in turn implies that one must have $K_X^{\otimes 2} \cong {\cal O}_X$, instead of the stronger Calabi-Yau condition $K_X \cong {\cal O}_X$. This condition has also been derived using a more sophisticated approach in annex A of \cite{other}.                 

Since ${\cal E} = TX$ at the $(2,2)$ locus, one can make the following field replacements: $\lambda_{a} \to \lambda_{i}$, $\lambda^a_z \to \lambda^i_z$, $A(\phi) \to \Gamma(\phi)$ and $F(\phi) \to R(\phi)$, where $A(\phi)$ and $F(\phi)$ are the connection and field strength of the gauge bundle $\cal E$, while $\Gamma(\phi)$ and $R(\phi)$ are the affine connection and Riemann curvature of $X$. In making these replacements in $S_{pert}$ of (\ref{actionpert}), we find that the action of the underlying twisted sigma model at the $(2,2)$ locus will be given by
\begin{eqnarray}        
S_{\mathrm (2,2)}& = & \int_{\Sigma} |d^2z| \left( g_{i{\bar j}} \partial_z \phi^{\bar j} \partial_{\bar z}\phi^i + \psi_{\bar z \bar j} D_z \psi^{\bar j} + \lambda_i D_{\bar z} \lambda^i_z  - {R^{\bar i k}{}_{j \bar l}} \hspace{0.05cm} {\psi}_{\bar z \bar i} {\lambda}_{k}  \lambda^j_z \psi^{\bar l}  \right),
\label{action(2,2)}     
\end{eqnarray} 
where $\psi_{\bar z \bar j}={g_{i \bar j} \psi^i_{\bar z} }$, $\lambda_i = g_{i \bar j} \lambda^{\bar j}$, and $i,j,k, l = 1, 2, \dots, {\textrm{dim}_{\mathbb C} X}$.  As usual, $R_{i {\bar k}  j {\bar l}}$ is the curvature tensor with respect to the Levi-Civita connection $\Gamma^i{}_{lj} = g^{i \bar k}\partial_lg_{j \bar k}$, and the covariant derivatives with respect to the connection induced on the worldsheet are given by 
\be
D_z\psi^{\bar j} = \partial_z \psi^{\bar j} + \Gamma^{\bar j}{}_{\bar i \bar k} \partial_z\phi^{\bar i}\psi^{\bar k}, \qquad  D_{\bar z}\lambda^{i}_z = \partial_{\bar z} \lambda^{i}_z + \Gamma^{i}{}_{j k} \partial_{\bar z} \phi^{j} \lambda^{k}_z.
\ee
Note that $S_{\mathrm (2,2)}$ is equivalent to the topological B-model action defined by Witten in \cite{mirror manifolds}.\footnote{The action $S_{\mathrm{(2,2)}}$ just differs from the explicit form of the B-model action in \cite{mirror manifolds} by a trivial redefinition of the fermi fields, and an integration by parts of the kinetic term of the right-moving fermions on a worldsheet $\Sigma$ without boundary.} Indeed, the theory defined by $S_{\mathrm (2,2)}$ exhibits a topological B-model anomaly as pointed out earlier.

Let us now discuss the classical symmetries of the action $S_{\mathrm (2,2)}$. Firstly, note that $S_{\mathrm (2,2)}$ has a left and right-moving ghost number symmetry whereby the left-moving fermionic fields transform as $\lambda^i_z \to e^{i\alpha}\psi^i$ and $\lambda_i \to e^{-i \alpha} \lambda_i$, and the right-moving fermionic fields transform as $\psi^{\bar i} \to e^{i \alpha}\psi^{\bar i}$ and $\psi_{\bar z \bar i} \to e^{-i \alpha}\psi_{\bar z \bar i}$, where $\alpha$ is real. In other words, the fields $\lambda^i_z$, $\lambda_i$, $\psi^{\bar i}$ and $\psi_{\bar z \bar i}$ can be assigned the $(g_L, g_R)$ left-right ghost numbers $(1,0)$, $(-1,0)$, $(0,1)$ and $(0,-1)$, respectively. The infinitesimal version of this symmetry transformation of the left-moving fermi fields read (after absorbing some trivial constants) 
\be
\delta\lambda^i_z = \lambda^i_z, \quad \delta \lambda_i = - \lambda_i,  
\label{ghostl}
\ee
while those of the right-moving fermi fields read
\be
\delta \psi^{\bar i} = \psi^{\bar i}, \quad \delta \psi_{\bar z \bar j} = - \psi_{\bar z \bar j}.
\label{ghostR}
\ee
The conserved holomorphic (i.e.\,left-moving) current associated with the transformation (\ref{ghostl}) will then be given by
\be
J(z) = \lambda^i_z \lambda_i.    
\label{J}
\ee 
$J(z)$ is clearly a dimension one bosonic current. (There is also an anti-holomorphic conserved current associated with the right-moving ghost symmetry. However, it is irrelevant to our discussion). Secondly, note that $S_{\mathrm (2,2)}$ is also invariant under the following field transformations:
\begin{eqnarray}
\delta \phi^{\bar i} = \lambda^{\bar i}, &\quad & \delta\phi^{i} = 0, \nonumber \\
\label{susyqlb}   
\delta\lambda^{i}_{z} = - \partial_z \phi^{i}, &  \quad & \delta \psi^{\bar i} = - \Gamma^{\bar i}{}_{\bar j \bar k} \lambda^{\bar j} \psi^{\bar k}, \\
\delta \psi^i_{\bar z} = 0, & \quad & \delta \lambda^{\bar i} = 0. \nonumber 
\end{eqnarray} 
The conserved, dimension one fermionic current in this case will be given by 
\be
Q(z) = - \lambda_k \partial_z \phi^{k}.
\label{Q}
\ee
For later convenience, let us label the charge corresponding to the current $Q(z)$ as $Q_L$. Note that $Q_L$ is sometimes written as ${\overline Q}_-$ in the physics literature.      
 
The third set of field transformations that leave $S_{\mathrm (2,2)}$ invariant is given by 
\begin{eqnarray}
\delta \phi^{\bar i} = \psi^{\bar i}, &\quad & \delta\phi^{i} = 0, \nonumber \\
\label{susyqrb}
\delta\psi^{i}_{\bar z} = - \partial_{\bar z} \phi^{i}, &  \quad & \delta \lambda^{\bar i} = - \Gamma^{\bar i}{}_{\bar j \bar k} \psi^{\bar j} \lambda^{\bar k}, \\
\delta \lambda^i_z = \delta \lambda_i = 0, & \quad & \delta \psi^{\bar i} = 0. \nonumber 
\end{eqnarray} 
The corresponding current of the above symmetry is given by $Q_R(\bar z) = g_{i \bar j}\psi^{\bar j}\partial_{\bar z} \phi^i$. Similarly, let us label the conserved charge of $Q_R(\bar z)$ as $Q_R$. Note that $Q_R$ is just ${\overline Q}_+$ of section 2.2 at the $(2,2)$ locus - from the supersymmetry variations in (\ref{txtwist}), and the action $S_{pert}$ in (\ref{actionpert}), we find that the supercurrent of the scalar supercharge ${\overline Q}_+$ is given by ${\overline Q}_+(\bar z) = g_{i \bar j} \psi^{\bar j} \partial_{\bar z} \phi^i$, which actually coincides with $Q_R(\bar z)$.

In Witten's topological B-model, the BRST-charge operator that defines the BRST cohomology is given by $Q_{BRST} = Q_L + Q_R$, where $Q_L$ and $Q_R$ are the above-mentioned left and right-moving (scalar) supercharges which generate the symmetry transformations in (\ref{susyqlb}) and (\ref{susyqrb}), respectively. However, in considering the cohomology of local operators with respect to only ${\overline Q}_+$, we are actually dealing with a greatly enriched variant in which one ignores $Q_L$ and considers $ Q_R$ as the sole effective BRST operator. We shall call this variant the half-twisted B-model. Since the cohomology of local operators is now defined with respect to a single, right-moving, scalar  supercharge $Q_R$, its classes need not be restricted to dimension $(0,0)$ operators (which correspond to ground states). In fact, the physical operators will have dimension $(n,0)$, where $n \geq 0$. Let us verify this important statement next.

From (\ref{action(2,2)}), we find that the anti-holomorphic stress tensor takes the form $ T_{\bar z \bar z} =g_{i \bar j}  \partial_{\bar z} \phi^i \partial_{\bar z} \phi^{\bar j} + g_{i \bar j}  \psi_{\bar z}^i \left ( \partial_{\bar z} \psi^{\bar j} + \Gamma^{\bar j}_{\bar l \bar k}\partial_{\bar z} \phi^{\bar l} \psi^{\bar k} \right)$. One can go on to show that $ T_{\bar z \bar z} = \{ Q_R , - g_{i \bar j} \psi_{\bar z}^i \partial_{\bar z} \phi^{\bar j} \}$, that is, $T_{\bar z \bar z}$ is trivial in $Q_R$-cohomology. Hence, from the arguments in section 3.1, we learn that operators which are non-trivial in the $Q_R$-cohomology must have scaling dimension $(n,0)$, where $n \geq 0$.

On the other hand, the holomorphic stress tensor is given by $T_{zz} =   g_{i \bar j} \partial_z \phi^i \partial_z \phi^{\bar j} + \lambda^i_z D_z \lambda_i$, and one can verify that it can be written as  $T_{zz} = \{ Q_L , -g_{i \bar j} \lambda^i_z \partial_z \phi^{\bar j} \}$,  that is, it is $Q_L$-exact. Since we are only interested in $Q_R$-closed modulo $Q_R$-exact operators, there is no restriction on the value that $n$ can take. These arguments persist in the quantum theory, since a vanishing cohomology in the classical theory continues to vanish when quantum effects are small enough in the perturbative limit.

Consequently, in contrast to the topological B-model, the BRST spectrum of physical operators and states in the half-twisted variant is infinite-dimensional. A specialisation of its genus one partition function, also known as the elliptic genus of $X$, is given by the index of the $Q_R$ operator. Indeed, the half-twisted  model is not a topological field theory, rather, it is a 2d conformal field theory - the full stress tensor derived from its action is exact with respect to the  combination $Q_L +Q_R$, but not $Q_R$ alone. 

In fact, more can be said about the observables of the half-twisted B-model. By the same argument in section 3.1, we can show that a local operator $\cal O$, as an element of the $ Q_R$-cohomology, varies holomorphically with $z$. Moreover, this observation will continue to hold at the quantum level. In addition, since the holomorphic stress tensor can be verified to be  $Q_R$-closed but not $Q_R$-exact (even at the quantum level), the space of local operators will be invariant under holomorphic reparameterisations of the coordinates on the worldsheet.  

Again, via the same arguments in section 3.1, one finds that the correlation functions of local physical operators are always holomorphic in $z$. One also finds that  because the trace of the stress tensor is also trivial in $Q_R$-cohomology, the correlation functions of operators will continue to be invariant under arbitrary scalings of $\Sigma$. Thus, the correlation functions are always independent of the K\"ahler structure on $\Sigma$ but vary holomorphically with its complex structure (as is familiar for chiral algebras). Since the correlation functions are holomorphic in the parameters of the theory, they are protected from perturbative corrections. 

Similar to the situation away from the $(2,2)$ locus as discussed in section 3.1, the $Q_R$-cohomology of holomorphic local operators has a natural structure of a holomorphic chiral algebra that we shall similarly denote as $\cal A$; in addition to having holomorphic expansion coefficients $f_k$, the OPE's of the local operators in the  chiral algebra also obey the usual relations of holomorphy, associativity, and invariance under scalings and arbitrary holomorphic reparameterisations of $z$. 

Last but not least, based on the discussion in sections 3.2 and 3.3, a moduli for the chiral algebra can be incorporated into the half-twisted B-model by introducing a non-K\"ahler deformation of $X$ via the addition of the $\cal H$-flux term (\ref{STex}) to the action $S_{\mathrm{(2,2)}}$. As argued in section 5.6, the moduli of the corresponding, globally-defined sheaf of CDO's on $X$ can then be represented by a class in $H^1(X, \Omega^{2,cl}_X)$ through this $\cal H$-flux term.

\bigskip\noindent{\it A Holomorphic Twisted  $N=2$ Superconformal Algebra}

We shall now examine the holomorphic structure of the half-twisted B-model with action $S_{\mathrm{(2,2)}}$. The reason for doing so is that some of its non-trivial aspects can be captured by the characteristics of the sheaf of chiral algebras describing the sigma model on $X$. Moreover, similar to what we had seen in section 5.7, one can also derive an interpretation of these non-trivial aspects purely in terms of mathematical data and vice-versa. We will demonstrate these claims shortly when we consider an example in section 7.1.

Let us write the conserved, dimension two holomorphic stress tensor  associated with the symmetry under holomorphic reparameterisations of the coordinates on the worldsheet  as $T(z) = - T_{zz}$ . Recall that it is given by 
\be
T(z) = - g_{i \bar j} \partial_z \phi^i \partial_z \phi^{\bar j} - \lambda^i_z D_z \lambda_i.
\label{T}
\ee
Also recall that one can write $T(z) = \{ Q_L, G(z) \} =\delta G(z)$, the variation of $G(z)$ under the field transformations  (\ref{susyqlb}), where 
\be
G(z) = g_{i \bar j} \lambda^i_z \partial_z \phi^{\bar j}.
\label{G}
\ee 
Hence, $G(z)$ is a conserved, dimension two fermionic current. Notice that the conserved currents and tensors $J(z)$, $Q(z)$, $T(z)$, $G(z)$ possess only holomorphic scaling dimensions. Thus, their respective spins will also be given by their dimensions.

One can verify that  $J(z)$, $Q(z)$, $T(z)$ and   $G(z)$ are all invariant under the field transformations of (\ref{susyqrb}). In fact, we find that $J(z)$, $Q(z)$, $T(z)$ and $G(z)$ are all $Q_R$-closed operators in the $Q_R$-cohomology of the half-twisted B-model, at least at the classical level. Also note that if $\cal O$ and $\cal O'$ are $Q_R$-closed operators in the $Q_R$-cohomology, i.e., ${\{Q_R, \cal O\}}={\{Q_R, {\cal O}'\}}= 0$, then $\{Q_R, {\cal O}{\cal O}'\} =0$. Moreover, if $\{Q_R, {\cal O}\}=0$, then ${\cal O}\{Q_R, W\}= \{Q_R, {\cal O}W\}$  for any operator $W$.  These two statements mean that the cohomology classes of operators that (anti)commute with $Q_R$ form a closed (and well-defined) algebra under operator products. One can indeed show that $J(z)$, $Q(z)$, $T(z)$ and $G(z)$ form a complete multiplet which generates a closed, holomorphic, ($\it{twisted}$) $N=2$ superconformal algebra with the following OPE relations: 
$$
T(z)T(w) \sim \frac{2T(w)}{(z-w)^2}+\frac{\partial T(w)}{z-w}
\eqno{(6.10a)}
$$
$$  
J(z)J(w) \sim \frac{d}{(z-w)^2};\ 
T(z)J(w) \sim \frac{d}{(z-w)^3}+\frac{J(w)}{(z-w)^2}+\frac{\partial J(w)}{z-w}
\eqno{(6.10b)}
$$
$$
G(z)G(w) \sim 0;\ T(z)G(w)\sim \frac{2G(w)}{(z-w)^2}+\frac{\partial G(w)}{z-w};\ 
J(z)G(w)\sim \frac{G(w)}{z-w}
\eqno{(6.10c)}
$$
$$
Q(z)Q(w)\sim 0; \ T(z)Q(w) \sim\frac{Q(w)}{(z-w)^2}+\frac{\partial Q(w)}{z-w};\ 
J(z)Q(w) \sim - \frac{Q(w)}{z-w}
\eqno{(6.10d)}
$$
$$
Q(z)G(w)\sim \frac{d}{(z-w)^3}+\frac{J(w)}{(z-w)^2}+\frac{T(w)}{z-w},
\eqno{(6.10e)}
$$ 
where $d = \textrm{dim}_{\mathbb C} X$. This structure is isomorphic to a structure of a topological vertex algebra of rank $d$ defined in the mathematical literature \cite{MSV1}. From  (6.10e) above, we see that $G(z)$ is a (worldsheet) superpartner of  $T(z)$ under the supersymmetry generated by the charge $Q_L$ of the supercurrent $Q(z)$. This observation will be relevant to our discussion momentarily. Also notice that the central charge in the stress tensor OPE (6.10a) is zero. This means that the Weyl anomaly vanishes and that the trace of the stress tensor is trivial in $Q_R$-cohomology at the quantum level. This simply reflects the invariance of the correlation functions under scalings of the worldsheet as noted earlier.    
          
The classical, holomorphic, OPE algebra of the half-twisted B-model given by (6.10a)-(6.10e) may or may not persist in the quantum theory. In fact, in a `massive' model where the first Chern class $c_1(X)$ is non-zero, the symmetry of the theory under arbitrary holomorphic reparameterisations of the worldsheet coordinates associated with $T(z)$ will be broken. Likewise for the symmetry associated with its superpartner $G(z)$. Hence, the generators $T(z)$ and $G(z)$ of the holomorphic, (twisted) $N=2$ superconformal algebra (which defines the B-model), will cease to remain as valid physical operators in the $Q_R$-cohomology at the quantum level. This is consistent with the fact that the conformal anomaly discussed in section 5.7 will be non-vanishing for $c_1(X) \neq 0$. We will examine this more closely from a different point of view when we consider an explicit example in section 7.1, where we describe the chiral algebra of the half-twisted B-model in terms of a sheaf of mirror CDR. Once again, we will be able to obtain a purely mathematical interpretation of this physical observation. In particular, we can interpret the non-vanishing beta-function solely in terms of holomorphic data.

\bigskip\noindent{\it $Q_R$-Cohomology Classes of Local Operators}  
 
We shall now discuss the $Q_R$-cohomology of local operators which furnish a holomorphic chiral algebra $\cal A$  of the half-twisted B-model. Note that we can describe the structure of the chiral operators in the half-twisted  B-model by specialising the arguments made in section 5.1 to the case where ${\cal E}=TX$. This can be achieved by making the field replacements $\lambda_a \to \lambda_i$ (where $\lambda_i \in \Phi^*(T^*X)$) and $\lambda^a_z \to \lambda^i_z$ (where $\lambda^i_z \in K \otimes \Phi^*(TX)$). In general, we find that a local operator $\cal F$ in the $Q_R$-cohomology of the half-twisted B-model will be given by ${\cal F}(\phi^i,\partial_z\phi^i,\partial_z^2\phi^i,\dots; \phi^{\bar i},\partial_z\phi^{\bar i},\partial_z^2\phi^{\bar i},\dots; \lambda_{i}, \partial_z\lambda_i, \partial_z^2 \lambda_{i} \dots; \lambda^i_z, \partial_z \lambda^i_z, \partial_z^2 \lambda^i_z  \dots ; \psi^{\bar i})$. If $\cal F$ is homogeneous of degree $k$ in $\psi^{\bar i}$, then it has ghost number $(g_L, g_R) =( p, k)$, where $p$ is determined by the net number of $\lambda^i_z$ over $\lambda_{i}$ fields (and/or of their corresponding derivatives) in $\cal F$. An operator   ${\cal F}(\phi^i,\partial_z\phi^i,\dots; \phi^{\bar i},\partial_z\phi^{\bar i},\dots; \lambda_{i}, \partial_z \lambda_{i}, \dots; \lambda^i_z, \partial_z \lambda^i_z, \dots ; \psi^{\bar i})$  with  $q_R= k$ can be interpreted as a $(0,k)$-form on $X$ with values in a certain tensor product bundle. Let us see this more explicitly.  

For example, a dimension $(0,0)$ operator will generally take the form ${\cal F}(\phi^i,\phi^{\bar i}; \lambda_j ; \psi^{\bar j})= f^{i_1, \dots, i_q }_{\bar j_1,\dots,\bar j_k}(\phi^k, \phi^{\bar k}) \lambda_{i_1} \dots \lambda_{i_q} \psi^{\bar j_i}\dots \psi^{\bar j_k}$. Such an operator will correspond to an ordinary $(0,k)$-form with values in $\Lambda^qTX$, the antisymmetric $q^{th}$ exterior power of the holomorphic tangent bundle of $X$, which one can write explicitly as ${\partial \over {\partial \phi^{i_1}}} \dots {\partial \over {\partial \phi^{i_q}}} f^{i_1, \dots, i_q }_{\bar j_1,\dots,\bar j_k} d\phi^{\bar j_1} \dots d \phi^{\bar j_k}$. In other words, a dimension $(0,0)$ operator will correspond to an element of the sheaf cohomology group $H^k (X, \Lambda^q TX)$. This observation will be important shortly when we discuss the topological chiral ring of ground operators. 

For dimension $(1,0)$ operators, one can have four cases. In the first case, we can have an operator ${\cal F} (\phi^l, \phi^{\bar l}; \partial_z \phi^{\bar i}, \lambda_m; \psi^{\bar j})=f^{j ; m_1, \dots, m_q}_{\bar j_1,\dots \dots \dots ,\bar j_k}(\phi^l, \phi^{\bar l})  g_{j \bar i}\partial_z \phi^{\bar i} \lambda_{m_1} \dots \lambda_{m_q} \psi^{\bar j_i}\dots \psi^{\bar j_k}$ that is linear in $\partial_z \phi^{\bar i}$. Such an operator will correspond to a $(0,k)$-form on $X$ with values in the tensor product of the holomorphic tangent bundle $TX$ and $\Lambda^q TX$, the antisymmetric $q^{th}$ exterior power of the same bundle. In the second case, we can have an operator ${\cal F}(\phi^l, \phi^{\bar l};\partial_z\phi^i, \lambda_m; \psi^{\bar j})=  f^{\bar j ; m_1, \dots, m_q}_{\bar j_1,\dots \dots\dots,\bar j_k}(\phi^l,\phi^{\bar l}) g_{\bar j i}\partial_z\phi^{i}\lambda_{m_1} \dots \lambda_{m_q} \psi^{\bar j_1}\dots\psi^{\bar j_k} $ that is linear in $\partial_z \phi^{i}$.  Such an operator will correspond to a $(0,k)$-form on $X$ with values in the tensor product of the bundle $\overline {TX}$ and $\Lambda^q TX$. In the third case, we can have an operator ${\cal F}(\phi^l, \phi^{\bar l};\lambda_m, \partial_z \lambda_i; \psi^{\bar j})=  f_{\bar j; \bar j_1,\dots,\bar j_k }^{m_1, \dots, m_q}(\phi^l,\phi^{\bar l}) g^{\bar j i}\partial_z\lambda_{i} \psi^{\bar j_1}\dots\psi^{\bar j_k} \lambda_{m_1} \dots \lambda_{m_q}$ that is linear in $\partial_z\lambda_{i}$ and does not depend on any other derivatives. Such an operator corresponds to a $(0,k)$-form on $X$ with values in the  (antisymmetric)   tensor product bundle of ${\overline {E}}$ with $\Lambda^q {TX}$, where the local holomorphic sections of the bundle $E$   are spanned by $\partial_z \lambda_i$, the $z$-derivative of the sections of the holomorphic  cotangent bundle of X. In the last case, we have an operator ${\cal F} (\phi^l, \phi^{\bar l}; \lambda_{k}, \lambda^i_z ; \psi^{\bar j})=f_{i; \bar j_1,\dots,\bar j_k}^{k_1, \dots, k_q}(\phi^l, \phi^{\bar l}) \lambda^i_{z}\psi^{\bar j_i}\dots \psi^{\bar j_k} \lambda_{k_1} \dots \lambda_{k_q}$; here, $\cal F$ may depend on $\phi^i$, $\phi^{\bar i}$, $\lambda_{k}$ and $\lambda^i_z$, but not on their derivatives. Such an operator corresponds to a $(0,k)$-form on $X$ with values in the (antisymmetric) tensor product bundle of $T^*X$ with $\Lambda^q{TX}$. In a similiar fashion, for any integer $n>0$, the operators of dimension $(n,0)$ and charge $q_R = k$ can be interpreted as $(0,k$)-forms with values in a certain tensor product   bundle over $X$. This structure persists in quantum perturbation theory, but there  may be perturbative corrections to the complex structure of the bundle. 

Based on the discussion in section 5.1, the action of $\overline Q_+ =Q_R$ on the above local operators can be succintly described as follows.  Firstly, at the classical level, $Q_R$ does not act as $\bar \partial = d\phi^{\bar i}  {\partial /{\partial \phi^{\bar i}}}$  on a general operator $\cal F$ that contains the derivatives  $\partial_z^m \phi^{\bar i}$ for $m>0$. However, it will act as such on dimension $(0,0)$ operators (since $m=0$ and $\delta \lambda_i =0$), in the absence of perturbative corrections. Secondly, if $X$ is flat, $Q_R$ will  act as the $\bar \partial$ operator on any $\cal F$ at the classical level - the equation of motion $D_z \psi^{\bar i} = 0$ ensures that the action of $Q_R$ on derivatives $\partial_z^m \phi^{\bar i}$ for $m>0$ can be ignored, and since $\delta \lambda_i = \delta \lambda^i_z =0$, one can also ignore the action of $Q_R$ on the $\lambda_i$ and $\lambda^i_z$ fields and their derivatives $\partial_z^m \lambda_i$ and $\partial_z^l \lambda^m_z$ with $m>0$. At the quantum level, for $X$ a flat manifold,  $Q_R$ may receive perturbative corrections from $\bar\partial$-cohomology classes that are constructed locally from the fields appearing in the action such as the class in $H^1(X, \Omega^{2,cl}_X)$.

\bigskip\noindent{\it A Topological Chiral Ring}

From the arguments in section 5.2, we learn that the $Q_R$-invariant ground (i.e. dimension $(0,0)$) operators $\widetilde {\cal F}$ define a topological chiral ring via their OPE  
\be
{{\widetilde {\cal F}}_a  {\widetilde {\cal F}}_b } = {\sum_{q_c =  q_a + q_b}  C_{abc} \ {\widetilde {\cal F}}_c},
\label{OPEgndCDR}
\ee
where $C_{abc}$ are structure constants, antisymmetric in their indices, and $q_a$ and $q_b$ represent the $(g_L, g_R)$ ghost number of $\widetilde {\cal F}_a$ and $\widetilde {\cal F}_b$ respectively.  The ring is effectively ${\mathbb Z_2} \times {\mathbb Z_2}$ graded in the absence of non-perturbative worldsheet instantons. At the classical level  (in the absence of perturbative corrections), $Q_R$ acts as $Q_{cl} = \bar \partial$ on any dimension $(0,0)$ operator $\widetilde {\cal F}$. As explained above, since an arbitrary dimension $(0,0)$ operator $\widetilde {\cal F}_d$ with $(g_L, g_R) = (-q,k)$ corresponds to an an element $f^{i_1, \dots, i_q }_{\bar j_1,\dots,\bar j_k} {\partial \over {\partial \phi^{i_1}}} \wedge \dots \wedge {\partial \over {\partial \phi^{i_q}}} \wedge d\phi^{\bar j_1} \wedge \dots  \wedge d \phi^{\bar j_k}$ of the sheaf cohomology group $H^k (X, \Lambda^q TX)$, the $\it{classical}$ ring is just the graded Dolbeault cohomology ring $H_{\bar \partial}^*(X, \Lambda^*TX)$. Alternatively, via the Cech-Dolbeault isomorphism in ordinary differential geometry, the classical ring can also be interpreted as the graded Cech cohomology ring $H^{*}(X, \Lambda^{*}{TX})$.  The operators $\widetilde {\cal F}$ will either be non-Grassmannian or Grassmannian, obeying either commutators or anti-commutators, depending on whether they contain an even or odd number of fermionic $\psi$ and $\lambda$ fields.

\newsubsection{Sheaf of Mirror Chiral de Rham Complex}

We shall now summarise the results of sections 5.3, 5.4 and 5.5 that have been specialised to the case where ${\cal E}=TX$. This will allow us to describe the appropriate sheaf of CDO's associated with the half-twisted B-model on a complex, hermitian manifold $X$.        

Firstly, note that as in the twisted heterotic sigma model of section 2, the perturbative chiral algebra $\cal A$ of local, holomorphic operators $\cal F$ in the $Q_R$-cohomology of the half-twisted B-model can also be described via Cech cohomology. (This is true because $Q_R$ also acts as the $\bar \partial$ operator on any $\cal F$ in the half-twisted B-model over an open set $U$ in $X$, and the results in section 5.3 are established based on this $\bar \partial$-action of the BRST supercharge on the local operators in $U \subset X$.)  In particular, let the sheaf $\widehat {\cal A}$ of chiral algebras have as its $\it{local}$ sections the $Q_R$-closed operators ${\widehat {\cal F}} (\phi^i,\partial_z\phi^i,\dots; \partial_z\phi^{\bar i},\dots; \lambda_{i}, \partial_z \lambda_{i}, \dots; \lambda^i_z, \partial_z \lambda^i_z, \dots)$ that  are $\psi^{\bar i}$-independent (i.e. $g_R =0$) $\it{and}$ $\phi^{\bar i}$-independent, with arbitrary integer values of $g_L$. Then, the $Q_R$-cohomology of local operators can be described in terms of the Cech cohomology of $\widehat {\cal A}$ for all $g_R$ in quantum perturbation theory; the perturbative chiral algebra $\cal A$ will thus be given by $\bigoplus_{g_R} H^{g_R}_{\textrm {Cech}} (X, {\widehat{\cal A}})$ as a vector space. 

\smallskip\noindent{\it The Local Action and its Holomorphic Structure} 

Next, we shall now describe the local structure of the sheaf  $\widehat {\cal A}$. Since ${\cal E} =TX$, the local action (derived from a flat hermitian metric) of the half-twisted B-model on a small open set $U \subset X$ will be given by 
\be
I = {1 \over 2 \pi} \int_{\Sigma} |d^2 z| \sum_{i, \bar j} \delta_{i \bar j} \left ( \partial_z \phi^{\bar j} \partial_{\bar z}\phi^i + \lambda^{\bar j} \partial_{\bar z} \lambda^i_z +  \psi ^i_{\bar z} \partial_z \psi^{\bar j}  \right),
\label{CDRlocalaction}
\ee       
where $\lambda^{\bar j}$ is a scalar on $\Sigma$ with values in the pull-back bundle $\Phi^*({\overline {TX}})$, and $\delta_{i \bar j} \lambda^{\bar j} =\lambda_i$. Note that locally on $U$, the Ricci tensor vanishes and the term containing the class $H^1(X, \Omega^{2,cl}_X)$ is also $Q_R$-trivial. Hence, via the same arguments found in section 5.4, we learn that $Q_R$ acts as $\psi^{\bar i} \partial / \partial \phi^{\bar i}$ on the local operators in $U$. Therefore, the $Q_R$-invariant operators of the local theory with action (\ref{CDRlocalaction}) take the form ${\widehat {\cal F}}(\phi^i,\partial_z\phi^i,\dots;\partial_z\phi^{\bar i},\partial_z^2\phi^{\bar i},\dots ; \lambda_{i}, \partial_z \lambda_{i}, \partial_z^2 \lambda_{i},  \dots; \lambda^i_z, \partial_z \lambda^i_z,  \partial_z^2 \lambda^i_z, \newline \dots)$. Note  also that the operators have to be  $\psi^{\bar i}$-independent on $U$ (see arguments in section 5.3), in addition to being $\phi^{\bar i}$-independent. Clearly, the operators, in their dependence on the center of mass coordinate of the string whose worldsheet theory is the half-twisted B-model, is holomorphic. Therefore,  the $Q_R$-cohomology of operators in the chiral algebra of the local half-twisted B-model with action (\ref{CDRlocalaction}),  are local sections of the sheaf of chiral algebras $\widehat {\cal A}$.    

The local theory with action (\ref{CDRlocalaction}) has an underlying, holomorphic, twisted $N=2$ superconformal structure as follows. Firstly, the action is invariant under the following field transformations
\be
\delta \lambda^i_z = \lambda^i_z, \quad \delta \lambda_{i} = - \lambda_{i}, \qquad \textrm{and} \qquad \delta \phi^{\bar i} = \lambda^{\bar i}, \quad \delta\lambda^{i}_{z} = - \partial_z \phi^{i},   
\label{CDRghostL}
\ee
where the corresponding conserved currents are given by the dimension one, bosonic and fermionic operators ${\widehat J}(z)$ and ${\widehat Q}(z)$ respectively. They can be written as 
\be
{{\widehat J}(z)} = \lambda^{k}_z \lambda_k \qquad \textrm{and} \qquad {{\widehat Q}(z)} = - \lambda_k \partial_z \phi^{k}. 
\label{CDRJ}
\ee 
Note that we also have the relation $[{\widehat Q}, {\widehat J}(z)] =  {\widehat Q}(z)$, where $\widehat Q$ is the charge of the current ${\widehat Q}(z)$.  Secondly, the conserved, holomorphic stress tensor is given by 
\be
{\widehat T}(z) = - \delta_{i \bar j} \partial_z \phi^i \partial_z \phi^{\bar j} - \lambda^{k}_z \partial_z \lambda_k,
\label{CDRT}
\ee
where one can derive another conserved, fermionic current ${\widehat G}(z)$, such that ${\widehat T}(z) = \{{\widehat Q}, {\widehat G}(z)\}$, and
\be
{\widehat G}(z) = {\delta}_{i \bar j} \lambda^{i}_z \partial_z \phi^{\bar j}.
\label{G}
\ee      
One can verify that  ${\widehat J}(z)$, ${\widehat Q}(z)$, ${\widehat T}(z)$ and   ${\widehat G}(z)$ satisfy the same OPE relations as that satisfied by $J(z)$, $Q(z)$, $T(z)$ and   $G(z)$ in (6.10a)-(6.10e). In other words, they furnish the same twisted $N=2$ superconformal algebra satisfied by $J(z)$, $Q(z)$, $T(z)$ and   $G(z)$  in the global version of the classical half-twisted B-model with action $S_{\mathrm{(2,2)}}$. In fact, ${\widehat J}(z)$, ${\widehat Q}(z)$, ${\widehat T}(z)$ and   ${\widehat G}(z)$ are simply local versions of  $J(z)$, $Q(z)$, $T(z)$ and   $G(z)$ respectively. Hence, if there is no obstruction to a global definition of  ${\widehat J}(z)$, ${\widehat Q}(z)$, ${\widehat T}(z)$ and   ${\widehat G}(z)$ in the quantum theory, the symmetries associated with ${ J}(z)$, ${ Q}(z)$, ${ T}(z)$ and  ${ G}(z)$ will persist in the non-linear half-twisted B-model at the quantum level.  Another way to see this is to first notice that ${J}(z)$, ${Q}(z)$, ${T}(z)$ and   ${G}(z)$ are $\psi^{\bar i}$-independent operators and as such, will correspond to classes in $H^0(X, \widehat {\cal A})$ (from our $Q_R$-Cech cohomology dictionary). Hence, these operators will exist in the $Q_R$-cohomology if they correspond to global sections of  $\widehat {\cal A}$. We will determine the specific type of vertex algebra that $\widehat {\cal A}$ represents shortly.

\bigskip\noindent{\it The Free $bc$-$\beta\gamma$ System} 

Let us now set $\beta_i  =   \delta_{i \bar j} \partial_z \phi^{\bar j}$, $\gamma^i = \phi^i$, $\lambda^i_z = b^i$ and $\lambda_i = c_i$, whereby $\beta_i$ and $\gamma^i$ are bosonic operators of dimension $(1,0)$ and $(0,0)$, while $b^i$ and $c_i$ are fermionic operators of dimension $(1,0)$ and $(0,0)$ respectively. Then, the $Q_R$-cohomology of operators regular in $U$ can be represented by arbitrary local functions of $\beta$, $\gamma$, $b$ and $c$, of the form ${\widehat {\cal F}} (\gamma, \partial_z \gamma, \partial_z^2 \gamma, \dots, \beta, \partial_z \beta, \partial_z^2 \beta, \dots, b, \partial_z  b, \partial_z^2 b, \dots, c, \partial_z c, \partial_z^2 c, \dots)$. The operators $\beta$ and $\gamma$ have the operator products of a standard $\beta\gamma$ system.  The products $\beta\cdot\beta$ and
$\gamma\cdot\gamma$ are non-singular, while
\be
\beta_i(z)\gamma^j(z') = -{\delta_{i}^j\over z-z'}+{\rm regular}.
\ee
Similarly, the operators $b$ and $c$ have the operator products of a standard $bc$ system. The products $b \cdot b$ and $c \cdot c$ are non-singular, while
\be
b^i(z) c_j(z')={\delta^i_{j}\over z-z'}+{\rm regular}.
\ee
These statements can be deduced from the flat action (\ref{CDRlocalaction}) by standard field theory methods. We can write down an action for the fields $\beta$, $\gamma$, $b$ and $c$, regarded as free elementary fields, which reproduces these OPE's.  It is simply the action of the following $bc$-$\beta\gamma$ system: 
\be
I_{bc \textrm{-} \beta\gamma}= {1\over
2\pi} \int_{\Sigma} |d^2z|  \sum_i \left ( \beta_i \partial_{\bar z}\gamma^i + b^i \partial_{\bar z} c_i \right).
\label{CDRbcaction}
\ee
Hence, we find that the linear (i.e.\,local) version of the $bc$-$\beta\gamma$ system above reproduces the $Q_R$-cohomology of $\psi^{\bar i}$-independent operators of the half-twisted B-model on $U$, i.e., the local sections of $\widehat {\cal A}$.

At this point, one can make an important observation about the conserved tensor and current ${\widehat T}(z)$ and ${\widehat G}(z)$ of the local half-twisted B-model with action (\ref{CDRlocalaction}), in the context of the local version of the $bc$-$\beta\gamma$ system above. Firstly, notice that the free $bc$-$\beta\gamma$ action (\ref{CDRbcaction}) is invariant under the following supersymmetric field variations $\delta b^i =-\partial_z \gamma^i$ and $\delta \beta_i = \partial_z c_i$, where the corresponding conserved, dimension one fermionic supercurrent is given by ${\cal Q}(z) = c_i \partial_z \gamma^i$. The holomorphic stress tensor of the local $bc$-$\beta\gamma$ system which generates the symmetries of the system under arbitrary holomorphic reparameterisations of the coordinates on $\Sigma$, is given by
\be
{\cal T}(z) = - \beta_i \partial_z \gamma^i - b^i \partial_z c_i.
\ee
Note that we also have the relation $\{ {\cal Q}, {\cal G}(z) \} = {\cal T}(z)$, where 
\be
{\cal G}(z) = b^i \beta_i 
\ee
 is a conserved, dimension two fermionic current that is the worldsheet superpartner of ${\cal T}(z)$. (Once again, we have omitted the normal-ordering symbol in writing the above conserved tensor and current for notational simplicity.) Via the respective identification of the fields $\beta_i$, $\gamma^i$, $c_k$ and $b^k$ with $g_{i \bar j}\partial_z \phi^{\bar j}$, $\phi^i$, $\lambda_{k}$ and $\lambda^k_z$, we find that ${\cal T}(z)$ and ${\cal G}(z)$ coincide with ${\widehat T}(z)$ and ${\widehat G}(z)$ respectively. This means that locally on $X$, the half-twisted B-model and the $bc$-$\beta\gamma$ system have the same generators of general holomorphic coordinate transformations on the worldsheet. This observation will be important in section 7.1, when we consider an explicit example.        

The $bc$-$\beta\gamma$ system above will certainly reproduce the $Q_R$-cohomology of $\psi^{\bar i}$-independent operators globally on $X$ if its non-linear version is anomaly-free. In order to ascertain the potential anomalies of the non-linear $bc$-$\beta \gamma$ system,  one must first make global sense of the non-linear $bc$-$\beta\gamma$ system with action (\ref{CDRbcaction}). To this end, one must interpret $\gamma$ as a map $\gamma:\Sigma\to X$, $\beta$ as a $(1,0)$-form on $\Sigma$ with values in the pull-back $\gamma^*(T^*X)$, the fermionic field $c$ as a scalar on $\Sigma$ with values in the pull-back $\gamma^*(T^*X)$, and the fermionic field $b$ as a $(1,0)$-form on $\Sigma$ with values in the pull-back $\gamma^* (TX)$. Next, expand around a classical solution of the non-linear $bc$-$\beta\gamma$ system, represented by a holomorphic map $\gamma_0:\Sigma \to X$, and a section $c_0$ of the pull-back $\gamma_0^*(T^*X)$. Setting ${\gamma} =\gamma_0 +\gamma'$, and $c = c_0 + c'$, the action, expanded to quadratic order about this solution, is $(1/2\pi) \left [ (\beta , {\overline D \gamma'}) + (b, \overline D c') \right]$. $\gamma'$, being a deformation of the coordinate $\gamma_0$ on $X$, is a section of the pull-back $\gamma_0^* (TX)$. Thus, the kinetic operator of the $\beta$ and $\gamma$ fields is the $\overline D$ operator on sections of $ \gamma_0^*(TX)$; it is the complex conjugate of the $D$ operator of the kinetic term of the right-moving fermions in $S_{\mathrm{(2,2)}}$.  Complex conjugation reverses the sign of the anomalies, but here the fields are bosonic, while in $S_{\mathrm{(2,2)}}$, they are fermionic; this gives a second sign change. Hence, the anomalies due to the $\beta\gamma$ kinetic operator are the same as those arising from the kinetic operator acting on the right-moving fermions in the half-twisted B-model.\footnote{Notice that the $D$ operator in $S_{\mathrm{(2,2)}}$  acts on sections of the pull-back of the anti-holomorphic bundle $\overline {TX}$ instead of the holomorphic bundle $TX$. However, this difference is irrelevant with regard to anomalies since $ch_2(\overline E) = ch_2(E)$ for any holomorphic vector bundle $E$.} Next, since $c'$ is a deformation of $c_0$, it will be a section of the pull-back $\gamma_0^*(T^*X)$. The kinetic operator of the $b$ and $c$ fields is therefore the $\overline D$ operator acting on sections of $\gamma_0^*(T^*X)$. Now, introduce a spin structure on $\Sigma$, so that we can equivalently interpret $\overline D$ as the complex conjugate of the Dirac operator acting on sections $K^{-1/2} \otimes \gamma_0^*(T^*X)$. Using the same argument found in section 4, we find that by tensoring $K^{-1/2}$ with $\gamma_0^*(T^*X)$, one will get an additional term ${1\over 2}c_1(\Sigma) c_1(T^*X)$. However, since $TX$ is a complex vector bundle, we will have $T^*X = {TX}^{\vee} \cong {\overline {TX}}$, and because $c_1(\overline {TX}) = - c_1(TX)$, the additional term can actually be written as $-{1\over 2} c_1(\Sigma)c_1(TX)$. Moreover, we also have $ch_2({TX}^{\vee}) = ch_2(\overline {TX}) = ch_2(TX)$. Thus, the anomalies due to the kinetic operator of the $b$ and $c$ fields, are the same as those due to the $\overline D$ operator acting on the left-moving fermions in $S_{\mathrm{(2,2)}}$. Hence, the non-linear $bc$-$\beta\gamma$ system has exactly the same anomalies as the underlying half-twisted B-model - the anomaly cancellation condition is $c_1(\Sigma) c_1(X) = 0$. If the anomaly vanishes, the $bc$-$\beta\gamma$ system will reproduce the $Q_R$-cohomology of $\psi^{\bar i}$-independent operators and their OPE's globally on $X$, i.e., one can find a global section of $\widehat {\cal A}$. In other words, for a general $\Sigma$ (where $c_1(\Sigma) \neq 0$), one can find global sections of $\widehat {\cal A}$ if only if $c_1(X) = 0$. Else, on any target space $X$ with $c_1(X) \neq 0$, one needs to work locally on $\Sigma$ (such that one can choose $c_1(\Sigma) = 0$). This last observation will be important when we consider an explicit example in section 7.1 where $c_1(X) \neq 0$.

Locally on $X$, the $Q_R$-cohomology of the half-twisted B-model is non-vanishing only for $g_R = 0$. However, there can generically be cohomology in higher degrees globally on $X$. Nevertheless, as explained in section 5.4, the $Q_R$-cohomology classes of positive degree (i.e.\,$g_R > 0$) can still be described in the framework of the free $bc$-$\beta\gamma$ system via Cech cohomology - the operators with degree $g_R > 0$ can be represented as Cech-$g_R$ cocycles that generate the ${g^{th}_R}$ Cech cohomology of the sheaf $\widehat {\cal A}$ of the chiral algebra of the linear (i.e. free) $bc$-$\beta\gamma$ system (with  action a linearised version of (\ref{CDRbcaction})).

As for the moduli of the theory, the complex structure is built into the definition of the fields in (\ref{CDRbcaction}). The moduli of the chiral algebra $\cal A$, given by a class in $H^1(X, \Omega^{2,cl}_X)$, is built into the definition of Cech 1-cocycles that represent the admissible automorphisms of the sheaves of  free conformal fields theories (see section 5.6). 

By specialising the arguments in section 5.4 to ${\cal E} =TX$, we shall now discuss the computation of a correlation function of  cohomology classes of local operators within the framework of the free $bc$-$\beta\gamma$ system. As explained in section 5.4, due to a right-moving ghost number anomaly, for generic correlation functions in perturbation theory to be non-vanishing, it is a requirement that some of the local operators have positive degrees. Hence, from our description above, the computation of the correlation functions will involve cup products of Cech cohomology groups and their corresponding maps into complex numbers. We can illustrate this scheme by computing a generic correlation function of dimension $(0,0)$ operators on a genus-zero Riemann surface such as a sphere. To this end, recall from section 6.1 that a dimension $(0,0)$ operator ${\cal O}_i$ with ghost number $(g_L, g_R) = (-p_i, q_i)$ can be interpreted as a $(0, q_i)$-form with values in the holomorphic bundle $\Lambda^{p_i}TX$. Thus, it represents a class in the Cech cohomology group $H^{q_i}( X, \Lambda^{p_i}{TX})$. Note that due to the additional left-moving ghost number anomaly, the correlation functions of our model must also satisfy $\sum_i p_i  =  \sum_i q_i = \textrm{dim}_{\mathbb C}X =n$ in perturbation theory, so as to be non-vanishing on the sphere. Since the half twisted B-model must be restricted to holomorphic  maps via the fixed-point theorem and the BRST field transformations in (\ref{susyqrb}),  the correlation function path integral will reduce to an integral over the moduli space of holomorphic maps. Because we are considering degree-zero maps in perturbation theory, the moduli space of holomorphic maps will be $X$ itself, i.e., the path integral reduces to an integral over the target space $X$. In summary, we find that a non-vanishing $\it{perturbative}$ correlation function involving $s$ dimension $(0,0)$ operators ${\cal O}_1$, ${\cal O}_2$, $\dots$, ${\cal O}_s$ on the sphere, can be computed as 
\be
\left < {\cal O}_1(z_1) \dots {\cal O}_s(z_s) \right>_0 = \int_X {\cal W}^{n,n}, 
\label{corr}
\ee                             
where  ${\cal W}^{n,n}$ is a top-degree form on $X$ which represents a class in the Cech cohomology group $H^n(X, K_X)$. This $(n,n)$-form  is obtained via the sequence of  maps
\be
{ H^{q_1} (X, \Lambda^{p_1} {TX}) \otimes \dots \otimes H^{q_s} (X, \Lambda^{p_s} {TX}) } \to H^n(X, \otimes_{i=1}^s \Lambda^{p_i} { TX}) \to H^n(X, K_X),
\label{classicalcorr}
\ee          
where $\sum_{i=1}^s q_i = \sum_{i=1}^s p_i = n$. The first map is given by the cup product of Cech cohomology classes  which represent  the corresponding dimension $(0,0)$ operators, while the second map is given by a wedge product of exterior powers of the holomorphic cotangent bundle. The third map is due to the constraint relation $\Lambda^n TX \cong K_X$. Similar procedures will apply in the computation of correlation functions of local operators with higher dimension.        

Note that in order to compute a $\it{non}$-$\it{perturbative}$ correlation function of dimension $(0,0)$ operators with $(g_L, g_R) = (-p_i, q_i)$, the operators must instead be represented by Cech cohomology classes $H^{q_i}({\cal M}, \Lambda^{p_i}T{\cal M})$ in the moduli space $\cal M$ of worldsheet instantons.\footnote{ This means that  the Cech cohomology classes in $X$ of (\ref{classicalcorr}),   will be replaced by Cech cohomology classes in the moduli space of worldsheet instantons      (See \cite{other}).}  An extension of this recipe to compute the non-perturbative correlation functions of local operators of higher dimension, will therefore serve as  the basis of a chiral version of quantum cohomology.

 \bigskip\noindent{\it The Sheaf \ $\widetilde{\Omega}^{ch}_X$ of Mirror Chiral de Rham Complex on $X$}    

In the case of the twisted heterotic sigma model, where ${\cal E} \neq TX$ but is equivalent to some arbitrary holomorphic vector bundle over $X$, we showed in section 5.5 that the relevant, free $bc$-$\beta\gamma$ system with action (\ref{bcactionalternative}) will reproduce the vertex superalgebras spanned by chiral differential operators on the exterior algebra $\Lambda {\widetilde {\cal E}}$. One may then ask the following question: in the case of the half-twisted B-model, what kind of vertex superalgebra does the free $bc$-$\beta\gamma$ system with action (\ref{CDRbcaction}) reproduce? In other words, what kind of sheaf does $\widehat {\cal A}$ mathematically describe in the case of the half-twisted B-model? 

In order to ascertain this, one must first and foremost determine the admissible automorphisms of the free $bc$-$\beta\gamma$ system with action (\ref{CDRbcaction}) (as was done for the ${\cal E} \neq TX$ case in section 5.5). Since we are considering ${\cal E} =TX$ or equivalently, $\widetilde {\cal E} = TX^{\vee}$, the components of the transition function matrix of the holomorphic vector bundle, given by $A_j{}^i$ in section 5.5, will now generate inverse holomorphic coordinate transformations on $X$. In other words, we must make the following replacements in (\ref{auto1})-(\ref{auto2}):
\begin{eqnarray}
A_n{}^l & = & {\partial \gamma^l \over {\partial g^n}}, 
\\
(\partial_k A^{-1})_l{}^m  & = &  {\partial^2 g^m \over{\partial \gamma^k \partial \gamma^l}},
\\
D^k{}_i & = &  {\partial \gamma^k \over {\partial g^i}},
\end{eqnarray}        
where $g^i(\gamma) = \widetilde {\gamma}^i$, and $i,,k,l,m, n = 1, 2, \dots, {\textrm{dim}_{\mathbb C} X}$.    As the obstruction to gluing of sheaves $ch_2(TX) - ch_2(\widetilde {\cal E})$ vanishes for any $X$ in the half-twisted B-model at the $(2,2)$ locus, it will mean that from our discussion in section 5.5 on the local symmetries of the associated free $bc$-$\beta\gamma$ system, the extension of groups given by (\ref{group extension}) will be trivial. Thus, the universal cohomology group $H^2({\widetilde H}, \Omega^{2,cl}_{\widetilde H})$ which characterises  the extension's non-triviality, will also vanish. As explained in section 5.5, this will mean that $\widetilde G = \widetilde H$, i.e., the admissible automorphisms of the associated free $bc$-$\beta\gamma$ system are solely generated by $\widetilde H$. Consequently, the last term on the RHS of (\ref{autobeta}) can be set to zero in the present computation. Hence, the admissible automorphisms of the free $bc$-$\beta\gamma$ system which describes the half-twisted B-model locally on $X$ will be given by:
\begin{eqnarray}
\label{autoCDRgamma}
{\tilde \gamma}^i & = & g^i (\gamma) ,\\
\label{autoCDRbeta}
{\tilde \beta}_i  & = &   {\partial \gamma^k \over {\partial{\widetilde \gamma^i}}}\beta_k   + { {{\partial \gamma^k}\over {\partial {\widetilde \gamma^l}}} {{\partial \over {\partial {\widetilde \gamma^i}}} {\left ( {\partial {\widetilde \gamma^l}} \over {\partial {\gamma^j}} \right )}} b^j c_k},  \\     
\label{autoCDRc}  
{\tilde c}_i & = & {\partial \gamma^k \over{\partial{\widetilde \gamma^i}}}c_k , \\
\label{autoCDRb}
{\tilde b}^i & = & {\partial{\widetilde \gamma^i} \over{\partial \gamma^k}}b^k,
\end{eqnarray} 
where $i,j,k, l = 1, 2, \dots, {\textrm{dim}_{\mathbb C} X}$. The field transformations for the $b$ and $c$ fields in (\ref{autoCDRgamma})-(\ref{autoCDRb}), are the inverse of the $b$ and $c$ field transformations in (3.17a)-(3.17d) of \cite{MSV1} which define the admissible automorphisms of a sheaf of conformal vertex superalgebras mathematically known as the chiral de Rham complex. This means that (\ref{autoCDRgamma})-(\ref{autoCDRb}) define the admissible automorphisms of an isomorphic sheaf  called the mirror chiral de Rham complex \cite{GMS1}. Indeed, from the explicit definition of the sheaf of mirror chiral de Rham complex in \cite{GMS1}, as we have shown  using purely physical arguments thus far, the $b$ and $c$ fields must take values in the bundles $\gamma^*(TX)$ and $\gamma^*(TX^{\vee})$ respectively. Hence, we learn that $\widehat {\cal A}$ is the sheaf of mirror chiral de Rham complex on $X$.  We shall henceforth label the sheaf of chiral algebras $\widehat {\cal A}$, associated with the half-twisted B-model on $X$, as the sheaf ${\widetilde {\Omega}}^{ch}_X$ of mirror chiral de Rham complex on $X$, or mirror CDR for short. Thus, the chiral algebra $\cal A$ of the half-twisted B-model is, as a vector space, given by $\bigoplus_{g_R} H^{g_R}_{\mathrm{Cech}} (X, {\widetilde {\Omega}}^{ch}_X)$. 

\newsubsection{The Elliptic Genus of the Half-Twisted B-Model}       

Physically, the elliptic genus is a certain specialisation of the partition function of the half-twisted B-model with worldsheet $\Sigma$ being a torus with complex structure $\tau$. It counts the number of  supersymmetric (BPS) states with ${\bar L}_0 = 0$ or rather, the right-moving ground states. These are simply the states in the $Q_R$-cohomology of the half-twisted B-model. (Recall that we discussed this in section 3.1.) The elliptic genus is also known to coincide with the Euler characteristic of $X$. Consequently, it is a topological invariant of $X$, and it can be written as a function of two variables $y$ and $q$ as \cite{Landweber, EG} 
\be
 \chi( X, y, q) = \textrm{Tr}_{\cal H} (-1)^F y^{J_L}q^{L_0 - {d\over 8}},
\label{elliptic genus}
\ee
 where $d = \textrm{dim}_{\mathbb C} X$, $q = e^{2 \pi i \tau}$ and $y = e^{2 \pi i  z}$,  with $z$ being a point in the torus Jacobian of the line-bundle over $\Sigma$ which the fermions of the theory are sections thereof. $F= F_L + F_R$ is the total fermion number, $\cal H$ is the Hilbert space obtained via quantising the loop space ${\cal L}X$, while $\textrm{Tr}_{\cal H} (-1)^F$ is the Witten index that counts the difference between the number of bosonic and fermionic states at each energy level $n$. The $U(1)$ charge $J_L$ is actually the left-moving ghost number $g_L$. (We have renamed it here to allow (\ref{elliptic genus}) to takes its standard form as found in the physics literature.)  

Notice that the above discussion on the elliptic genus involves the states but not the operators in the half-twisted B-model. When and how do the local operators come into the picture? In order to associate the elliptic genus with the local operators in the chiral algebra of the sigma model, one has to consider the canonical quantisation of the sigma model on an infinitely long cylinder ${\mathbb R} \times S^1$. If $c_1(X) = 0$, one can  proceed to employ the CFT state-operator isomorphism, from which one can then obtain a correspondence between the above states and operators. The elliptic genus can thus be expressed in terms of the difference between the number of bosonic and fermionic operators in the $Q_R$-cohomology, such that the holomorphic (i.e.\,left-moving) dimension  of the operators $n$, will now correspond to the energy level $n$ of the supersymmetric states that the operators are isomorphic to. Note that if $c_1(X) \neq 0$, the state-operator correspondence will $\it{not}$ be an isomorphism. Rather, the states just furnish a module $\cal V$ of the chiral algebra $\cal A$ of local operators, whereby $\cal V$ is only isomorphic to $\cal A$ if $c_1(X) = 0$. 
Based on the above correspondence, the description of $\cal A$ in terms of the Cech cohomology of ${\widetilde{\Omega}}^{ch}_X$, and the fact that bosonic and fermionic operators have even and odd total ghost numbers $g_L + g_R$ respectively, we find that in the smooth Calabi-Yau case (i.e. $c_1(X) = 0$), the elliptic genus in (\ref{elliptic genus}) can be written as
\be
\chi_B(X, q, y) = q^{-{d/8}}\sum_{{g_L,  g_R}} \sum_{n=0}^{\infty} (-1)^{g_L + g_R} {\mathrm{dim}} {H}^{g_R}(X, {\widetilde {\Omega}}^{ch; g_L}_{X; n}) y^{g_L} q^n,
\label{elliptic genus mirrorCDR}      
\ee 
where ${\widetilde {\Omega}}^{ch, g_L}_{X, n}$ is a sheaf of $\it{mirror}$ CDR on $X$ whose local sections correspond to the $\psi^{\bar i}$-independent  $Q_R$-cohomology classes with dimension $(n,0)$ and left-moving ghost number $g_L$. 

Mathematically, the elliptic genus can be understood as the $S^1$-equivariant Hirzebruch $\chi_y$-genus of the loop space of $X$. Since we have assumed $X$ to be Calabi-Yau in deriving (\ref{elliptic genus CDR}), the elliptic genus $\chi_B(X, q, y)$ will have nice modular properties under $SL(2, \mathbb Z)$. Notice also that $\chi_B(X, q, y)$ is  ${\mathbb Z_{\geq 0}} \times \mathbb Z$  graded by the holomorphic dimension $n$ and left-moving ghost number $g_L$ of the $Q_R$-invariant operators, respectively. The grading by dimension follows naturally from the scale invariance of the  correlation functions and the chiral algebra $\cal A$ of the half-twisted B-model. Note that $\chi_B(X, q, y)$ has no perturbative quantum corrections.\footnote{Absence of quantum corrections can be inferred from the fact that both the energy and the $(-1)^F$ operator that distinguishes the bosonic and fermionic states are exactly conserved quantum mechanically.} However, if $c_1(X) \neq 0$, non-perturbative worldsheet instanton corrections may violate the scale invariance of the correlation functions and hence, the grading by dimension of the operators in $\cal A$.\footnote{In the non-perturbative small radius limit, if $c_1(X) \neq 0$, the contribution from worldsheet instantons (resulting from a pull-back of the $(1,1)$-form $\omega_T$ on holomorphic curves) will serve to renormalise $\omega_T$. This gives rise to dimensional transmutation, whereby the exponential of $\omega_T$ which appears in the non-perturbative correlation functions, will be replaced by a dimensionful scale parameter $\Lambda$.}  Consequently, supersymmetry may be spontaneously broken, thus rendering $\cal V$ empty, as all the bosonic and fermionic operators pair up. 

\newsubsection{Relation to the Sheaf of CDR via Mirror Symmetry}

Note that the elliptic genus is also a specialisation of the partition function of the $\it{untwisted}$ sigma model on a worldsheet $\Sigma$ of genus one. Since the genus one partition functions of a pair of mirror symmetric $(2,2)$ sigma models must be equivalent, it will then mean that the elliptic genus of the half-twisted A-model on $\widetilde X$ is the same as the elliptic genus of the half-twisted B-model on $X$, where $X$ and $\widetilde X$ are a mirror pair of Calabi-Yau's. 

From \cite{MC}, we find that the elliptic genus of the half-twisted A-model on ${\widetilde X}$ will be given by
\be
\chi_A({\widetilde X}, {\tilde q}, {\tilde y}) = {\tilde q}^{-d/8}\sum_{{{\widetilde g_L},  {\widetilde g_R}}} \sum_{m=0}^{\infty} (-1)^{{\widetilde g_L} + {\widetilde g_R}} {\mathrm{dim}} {H}^{\widetilde g_R}({\widetilde X}, {\widehat {\Omega}}^{ch; {\widetilde g_L}}_{{\widetilde X}; m}) {\tilde y}^{{\widetilde g_L}} {\tilde q}^m,    
\label{elliptic genus CDR}      
\ee    
where $d =\textrm{dim}_{\mathbb C} {\widetilde X}$, and ${\widehat {\Omega}}^{ch; {\widetilde g_L}}_{{\widetilde X}; m}$ is a sheaf of CDR on $\widetilde X$ whose local sections correspond to the $\psi^{\bar i}$-independent $Q_R$-cohomology classes of the half-twisted A-model with dimension $(m,0)$ and left-moving ghost number $\widetilde g_L$. 

Note that between the half-twisted A and B models, there is a sign difference in the left-moving ghost number current $J(z)$ - in making the substitution $J(z) \to -J(z)$ in the OPE algebra of (6.10a)-(6.11e), one will get the OPE algebra of the half-twisted A-model defined in (6.10a)-(6.10e) of \cite{MC}. On the other hand, the right-moving ghost number current is the $\it{same}$ for both models, and its charge on a local operator counts the number of $\psi^{\bar i}$ fields it contains, where this number must be less than or equal to $d$ because of the Grassmannian nature of the $\psi^{\bar i}$ fields. In addition, note that ${\tilde q} = e^{2 \pi i {\tilde \tau}}$ is arbitrary, where $\tilde \tau$ is the complex structure of the genus one wordsheet of the half-twisted A-model.  However, in equating the underlying, untwisted $(2,2)$ sigma models on mirror Calabi-Yau pairs, one necessarily works with equivalent worldsheets on both sides of the duality, i.e., $\tau = {\tilde \tau}$. Hence, in equating $\chi_A$ and $\chi_B$ under mirror symmetry, one can set $q = {\tilde q}$. Lastly, notice that the worldsheet fermions of the A and B models are sections of $\it{different}$ line bundles over $\Sigma$. This is due to the inequivalent twists of the A and B models. Thus, $y \neq {\tilde y}$. In summary, we find that           
\be
\sum_{{{\widetilde g_L}}} \sum_{k \geq 0}^{d}  (-1)^{ k+ {\widetilde g_L}} {\mathrm{dim}} {H}^{k}({\widetilde X}, {\widehat {\Omega}}^{ch; {\widetilde g_L}}_{{\widetilde X}; w}) {\tilde y}^{{\widetilde g_L}} = \sum_{{g_L}} \sum_{l \geq 0}^{d} (-1)^{l +g_L} {\mathrm{dim}} {H}^{l}(X, {\widetilde {\Omega}}^{ch; g_L}_{X; w}) y^{g_L}    
\label{relation to CDR}
\ee        
for any $w \geq 0$.     
    
Hence, via (\ref{relation to CDR}), we have an expression which relates the sheaf cohomology of the CDR on $\widetilde X$, to the sheaf cohomology of the mirror CDR on $X$, the  Calabi-Yau mirror of $\widetilde X$. It would certainly be interesting to prove (\ref{relation to CDR}) from a purely mathematical  point of view.            

\newsection{Examples of Sheaves of Mirror CDR}   

In this section, we study in detail, examples of sheaves of mirror CDR and their cohomologies on two different smooth manifolds. Our main objective is to 
illustrate the rather abstract discussion in section 6. In the process, we will again obtain an interesting and novel understanding of the relevant physics in terms of pure mathematical data.

\newsubsection{The Sheaf of Mirror CDR on $\mathbb{CP}^1$}
   
For our first example, following \cite{MC}, we take $X=\mathbb{CP}^1$.  In other words, we will be exploring and analysing the chiral algebra $\cal A$ of  operators in the half-twisted B-model on $\mathbb {CP}^1$. Before we proceed further, recall from our earlier discussion that there is a B-model anomaly quantified by $c_1(\Sigma)c_1(X)$. Since $c_1(\mathbb {CP}^1) \neq 0$, one cannot consistently define the theory on $\mathbb {CP}^1$ unless $c_1(\Sigma) = 0$. This can be achieved either by working on a flat $\Sigma$, or by working locally on a general, possibly curved $\Sigma$. Since our main interest will be the OPE algebras that the various operators satisfy, we shall work locally on a general $\Sigma$, choosing a local complex parameter $z$. 

As mentioned, $\mathbb{CP}^1$ can be regarded as the complex $\gamma$-plane plus a point at infinity.  Thus, we can cover it by two open sets, $U_1$ and $U_2$, where $U_1$ is the complex $\gamma$-plane, and $U_2$ is the complex $\tilde \gamma$-plane, where $\tilde\gamma=1/\gamma$. 

Since $U_1$ is isomorphic to $\mathbb{C}$, the sheaf of mirror CDR in $U_1$ can be described by a single free $bc$-$\beta\gamma$ system with action
\be
I={1\over 2\pi}\int|d^2z| \ \beta \partial_{\bar z} \gamma + b \partial_{\bar z} c.
\label{actionU1}
\ee 

Here  $\beta$, $b$, and $c$, $\gamma$,  are fields of dimension $(1,0)$ and $(0,0)$ respectively. They obey the usual free-field OPE's; there are no singularities in the operator products $\beta(z)\cdot \beta(z')$, $b(z) \cdot b(z')$, $\gamma(z)\cdot\gamma(z')$ and $c(z) \cdot c(z')$,  while 
\be
\beta(z) \gamma(z')  \sim  -{1\over z-z'} \quad  \textrm{and} \quad  b(z) c(z')  \sim  {1\over z-z'}.
\ee

Similarly, the sheaf of mirror CDR in $U_2$ is described by a single free ${\tilde b} {\tilde c}$-$\tilde\beta\tilde\gamma$ system with action 
\be
I= {1\over 2\pi}\int|d^2z| \ \tilde
\beta \partial_{\bar z} \tilde\gamma +  \tilde b \partial_{\bar z} \tilde c, 
\label{actionU2}
\ee
where the fields $\tilde \beta$, $\tilde b$, $\tilde \gamma$ and $\tilde c$ obey the same OPE's as $\beta$, $b$, $\gamma$ and $c$. In other words, the non-trivial OPE's are given by 
\be
\tilde \beta(z) \tilde \gamma(z')  \sim  -{1\over z-z'} \quad  \textrm{and} \quad  \tilde b(z) \tilde c(z')  \sim  {1\over z-z'}.
\ee 

In order to describe a globally-defined sheaf of mirror CDR, one will need to glue the free conformal field theories with actions (\ref{actionU1}) and (\ref{actionU2}) in the overlap region $U_1 \cap U_2$. To do so,  one must use the admissible automorphisms of the free conformal field theories defined in (\ref{autoCDRgamma})-(\ref{autoCDRb}) to glue the free-fields together. In the case of $X = \mathbb {CP}^1$, the automorphisms will be given by 
\begin{eqnarray}
\label{autoCP1gamma}
{\tilde \gamma} & = & {1 \over \gamma},\\
\label{autoCP1beta}
{\tilde \beta} & = &   - \gamma^2 \beta   + 2 \gamma b c,  \\
\label{autoCP1c}
{\tilde c} & = & -\gamma^2 c, \\
\label{autoCP1b}
{\tilde b} & = & - {b \over {\gamma^2}}.
\end{eqnarray} 
As there is no obstruction to this gluing in the half-twisted B-model, a sheaf of mirror CDR can be globally-defined on the target space $\mathbb {CP}^1$ (but only locally-defined on the worldsheet $\Sigma$ of the conformal field theory, because we are using a local complex parameter $z$ in the above).

\bigskip\noindent{\it Global Sections of the Sheaf}

Recall that for a general manifold $X$ of complex dimension $n$, the chiral algebra $\cal A$ will be given by ${\cal A} = \bigoplus_{g_R = 0}^{g_R = n} H^{g_R}( X, {\widetilde \Omega}^{ch}_X)$ as a vector space. Since $\mathbb {CP}^1$ has complex dimension 1, we will have, for $X=\mathbb {CP}^1$, the relation ${\cal A} = \bigoplus_{g_R = 0}^{g_R = 1} H^{g_R}( \mathbb {CP}^1, {\widetilde \Omega}^{ch}_{\mathbb P^1})$. Thus, in order to understand the chiral algebra of the half-twisted B-model, one needs only to study the global sections of the sheaf ${\widetilde \Omega}^{ch}_{\mathbb P^1}$, and its first Cech cohomology $H^1( \mathbb {CP}^1, {\widetilde \Omega}^{ch}_{\mathbb P^1})$.        
          
First, let us consider $H^0( \mathbb {CP}^1, {\widetilde \Omega}^{ch}_{\mathbb P^1})$, the global sections of ${\widetilde \Omega}^{ch}_{\mathbb P^1}$. At dimension 0, the space of global sections $H^0( \mathbb {CP}^1, {\widetilde \Omega}^{ch;* }_{\mathbb P^1; 0})$ must be spanned by functions of $\gamma$ and/or $c$ only. Note that it can be a function of higher degree in $\gamma$, but only a function of single degree in $c$ - higher powers of $c$ vanish (since $c^2 = 0$) because it is fermionic. In other words, the global sections are given by $H^0( \mathbb {CP}^1, {\widetilde \Omega}^{ch ; g_L}_{\mathbb P^1; 0})$, where $g_L = 0$ or $-1$. Notice that ${\widetilde \Omega}^{ch ; {0}}_{{\mathbb {P}}^1; 0}$ is just the sheaf  $\cal O$ of holomorphic functions in $\gamma$ on $\mathbb {CP}^1$, and that classically (from ordinary algebraic geometry), we have the result $H^1(\mathbb {CP}^1, {\cal O}) = 0$. Since a vanishing cohomology in the classical theory continues to vanish in the quantum theory,  $H^1(\mathbb {CP}^1,{\widetilde \Omega}^{ch ; {0}}_{{\mathbb {P}}^1; 0}) = 0$ will hold in quantum perturbation theory. As a relevant digression, notice that from chiral Poincar\'e duality \cite{CPD},\footnote{Note that the chiral Poincar\'e duality was originally formulated in the context of the sheaf of CDR. However, since the sheaf of mirror CDR is isomorphic to the sheaf of CDR, the duality principle should apply in the case of the mirror sheaf as well. The author wishes to thank F. Malikov for verifying this point.} we have the relation $H^0 ( {\mathbb {CP}}^1, {\widetilde \Omega}^{ch ; 1}_{{\mathbb {P}}^1 ; 0})^* = H^{1} ({\mathbb {CP}}^1, {\widetilde \Omega}^{ch ; {0}}_{{\mathbb {P}}^1; 0})$. This means that $H^0 ( {\mathbb {CP}}^1, {\widetilde \Omega}^{ch ; 1}_{{\mathbb {P}}^1 ; 0}) =0$, or rather, the global sections at dimension 0 do not contain fields with $g_L =1$; since the only field that has $g_L =1$ is the $b$ field with dimension 1, this observation is consistent. On the other hand, there is no such restriction on polynomials with $g_L = -1$ to span the space of global sections at dimension 0. In fact, from the automorphism relation of  (\ref{autoCP1c}), we find an immediate example, since its LHS, given by $\tilde c$, is by definition regular in $U_2$, while the RHS, being polynomial in $\gamma$ and $c$, is manifestly regular in $U_1$. Their being equal means that they represent a dimension 0 global section of ${\widetilde \Omega}^{ch}_{\mathbb {P}^1}$ that we will call $j_+$:
\be
j_+ = - \gamma^2 c = {\tilde c}.
\label{j_+}
\ee
The construction is completely symmetric between $U_1$ and $U_2$, with $\gamma\leftrightarrow \tilde\gamma$ and $c \leftrightarrow \tilde c$, so a reciprocal formula gives
another dimension 0 global section $j_-$:
\be
j_- = c = - {\tilde \gamma}^2 {\tilde c}.
\label{j_+}
\ee 
(Note that normal-ordering is understood for all operators above and below). Since these are global sections of a sheaf of chiral vertex operators, we can construct more of them from their OPE's. However, there are no singularities in the $j_a \cdot  j_b$ operator products for $a,b = + \  \textrm{or} \ -$, and as such, we cannot construct any more global sections from their OPE's with each other. Nevertheless, we will be able to find another dimension 0 global section $j_3$, when we discuss the dimension 1 global sections and their OPE's with $j_+$ and $j_-$.      
   
Note that in contrast to the half-twisted A-model on $\mathbb {CP}^1$ of \cite{MC}, where we discussed the chiral algebra of local operators corresponding to global sections of the sheaf of CDR at dimension 0, the space of global sections of the sheaf of mirror CDR corresponding to the chiral algebra of local operators in the half-twisted B-model at dimension 0, is $\it{not}$ one-dimensional and generated by 1. Instead, we can have global sections $j_+$, $j_-$, 1 etc.\footnote{Note that the operator 1 generates the one-dimensional class of $H^0( \mathbb{CP}^1, {\widetilde \Omega}^{ch ; 0}_{\mathbb {P}^1 ; 0})$. This is because ${\widetilde \Omega}^{ch ; 0}_{\mathbb {P}^1 ; 0}$ corresponds to the sheaf $\cal O$ of holomorphic functions in $\gamma$ on $\mathbb{CP}^1$, and $H^0(\mathbb {CP}^1, {\cal O}) \cong {\mathbb C}$.}         

Let us now ascertain the space $H^0( \mathbb {CP}^1, {\widetilde \Omega}^{ch ;* }_{\mathbb P^1; 1})$ of global sections of dimension 1. In order to get a global section of ${\widetilde \Omega}^{ch}_{\mathbb P^1}$ of dimension 1, we can act on a global section of ${\widetilde \Omega}^{ch}_{\mathbb P^1}$ of dimension 0 with the partial derivative $\partial_z$. Since $\partial_z 1 = 0$, this prescription will not apply to the operator 1. One could also consider operators of the form $f(\gamma) \partial_z \gamma$, where $f(\gamma)$ is a holomorphic function of $\gamma$.  However,  there are no such global sections either -  such an operator, by virtue of the way it transforms purely geometrically under (\ref{autoCP1gamma}), would correspond to a section of $\Omega^1(\mathbb {CP}^1)$, the sheaf of holomorphic differential forms $f(\gamma) d\gamma$ on $\mathbb {CP}^1$, and from the classical result $H^0(\mathbb {CP}^1, \Omega^1(\mathbb {CP}^1)) = 0$, which continues to hold in the quantum theory, we see that $f(\gamma) \partial_z \gamma$ cannot be a dimension  1 global section of ${\widetilde \Omega}^{ch}_{\mathbb P^1}$.

Other possibilities include operators which are linear in $b$, $\partial_z c$ or $\beta$. In fact, from the automorphism relation of (\ref{autoCP1beta}), we find an immediate example as the LHS, $\tilde \beta$, is by definition regular in $U_2$, while the RHS, being polynomial in $\gamma$, $b$ and $c$, is manifestly regular in $U_1$. Their being equal means that they represent a dimension 1 global section of ${\widetilde \Omega}^{ch}_{\mathbb {P}^1}$ that we will call $J_+$:
\be
J_+  = - \gamma^2 \beta + 2 \gamma b c  = {\tilde \beta}.
\label{J_+}
\ee   
The construction is completely symmetric between $U_1$ and $U_2$, with $\gamma\leftrightarrow \tilde\gamma$, $\beta\leftrightarrow\tilde\beta$, $b \leftrightarrow \tilde b$ and $c \leftrightarrow \tilde c$, so a reciprocal formula gives
another dimension 1 global section $J_-$:
\be
J_- =\beta = -\tilde\gamma^2\tilde \beta + 2\tilde\gamma \tilde b \tilde c.
\label{J_-}
\ee
Hence,  $J_+$ and $J_-$ give us two dimension 1 global sections of the sheaf ${\widetilde \Omega}^{ch}_{\mathbb P^1}$.   Since these are global sections of a sheaf of chiral vertex operators, we can construct more of them from their OPE's. There are no singularities in the $J_+ \cdot J_+$ or $J_-\cdot J_-$ operator products, but 
\be
J_+ J_- \sim {2J_3\over z-z'},
\ee
where $J_3$ is another global section of dimension 1 given by 
\be
J_3 = -\gamma\beta + bc.
\label{J_3}
\ee
What about the other dimension 0 global section that we mentioned earlier? Well, note that we also have the OPE    
\be
J_+ j_- \sim {2j_3\over z-z'},
\ee
where $j_3$ is the global section of dimension 0 that we are looking for. It is given by 
\be
j_3 = - \gamma c. 
\label{j_3}
\ee
 
Notice that $\{J_+, J_-, J_3 \}$ are bosonic operators that belong in $H^0(\mathbb {CP}^1, {\widetilde \Omega}^{ch ; 0}_{\mathbb{P}^1 ; 1})$, while $\{ j_+, j_-, j_3 \}$ are fermionic operators that belong in $H^0(\mathbb {CP}^1, {\widetilde \Omega}^{ch ; -1}_{\mathbb{P}^1 ; 0})$. (Again, this is in contrast to the A-model/sheaf of CDR case considered in \cite{MC}, whereby the counterparts of the fermionic global sections $\{ j_+, j_-, j_3 \}$  are of dimension 1.) One can compute that they satisfy the following closed OPE algebra:
\begin{eqnarray}
{ J}_a (z) { J}_a (z') \sim \textrm{regular}, & \quad &  {J}_3 (z) {J}_+ (z') \sim  {{+{ J}_+ (z')} \over z-z'},  \\
{J}_3 (z) {J}_- (z') \sim  {{-{ J}_- (z')} \over z-z'}, & \quad & {J}_+ (z) {J}_- (z') \sim {{2{ J}_3 (z')} \over z-z'}, \\
{J}_3 (z) {j}_- (z') \sim  {{-{ j}_- (z')} \over z-z'}, & \quad & {J}_3 (z) {j}_+ (z') \sim  {{+{ j}_+ (z')} \over z-z'}, \\
 {J}_+ (z) {j}_- (z') \sim {{2{ j}_3 (z')} \over z-z'}, & \quad & {J}_+ (z) {j}_3 (z') \sim {{-{ j}_+ (z')} \over z-z'}, \\
{J}_- (z) {j}_+ (z') \sim {{-2{ j}_3 (z')} \over z-z'}, & \quad & {J}_- (z) {j}_3 (z') \sim {{{ j}_- (z')} \over z-z'}, \\
{j}_a (z) {j}_b (z') \sim \textrm{regular}, & \quad & {J}_a (z) {j}_a (z') \sim \textrm{regular}, 
\end{eqnarray}
where $a,b = +, - \ \textrm{or} \ 3$. From the above OPE algebra, we learn that the $J$'s and $j$'s together generate a super-affine algebra of  $SL(2)$ at level 0, which here, appears in the Wakimoto free-field representation \cite{CDO28}. Indeed, these chiral vertex operators are holomorphic in $z$, which means that one can expand them in a Laurent series that allows an affinisation of the $SL(2)$ superalgebra generated by their resulting zero modes. Thus, the space of global sections of  ${\widetilde \Omega}^{ch}_{\mathbb {P}^1}$ is a module for the super-affine algebra of $SL(2)$ at level 0. (Given that the sheaf ${\widetilde \Omega}^{ch}_{X}$ is supposed to be isomorphic to the sheaf ${\widehat \Omega}^{ch}_{X}$, this observation should not come as a surprise, since as shown in \cite{MC}, the space of global sections of ${\widehat \Omega}^{ch}_{\mathbb {P}^1}$ is also a module for a super-affine algebra of $SL(2)$ at level 0.) 

It is shown in \cite{kac} that for a general representation of a super-affine algebra of $SL(2)$ at level $k$, one can write its (bosonic) current generators as  
\be
J_a (z) = J^f_a (z) +  {\hat J}_a (z).
\ee
The current $J^f_a (z)$, constructed from Majorana-Weyl free fermions, defines a representation of the super-affine algebra of $SL(2)$ at level $2$, while the current ${\hat J}_a (z)$ defines a representation at level $k-2$. The current ${\hat J}_a (z)$ may be thought as that of a WZW theory, and it obeys ${\hat J}_a (z) J^f_a (z) \sim \textrm{reg}$. A stress tensor can then be defined as \cite{kac}  
\be
T(z) = {  {: {\hat J}_+ {\hat J}_- + {\hat J}^2_3 : -   : ( j_+ \partial_z j_+ + j_- \partial_z j_-  +  j_3 \partial_z j_3 ) :} \over{k} }.
\label{SST}
\ee  
For every $k \neq 0$, $T(z)$ generates a (super)-Virasoro algebra. Similarly, one can define its superpartner current
\be
G(z) = {  {  k : (j_+ {\hat J}_- + j_3 {\hat J}_3) : -  ({i \over {6}}) f^{abc} : j_a j_b j_c :} \over{k^2} }.
\label{SSG}
\ee 
However, notice that the above definitions of $T(z)$ and $G(z)$ break down for $k=0$. This implies that $T(z)$ and $G(z)$ do not exist as physical operators in the half-twisted B-model on $\mathbb {CP}^1$. Consequently, the space of local operators has a structure of a chiral algebra only in a $\it{partial}$ physical sense; it obeys all the physical axioms of a chiral algebra, except for reparameterisation invariance on the $z$-plane or worldsheet $\Sigma$. We will substantiate this last statement momentarily by demonstrating an $\it{absence}$ of the holomorphic stress tensor (and its superpartner) in the $Q_R$-cohomology.

Notice that in order to obtain operators that make sense at $k=0$, we can remove the factors of $1/k$ and $1/{k^2}$ from (\ref{SST}) and (\ref{SSG}), and in doing so, we get       
\be
S(z) = {: {\hat J}_+ {\hat J}_- + {\hat J}^2_3 : -   : ( j_+ \partial_z j_+ + j_- \partial_z j_-  +  j_3 \partial_z j_3 ) :}
\label{SSS}
\ee  
and
\be
R(z) = {  k : (j_+ {\hat J}_- + j_3 {\hat J}_3) : -  ({i \over {6}}) f^{abc} : j_a j_b j_c :},
\label{SSR}
\ee 
where $S(z) = k T(z)$ and $R(z) = k^2  G(z)$ are well-defined operators for any $k \neq \infty$. Hence, we see that  $S(z)$ generates $k$ times the symmetry generated by $T(z)$, and $R(z)$ generates $k^2$ times the symmetry generated by $G(z)$. For $k =0$, $S(z)$ and $G(z)$ generate no symmetries at all - the OPE's of all fields with $S(z)$ and $G(z)$ are regular. Thus, in an irreducible representation of the super-affine algebra, $S(z)$ and $G(z)$ can be represented by $c$-numbers, and might vanish.      

One can actually go on to say more about $S(z)$ as follows. First, note that $T(z)$ in (\ref{SST}) will generate a (super)-Virasoro algebra with central charge $c_k= 3(k-2) / k + 3/2$ \cite{kac}. Second, note that since $S(z) = k T(z)$, under a finite conformal transformation $z \to w(z)$, we will have 
\be
(\partial_z w)^2 S'(w) = S(z) -  k ({ {c_k} \over 12}) {\mathfrak S}(w,z),    
\ee                    
where $S'(w)$ is the transformed operator, and    
\be
{\mathfrak S}(w,z) = {{2 (\partial_z w) (\partial_z^3 w) - 3 (\partial_z^2 w)^3} \over {2 (\partial_z w)^2}} 
\ee
is the $\it{Schwarzian}$ $\it{derivative}$. Thus, for $k=0$, we have the transformation 
\be
S(z) =  (\partial_z w)^2 S'(w) -  {1 \over 2} {\mathfrak S}(w,z).   
\label{Stx}
\ee
Third, note that (\ref{Stx}) coincides with the transformation formula for self-adjoint differential operators acting from $K^{-1/2}$ to $K^{3/2}$, where $K$ is the canonical line bundle on $\Sigma$. Consequently, the $S(z)$ operator is a $\it{projective}$ $\it{connection}$ on $\Sigma$ \cite{GL}. Projective connections are important in the conformal field-theoretic approach to the geometric Langlands program as reviewed in \cite{GL}. However, the relevant projective connections in that context were related to affine instead of super-affine Lie algebras. Hence, $S(z)$ may potentially find its place in a conformal field-theoretic approach to a $\it{supersymmetric}$ $\it{extension}$ of the geometric Langlands conjecture, if the extension should exist at all.

Still on the subject of global sections, recall from sections 6.1 and 6.2, and our $Q_R$-Cech cohomology dictionary, that there will be $\psi^{\bar i}$-independent operators ${T}(z)$ and ${G}(z)$ in the $Q_R$-cohomology of the underlying half-twisted B-model on $\mathbb {CP}^1$ if and only if the corresponding ${\widehat T}(z)$ and ${\widehat G}(z)$ operators can be globally-defined, i.e., the ${\cal T}(z)$ and ${\cal G}(z)$ operators of the free $bc$-$\beta\gamma$ system belong in $H^0(\mathbb {CP}^1, {\widetilde \Omega}^{ch}_{\mathbb P^1})$ - the space of global sections of  ${\widetilde \Omega}^{ch}_{\mathbb P^1}$. Let's look at this more closely. 

Note that  for $X= \mathbb {CP}^1$, we have 
\begin{eqnarray} 
{\cal T}(z) & = &  -: \beta \partial_z \gamma: (z) - : b\partial_z c: (z), \\
{\cal G}(z) & = & : b \beta:(z),   
\end{eqnarray}   
where the above operators are defined and regular in $U_1$. Similarly, we also have 
\begin{eqnarray} 
\label{A1}
{\widetilde{\cal T}(z)} & = &  -: \tilde\beta \partial_z \tilde\gamma: (z) - : \tilde b\partial_z \tilde c: (z), \\
\label{A4}
{\widetilde{\cal G}(z) }& = & :\tilde b \tilde \beta:(z),   
\end{eqnarray}   
where the above operators are defined and regular in $U_2$. By substituting the automorphism relations (\ref{autoCP1gamma})-(\ref{autoCP1b}) into (\ref{A1})-(\ref{A4}), a small computation shows that in $U_1 \cap U_2$, 
\be 
{\widetilde{\cal T}(z)} - {\cal T}(z) = 4 \left( {{\partial_z \gamma} \over \gamma} \right)^2 (z) , 
\label{B1}
\ee   
\be
{\widetilde{\cal G}(z)}  - {\cal G}(z) = 2  \partial_z \left({b \over \gamma} \right) (z), 
\label{B3}
\ee
\hspace{-0.0cm} where an operator that is a global section of ${\widetilde \Omega}^{ch}_{\mathbb P^1}$ must agree in $U_1 \cap U_2$.

Notice that in $U_1 \cap U_2$, we have ${\widetilde {\cal T}} \neq {\cal T}$ and ${\widetilde {\cal G}} \neq {\cal G}$. The only way to consistently modify $\cal T$ and $\widetilde {\cal T} $ so as to agree on $U_1\cap U_2$, is to shift them by a multiple of the term ${(\partial_z \gamma)^2 / \gamma^2} = -{(\partial_z \tilde \gamma)^2 / \tilde \gamma}^2$. However, this term has a double pole at both $\gamma=0$ and $\tilde\gamma=0$. Thus, it cannot be used to redefine $\cal T$ or $\widetilde {\cal T} $ (which has to be regular in $U_1$ or $U_2$, respectively). The only way to consistently modify $\cal G$ and $\widetilde {\cal G} $ so as to agree on $U_1\cap U_2$, is to shift them by a linear combination  of the terms $({\partial_z b}) / \gamma = - \tilde \gamma \partial_z (\tilde b / {\tilde \gamma}^2)$, and $({b \partial_z \gamma}) / \gamma^2 = ({\tilde b \partial_z \tilde \gamma}) / {\tilde \gamma}^2$. Similarly, these terms have poles at both $\gamma = 0$ and $\tilde\gamma = 0$, and hence, cannot be used to redefine $\cal G$ or $\widetilde {\cal G}$ (which also has to be regular in $U_1$ or $U_2$ respectively). 

Therefore, we conclude that ${\cal T}(z)$ and ${\cal G}(z)$ $\it{do}$ $\it{not}$ belong in $H^0(\mathbb {CP}^1, {\widehat \Omega}^{ch}_{\mathbb P^1})$. Since $c_1(\mathbb {CP}^1) \neq 0$, this conclusion is consistent with the conformal anomaly cancellation condition discussed in section 5.7, where for ${\cal E} = TX$ at the $(2,2)$ locus, tells us that  ${\cal T}(z) \neq \widetilde {\cal T}(z)$ unless $c_1(X) = 0$. Again from our $Q_R$-Cech cohomology dictionary, this means that $T(z)$ and $G(z)$ are not in the $Q_R$-cohomology of the underlying half-twisted B-model on $\mathbb {CP}^1$. This last statement is in perfect agreement with the physical picture presented in section 6.1, which tells us that since $c_1(\mathbb {CP}^1) \neq 0$, the symmetries associated with $T(z)$ and $G(z)$ ought to be broken such that  $T(z)$ and $G(z)$ will cease to exist in the $Q_R$-cohomology at the quantum level. Notice then that (\ref{B1}) and (\ref{B3}) actually provide us with a purely mathematical interpretation of the absence of the stress tensor $T(z)$ and its superpartner $G(z)$, as an obstruction to gluing the ${\cal T}(z)$'s and the ${\cal G}(z)$'s (on overlaps) into global sections of the sheaf ${\widetilde \Omega}^{ch}_{\mathbb P^1}$ of mirror CDR on $\mathbb {CP}^1$. Last but not least, our findings also imply that unlike the sheaf ${\widehat \Omega}^{ch}_{X}$ of CDR on $X$, the sheaf ${\widetilde \Omega}^{ch}_{X}$ of mirror CDR on $X$ will only have a structure of a conformal vertex superalgebra if $c_1(X) = 0$.      

As mentioned in section 6.1, the symmetries associated with the stress tensor $T(z)$ and its superpartner $Q(z)$ will remain unbroken in the conformal limit where $c_1(X) = 0$, i.e., if the sigma model one-loop beta function vanishes. Thus, one is led to the following question:  is the non-vanishing of the obstruction terms on the RHS of (\ref{B1}) and (\ref{B3})  due to a non-zero one-loop beta function? And will they vanish if the one-loop beta function is zero? 

In order to answer this question, first recall from the $\mathbb {CP}^1$ example in section 5.7 that we have a correspondence between the holomorphic term $(\partial_z\gamma) / \gamma$ and the sigma model operator $R_{i \bar j} \partial_z \phi^i \psi^{\bar j}$. Hence, since $R_{i \bar j}$ is proportional to the one-loop beta function, we find that the RHS of (\ref{B1}) will be zero if the one-loop beta function vanishes. Consequently, ${\cal T}(z)$ will be a global section of ${\widetilde \Omega}^{ch}_{\mathbb P^1}$ and $T(z)$ will hence be in the $Q_R$-cohomology of the half-twisted B-model on $\mathbb {CP}^1$, if and only if the one-loop beta function vanishes. 

What about ${\cal G}(z)$ and $G(z)$? Firstly, the identification $\gamma^i = \phi^i$ further implies a correspondence between the term $1/ \gamma$ and the sigma model operator $R_{i \bar j}\psi^{\bar j}$. Secondly, notice that the RHS of (\ref{B3}) is given by $2[ (\partial_z b) / \gamma - (b \partial_z \gamma) / \gamma^2]$. Thus, via the above-mentioned correspondence between the holomorphic terms and operators, the identification $b^i = \lambda^i_z$, and the fact that $\partial_{\gamma}( 1/ \gamma) = - 1/ \gamma^2$, we find that the physical counterpart of the RHS of (\ref{B3}) will be given by the sigma model operator $2[ R_{i \bar j} \partial_z \lambda^i_z \psi^{\bar j} + R_{i \bar j, k}\lambda^k_z \partial_z \phi^i \psi^{\bar j}]$. Therefore, we see that the RHS of (\ref{B3}) is proportional to the one-loop beta function.  Consequently, ${\cal G}(z)$ will be a global section of ${\widetilde \Omega}^{ch}_{\mathbb P^1}$ and $G(z)$ will hence be in the $Q_R$-cohomology of the half-twisted B-model on $\mathbb {CP}^1$, if and only if the one-loop beta function vanishes. 

From our above discussion, we have once again obtained an interpretation of the one-loop beta function solely in terms of holomorphic data.

\bigskip\noindent{\it The First Cohomology}

We shall now proceed to investigate the first cohomology group $H^1(\mathbb{CP}^1, {\widetilde \Omega}^{ch}_{\mathbb P^1})$. 
   
In dimension 0, we again have, as possible candidates, functions that are of a higher degree in $\gamma$ but of a single degree in $c$. However, from ordinary algebraic geometry, we have the classical result that $H^1(\mathbb {CP}^1, {\cal O}) = 0$, where $\cal O$ is the sheaf of functions over $\mathbb {CP}^1$ which are holomorphic in $\gamma$. Since a vanishing cohomology at the classical level continues to vanish at the quantum level, we learn that we cannot have functions which are monomials in $\gamma$. 

That leaves us to consider polynomials of the form $f(\gamma)c$ or the monomial $c$. In order to determine if they span the first cohomology, first notice that the polynomials of the form $f(\gamma) c$ or the monomial $c$, are simply sections of the sheaf ${\widetilde \Omega}^{ch ; -1}_{\mathbb P^1 ; 0}$. From chiral Poincar\'e duality \cite{CPD}, we have the relation $H^0(\mathbb {CP}^1, {\widetilde \Omega}^{ch ; p}_{\mathbb P^1 ; n})^* = H^1(\mathbb {CP}^1, {\widetilde \Omega}^{ch ; 1-p}_{\mathbb P^1 ; n})$. Hence, $H^0(\mathbb {CP}^1, {\widetilde \Omega}^{ch ; 2}_{\mathbb P^1 ; 0})^* = H^1(\mathbb {CP}^1, {\widetilde \Omega}^{ch ; -1}_{\mathbb P^1 ; 0})$. Since one can certainly find global sections of the sheaf ${\widetilde \Omega}^{ch ; 2}_{\mathbb P^1 ; 0}$, $H^1(\mathbb {CP}^1, {\widetilde \Omega}^{ch ; -1}_{\mathbb P^1 ; 0})$ will be non-zero. Now, let us recall that $c$ is a local section of the pull-back of the holomorphic cotangent bundle of $\mathbb {CP}^1$, i.e.,  $c \in {\cal O}(\gamma^*(T^*{\mathbb {P}^1}))$. Hence, polynomials of the form $f(\gamma) c$ are sections of the sheaf ${\cal O} \otimes {\cal O}(\gamma^*(T^*{\mathbb {P}^1}))$. Note that since $H^1(\mathbb {CP}^1, {\cal O}) = 0$, and $H^0(\mathbb {CP}^1, {\cal O})$ is generated by 1, we effectively have the map $H^1(\mathbb {CP}^1, {\cal O}(\gamma^*(T^*{\mathbb {P}^1})))  \to H^1(\mathbb {CP}^1, {\cal O} \otimes {\cal O}(\gamma^*(T^*{\mathbb {P}^1})))$.\footnote{From the cup product map, we have $ [H^0(\mathbb {CP}^1, {\cal O}) \otimes H^1(\mathbb {CP}^1, {\cal O}(\gamma^*(T^*{\mathbb {P}^1})))]  \oplus  [H^1(\mathbb {CP}^1, {\cal O}) \otimes H^0(\mathbb {CP}^1, {\cal O}(\gamma^*(T^*{\mathbb {P}^1})))] \to H^1(\mathbb {CP}^1, {\cal O} \otimes {\cal O}(\gamma^*(T^*{\mathbb {P}^1})))$. Since $H^1(\mathbb {CP}^1, {\cal O}) = 0$, and $ H^0(\mathbb {CP}^1, {\cal O})$ is generated by 1, we effectively have the map $H^1(\mathbb {CP}^1, {\cal O}(\gamma^*(T^*{\mathbb {P}^1})))  \to H^1(\mathbb {CP}^1, {\cal O} \otimes {\cal O}(\gamma^*(T^*{\mathbb {P}^1})))$.} Therefore, we find that just as in the case of the A-model/sheaf of CDR on $\mathbb {CP}^1$, the first cohomology group at dimension 0 or  $H^1(\mathbb {CP}^1, {\widetilde \Omega}^{ch ;*}_{\mathbb P^1 ; 0})$, is one-dimensional and generated by $c$.

In dimension 1, we will need to consider functions which are linear in $\beta$, $b$, $\partial_z \gamma$ or $\partial_z c$. One clue that we have is the standard result from algebraic geometry that $H^1( \mathbb {CP}^1, K ) \neq 0$, where $K$ is the sheaf of holomorphic differentials $d\gamma /{\gamma}$. This implies that  ${\partial_z \gamma} / \gamma$ ought to generate a dimension 1 class of the cohomology group  $H^1(\mathbb {CP}^1, {\widetilde \Omega}^{ch }_{\mathbb {P}^1})$. However, this classical result may be violated by quantum effects in perturbation theory. How so, one may ask? 

To understand this, let us first consider the operator ${\cal J}(z) = {:bc:(z)}$ on an open set $U_1$. From the fields correspondence between the free $bc$-$\beta \gamma$ system and the local half-twisted B-model in section 6.2, we find that ${\cal J}(z)$ just corresponds to $\widehat{J}(z)$, the left-moving ghost number current of the local half-twisted B-model on $U_1$. Next, let ${\widetilde {\cal J}}(z) = : {\tilde b} {\tilde c}:(z)$ define the same operator on another open set $U_2$. By using the automorphism relations (\ref{autoCP1gamma})-(\ref{autoCP1b}), one can compute that on $U_1 \cap U_2$,            
\be
{\widetilde {\cal J}}(z) - {\cal J}(z) = 2 \left({\partial_z \gamma \over \gamma}\right) (z). 
\label{calJ-calJ}
\ee   
Note that one could attempt to make ${\widetilde {\cal J}}(z)$ and ${\cal J}(z)$ agree on $U_1 \cap U_2$ by removing the term on the RHS of (\ref{calJ-calJ}). The only way to do this consistently (such that the symmetries on both sides of (\ref{calJ-calJ}) continue to be respected) is to add to ${\cal J}(z)$ a term  that is linear in ${\partial_z \gamma} /{\gamma}$, or to add to ${\widetilde {\cal J}}(z)$ a term  that is linear in $-{\partial_z {\tilde \gamma}} / {{\tilde \gamma}}$. However, note that these two terms have a pole at $\gamma = 0$ and $\tilde \gamma = 0$ respectively, and since ${\widetilde {\cal J}}(z)$ and ${\cal J}(z)$ are defined to be regular in $U_2$ and $U_1$, we cannot use these terms to modify ${\widetilde {\cal J}}(z)$ and ${\cal J}(z)$. This means that the RHS of (\ref{calJ-calJ}) $\it{cannot}$ be set to zero, and that ${\cal J}(z)$ will fail to be a global section of the sheaf of mirror CDR, i.e., ${\cal J}(z) \notin H^0( \mathbb {CP}^1, {\widetilde \Omega^{ch}_{\mathbb P^1}})$. Therefore, from our $Q_R$-Cech cohomology dictionary, this translates to the fact that $J(z)$ of the underlying  half-twisted sigma model on $\mathbb{CP}^1$ is absent in the $Q_R$-cohomology. As explained in section 6.1, this is due to the quantum perturbative effects of a non-zero one-loop beta function arising from a non-vanishing first Chern class on $\mathbb{CP}^1$. Now, since ${\widetilde {\cal J}}$ and ${\cal J}$ are by definition holomorphic in $U_2$ and $U_1$ respectively, it will mean from (\ref{calJ-calJ}) that ${\partial_z \gamma} / \gamma$ cannot be a dimension 1 element of the group $H^1(\mathbb {CP}^1, {\widetilde \Omega}^{ch }_{\mathbb {P}^1})$. This is because it can be written as a difference between a term that is holomorphic in $U_2$ and a term that is holomorphic in $U_1$. It is in this collective sense that we understand the following - although $H^1( \mathbb {CP}^1, K )$ is non-vanishing classically, ${\partial_z \gamma / {\gamma}} \notin H^1(\mathbb {CP}^1, {\widetilde \Omega}^{ch ; 0}_{\mathbb {P}^1 ; 1})$ due to quantum effects in perturbation theory.  One can go on to consider other classical, dimension 1 operators using standard results in ordinary algebraic geometry, where the existence of such operators at the quantum level can be checked against  conditions analogous to (\ref{calJ-calJ}) that one might have. We shall omit the computation of these operators for brevity.    

Another important point to note is that from chiral Poincar\'e duality, we have the relations $H^0(\mathbb {CP}^1, {\widetilde \Omega}^{ch ; 0}_{\mathbb P^1; 1})^* = H^1(\mathbb {CP}^1, {\widetilde \Omega}^{ch ; 1}_{\mathbb P^1 ; 1})$, and  $H^0(\mathbb {CP}^1, {\widetilde \Omega}^{ch ; -1}_{\mathbb P^1 ; 0})^* = H^1(\mathbb {CP}^1, {\widetilde \Omega}^{ch ; 2}_{\mathbb P^1 ; 0})$. Since $\{J_+, J_-, J_3 \} \in H^0(\mathbb {CP}^1, {\widetilde \Omega}^{ch ; 0}_{\mathbb P^1 ; 1})$, and $\{j_+, j_-, j_3 \} \in H^0(\mathbb {CP}^1, {\widetilde \Omega}^{ch ; -1}_{\mathbb P^1 ; 0})$, we find that the space $H^1(\mathbb {CP}^1, {\widetilde \Omega}^{ch}_{\mathbb P^1})$ is also a module for a super-affine algebra of $SL(2)$ at level 0. This observation is consistent with the fact that the sheaf of mirror CDR is isomorphic to the sheaf of CDR. 

To ascertain the operators of dimension 2, we can follow the same prescription employed in ascertaining the operators of dimension 0 and 1 - we could start off by first using the results from standard algebraic geometry to ascertain, at the classical level, the relevant operators of dimension 2 in the first cohomology, and then proceed to check for their existence at the quantum level by comparing against conditions analogous to (\ref{calJ-calJ}) that one might have. In light of this prescription, note however that from the conditions (\ref{B1}) and (\ref{B3}), one can infer that $({{\partial_z \gamma} / \gamma})^2 \notin H^1(\mathbb{CP}^1, {\widetilde \Omega}^{ch ; 0}_{{\mathbb P}^1 ; 2})$, and $\partial_z ({b / \gamma}) \notin H^1(\mathbb{CP}^1, {\widetilde \Omega}^{ch ; 1}_{{\mathbb P}^1 ; 2})$, regardless of the results from algebraic geometry of the first cohomology on $\mathbb{CP}^1$ at dimension 2. For brevity, we shall omit the computation of these operators.    

We can do likewise to ascertain the operators of dimension 3 and higher. But in these higher dimensional cases, we do not have relations that are analogous to (\ref{calJ-calJ}) in dimension 1, and (\ref{B1})-(\ref{B3}) in dimension 2. Thus, we could very well borrow the results from standard algebraic geometry to ascertain the relevant operators of dimension 3 and higher in the first cohomology. In view of the length of this paper, we will again omit the computation of these operators for brevity.

\newsubsection {\it {The Half-Twisted B-Model on $\bf{S}^3 \times \bf{S}^1$}}

As shown in section 3.3 of \cite{MC}, a twisted version of the usual $(0,2)$ heterotic sigma model can be related to a unitary model with $(0,2)$ supersymmetry. Likewise on the $(2,2)$ locus, a half-twisted $(2,2)$ model (such as our half-twisted B-model) can be related to a  unitary model with $(2,2)$ supersymmetry. Thus, if we are to allow for the possibility of constructing a $\it{family}$ of sheaves of mirror CDR on the target space $X$, $X$ should be non-K\"ahler with torsion, just as in the $(0,2)$ case.\footnote{Recall from our discussion in section 3.3 that the non-K\"ahlerity and torsion of the target space are required to define the moduli of the sheaves of CDO's.}

It is commonly known that a $(2,2)$ model formulated using only chiral superfields does not admit non-K\"ahler target spaces \cite{west}. However, if the model is being formulated in terms of chiral $\it{and}$ twisted chiral superfields, one can allow for non-K\"ahler target spaces with torsion \cite{gates}. An example of a non-K\"ahler complex manifold that exists as the target space of a $(2,2)$ sigma model is $X= \bf{S}^3 \times \bf{S}^1$. In fact, an off-shell construction of this model has been given in \cite{rocek}, where it is also shown that for a $(2,2)$ sigma model on a group manifold, the only example amenable to such a formulation is the parallelised group manifold $X = SU(2) \times U(1) \cong \bf{S}^3 \times \bf{S}^1$. The essential properties of $\bf{S}^3 \times \bf{S}^1$ have been discussed in section 3.4, where the hermitian form $\omega_T$ which defines its torsion whilst obeying the weaker  condition $\partial \bar{\partial} \omega_T  = 0$, has been given in (\ref{omegaT}). Let us therefore explore this model further.

\bigskip\noindent{\it The WZW Model}

As explained in \cite{rocek}\cite{spindel}, the $(2,2)$  model on $\bf{S}^3 \times \bf{S}^1$ is a tensor product of an $SU(2)$ WZW model, times a free field theory on $\bf{S}^1$, times four free left $\it{and}$ right-moving real fermions. The real fermions combine into four complex fermions which transform in the adjoint representation of $SU(2) \times U(1)$, i.e., {\bf{3}} of $SU(2)$ and {\bf{1}} of  $U(1)$. The $SU(2)$ fermions are free because the connection on $SU(2)$, which follows from the $(2,2)$ model on $\bf{S}^3 \times \bf{S}^1$, has torsion and is parallelised. There is thus a shift in the level of the $SU(2)$ WZW model due to a relevant redefinition of these fermionic fields. This will be apparent shortly.

On the (2,2) locus, the free left-moving fermions are equivalent to a set of fermionic $bc$ fields (labelled by $\lambda_i$ and $\lambda^i_z$ in section 6.1) with spins 0 and 1, while the free right-moving fermions are equivalent to a set of $\tilde b  \tilde c$ fields (labelled by $\psi^{i}_{\bar z}$ and $\psi^{\bar i}$ in section 6.1) with spins $1$ and $0$. The $b c$ and $\tilde b  \tilde c$ systems have left and right central charges $(-2,0)$ and $(0,-2)$. On a manifold such as   $\bf{S}^3 \times \bf{S}^1$ with complex dimension 2, there will be 2 sets of left and right-moving fermions. Hence, the fermions contribute a total of $(-4,-4)$ to the left and right central charges of the model.  The $SU(2)$ WZW model at level $k$ contributes $(3k/(k+2),3k/(k+2))$ to the central charges, and the free theory on ${\bf{S}}^1$ contributes $(1,1)$. The total left and right central charges are therefore $(3k/(k+2)-3,3k/(k+2)-3)$. The difference between the left and right central charges remains the same in passing from the physical theory to the $Q_R$-cohomology. In this example, it is given by $c=0$. This should be the central charge of the stress tensor which will appear as a global section of the sheaf ${\widetilde \Omega}^{ch}_X$ of mirror CDR on $X = \bf{S}^3 \times \bf{S}^1$. 

Similarly, we can pre-ascertain the central charges of the current
algebra which will be furnished by the appropriate global sections of the sheaf
of  mirror CDR on $\bf{S}^3 \times \bf{S}^1$.  The underlying $SU(2)$ WZW model has an $SU(2)$-valued
field $g$, with symmetry $SU(2)_L\times SU(2)_R$  (to be precise, it is
$(SU(2)_L\times SU(2)_R)/\Bbb{Z}_2$, where $\Bbb{Z}_2$ is the
common center of the two factors). The symmetry acts by $g\to
agb^{-1}$, $a,b \in SU(2)$. In the WZW model, the $SU(2)_L$
symmetry is part of a holomorphic $SU(2)$ current algebra of level
$k+2$, while $SU(2)_R$ is part of an antiholomorphic $SU(2)$ current
algebra of level $k+2$.  As mentioned earlier, the shift by ``2" in the level of the $SU(2)$ current algebra is expected, and it is due to the fact that the complex fermions transform freely in the adjoint representation of $SU(2)$. The left and right central charges are therefore $(k+2,0)$
for $SU(2)_L$ and $(0,k+2)$ for $SU(2)_R$. 

Next, notice that the (right-moving) supersymmetry generator $Q_R = \overline Q_+$, although invariant under a left-moving $U(1)$ current, is nevertheless charged under a right-moving one. (Recall from section  2.2 that $\overline Q_+$ has charge $(q_L, q_R) =(0,+1)$.) Hence, the physical characteristics of $Q_R$, and the symmetry of the $Q_R$-cohomology that it defines, will depend on the twist one makes on the right-moving fields. Since the twisting of the four real right-moving fermions of the underlying $(2,2)$ model on $\bf{S}^3 \times \bf{S}^1$ reduces the $SU(2)_R$ symmetry to its maximal torus $U(1)_R$, the symmetry that should survive at the level of the $Q_R$-cohomology or sheaf of mirror CDR is $(SU(2)_L\times U(1)_R)/\Bbb{Z}_2=U(2)$.

The difference between the left and right central charges remains the same in passing to the $Q_R$-cohomology or sheaf of CDR. Hence, the expected levels of the $SU(2)_L$ and $U(1)_R$ current algebras, furnished by global sections of the sheaf of mirror CDR, should be given by $k+2$ and $-k-2$ respectively. The only case in which they are equal is $k=-2$, for which the levels are both $0$. This is not really a physically sensible value for the WZW model; physically sensible, unitary WZW models with convergent path integrals must be restricted to integer values of $k$ with $k \geq 0$.  However, as we will see shortly, $k$ is, in our case at hand, an arbitrary complex parameter that is directly related to the moduli of the sheaves of mirror CDR (that are in turn represented by $H^1( {\bf{S}^3} \times {\bf{S}^1}, \Omega ^{2,  cl}) \cong \mathbb C$). 

In the sheaf of mirror CDR, the symmetries are readily complexified, so that the symmetry of the corresponding current algebra which appears, should be at the Lie algebra level $GL(2)$ instead of $U(2)$. Likewise, with respect to the $SU(2)$ and $U(1)$ subgroups of $GL(2)$, the symmetry of the corresponding current algebra that will appear should be given by $SL(2)$ and $GL(1)$ respectively. In addition,  the $U(1)_R$ (which acts on the coordinate variables $v^i$, to be introduced shortly, by $v^i \to e^{i\theta} v^i$) and the rotation of ${\bf{S}}^1$ (which acts by $v^i \to e^{\chi} v^i$ with real $\chi$) combine together to generate the center of $GL(2)$. At the Lie algebra level, the center is $GL(1)$. This is the symmetry that we will expect to see as well. Note that the rotation of ${\bf{S}}^1$ will always corresponds to a $U(1)$ current algebra with equal left and right central charges. Thus, it will not affect our above discussion whereby only the differences between the left and right central charges are important.

\bigskip\noindent{\it Constructing a Sheaf of Mirror CDR on $\bf{S}^3 \times \bf{S}^1$}

We now proceed towards our main objective of constructing a family of sheaves of mirror CDR on $\bf{S}^3 \times \bf{S}^1$. As a starter, we will first construct a sheaf of mirror CDR without introducing any modulus. At this point, one would already be able to see, within the current algebras derived, the expected symmetries discussed above. Thereafter, we will generalise the construction and introduce a variable parameter which will serve as the modulus of the sheaves of mirror CDR. It is at this juncture that we find an explicit relation between the modulus of the sheaves and the level of the underlying $SU(2)$ WZW model.       

Let us begin by noting that $\bf{S}^3 \times \bf{S}^1$ can be expressed as $(\Bbb{C}^2-\{0\})/\Bbb{Z}$, where $\Bbb{C}^2$ has coordinates $v^1,$ $v^2$, and $\{0\}$ is the origin in $\Bbb{C}^2$ (the point $v^1=v^2=0$) which should be removed before dividing by $\Bbb{Z}$. Also, $\Bbb{Z}$ acts by $v^i \to \lambda^n v^i$, where $\lambda$ is a nonzero complex number of modulus less than 1, and $n$ is any integer.  $\lambda$ is a complex modulus of $\bf{S}^3 \times \bf{S}^1$ that we shall keep fixed.

To construct the most basic sheaf of mirror CDR with target $\bf{S}^3 \times \bf{S}^1$, one simply defines the scalar coordinate variables $v^i$ as free bosonic fields of spin 0, with conjugate spin 1 fields $V_i$. From our earlier discussions, one will also need to introduce fermionic fields $w_i$ of spin 0, with conjugate spin 1 fields $W^i$. Since $\bf{S}^3 \times \bf{S}^1$ has complex dimension 2, the index $i$ in all fields will run from 1 to 2. Therefore, the free field action that one must consider is given by    

\be
{I={1\over
2\pi}\int|d^2z| \left(\ V_1\bar\partial v^1+ V_2 \bar \partial
v^2 +  W^1\bar\partial w_1+ W^2 \bar \partial
w_2   \right).  } 
\label{lagrangian}
\ee
Notice that the above $Vv$-$Ww$ system is just a $\beta\gamma$-$bc$ system with nontrivial OPE's $V_i(z)v^j(z')\sim -\delta^i_j/(z-z')$ and $W^i(z)w_j(z')\sim \delta^i_j/(z-z')$.

In the above representation of $\bf{S}^3 \times \bf{S}^1$, the action of $\mathbb Z$ represents a geometrical symmetry of the system. Thus, the only allowable operators are those which are invariant under the finite action of $\mathbb Z$. These operators will therefore span the space of global sections of the sheaf of mirror CDR. Under this symmetry, $v^i$ transforms as $v^i \to {\tilde v}^i = \lambda v^i$.  In order to ascertain how the rest of the fields ought to transform under this symmetry, we simply substitute $v^i$ and ${\tilde v}^i$ (noting that it is equivalent to $\gamma^i$ and ${\tilde \gamma}^i$ respectively) into (\ref{autoCDRgamma})-(\ref{autoCDRb}).  In short, the only allowable operators are those which are invariant under $v^i \to \lambda v^i$, $V_i \to \lambda^{-1} V_i$, $w_i \to \lambda^{-1} w_i$ and $W^i \to \lambda W^i$.  
            
One operator that possesses this invariance is the stress tensor      
\be
{ T_{zz}={1 \over {2 \pi}}\sum_i (V_i\partial v^i + W^i\partial w_i).}                
\ee      
Contrary to the $\mathbb {CP}^1$ example, a stress tensor exists in the chiral algebra of the model on $\bf{S}^3 \times \bf{S}^1$ - as explained previously, one can find, in the mirror CDR case, a global definition for the stress tensor, because $\bf{S}^3 \times \bf{S}^1$ is parallelised and its first Chern class vanishes. Hence, the chiral algebra of the described theory is conformally invariant, and the sheaf of mirror CDR has a structure of a topological vertex (super)algebra. This reflects the superconformal invariance of the underlying $(2,2)$ model on $\bf{S}^3 \times \bf{S}^1$. A bosonic $\beta\gamma$ system of spins 1 and 0 has central charge $c=2$, while a fermionic $bc$ system of spins 1 and 0 has $c=-2$. Thus, the stress tensor $T$ has $c=0$, in agreement with what we had anticipated from the underlying WZW model.
                    
The chiral algebra of the underlying model also contains the dimension $1$ currents $J^i_j= -(V_jv^i + W^i w_j)$.  As required, these (bosonic) current operators are invariant under the field transformations $v^i \to \lambda v^i$, $V_i \to \lambda^{-1} V_i$, $w_i \to \lambda^{-1} w_i$ and $W^i \to \lambda W^i$. They obey the OPE's
\be
{J^i_j(z)J^m_l(z')\sim {\delta_j^mJ^i_l-\delta^i_lJ^m_j\over z-z'}.} 
\ee
We recognise this as a $GL(2)$ current algebra at level $0$.

When we proceed to generalise the above construction by introducing a variable parameter to serve as the modulus of the sheaves of mirror CDR, it will not be possible to maintain manifest $GL(2)$ symmetry. Hence, it will be useful to pick a basis in the current algebra now. The $SL(2)$ subgroup is generated by $J_3=- {1 \over 2} (V_1v^1 + W^1w_1 -V_2v^2 - W^2 w_2)$, $J_+=- (V_2v^1 + W^2 w_1)$, $J_- = -(V_1v^2 + W^1w_2)$, with nontrivial
OPE's
\begin{eqnarray}
 J_3(z)J_3(z') &  \sim & \mathrm{reg.} \nonumber \\
J_3(z)J_{\pm}(z') & \sim & {\pm {J_{\pm}(z') \over {z-z'}}} \\
J_+(z)J_-(z') & \sim & {2J_3(z')\over z-z'}. \nonumber  
\end{eqnarray}
Notice that this is just an $SL(2)$ current algebra at level 0. The centre of $GL(2)$ (at the Lie algebra level) is given by $GL(1)$. The corresponding current algebra is generated by $K=-{1\over 2}\left(V_1v^1+ W^1w_1 + V_2v^2 + W^2w_2 \right)$, with OPE given by
\be
K(z)K(z')  \sim \mathrm{reg.}
\ee   
This is just a $GL(1)$ current algebra at level 0. 
 
\bigskip\noindent{\it The Modulus of Mirror CDR}

Let us now generalise the above construction of the sheaf of mirror CDR on $\bf{S}^3 \times \bf{S}^1$. In order to do so, we must invoke a modulus that will enable us to obtain a family of sheaves of mirror CDR. Recall that the modulus is represented by the Cech cohomology group $H^1({\bf{S}^3 \times \bf{S}^1},\Omega^{2,cl})\cong \Bbb{C}$.\footnote{Since we are computing the short distance operator product expansion of fields in the present context, it suffices to work locally on $\Sigma$. Hence, the modulus will be represented by $H^1(X ,\Omega^{2,cl}_X)$ (where $X = {\bf{S}^3 \times \bf{S}^1}$) instead of $H^1(X \times \Sigma ,\Omega^{2,cl}_{X \times \Sigma})$ as stated at the end of section 5.6.} To model the modulus, one simply needs to introduce a variable complex parameter associated with $H^1({\bf{S}^3 \times \bf{S}^1}, \Omega^{2,cl})$. 

Before we proceed any further, it will first be necessary for us to know how the relevant fields will transform under a variation of the modulus. Recall from our discussion on local symmetries in section 5.5 that $\Omega^{2,cl}$, the sheaf of closed, holomorphic $(2,0)$-forms on a manifold $X$, is associated with a non-geometrical symmetry of the free $\beta\gamma$-$bc$ system on $X$. Consider a general system of $n$ conjugate $\beta\gamma$ and $bc$ systems, with nontrivial OPE's $\beta_i(z)\gamma^j(z')\sim -\delta_i^j/(z-z')$ and $b^i(z) c_j(z') \sim \delta^i_j/(z-z')$ respectively. Let $F={1\over 2}f_{ij}(\gamma)d\gamma^i\wedge d\gamma^j$ be a closed holomorphic two-form.  Under the symmetry associated with $F$, the fields
transform as 
\begin{eqnarray}
\gamma^j & \to & \gamma^j \nonumber \\
\beta_i& \to &\beta_i'= \beta_i+f_{ij}\partial\gamma^j \\
c_j & \to & c_j \nonumber \\
b^i & \to & b^i. \nonumber
\end{eqnarray}       
In the spirit of section 5.5, one can verify the above transformations by locally
constructing a holomorphic one-form $A=A_i(\gamma) d \gamma^i$, with
$dA=F$ so that $F$ is closed, and then computing the relevant OPE's which determine how the fields transform under the action of
the conserved charge $\oint A_i \partial_z\gamma^i dz$. 

To apply the above discussion to the present case where $X = \bf{S}^3 \times \bf{S}^1$,  let us first make a cover of $\bf{S}^3 \times \bf{S}^1$ by two open sets $U_1$ and $U_2$, where $U_1$ is characterized by the condition $v^1\not=0$, and $U_2$ by $v^2\not= 0$.  Note that this is not a ``good cover,'' as $U_1$ and $U_2$ are topologically complicated (each being isomorphic to $\Bbb{C}\times E$, where $E$ is an elliptic curve).  As such, one cannot, in general, be guaranteed that  an arbitrary cohomology class can be represented by a Cech cocycle with respect to this cover. However, in the present context, we have on $U_1\cap U_2$, a holomorphic section of $\Omega^{2,cl}$ given by
\be
F={dv^1\wedge dv^2\over v^1 v^2}. 
\ee
Since $F$ cannot be ``split'' as a difference of a form holomorphic in $U_1$ and one holomorphic in $U_2$, it thus represents an element of $H^1({\bf{S}^3 \times \bf{S}^1},\Omega^{2,cl})$. From the correspondence between the $Vv$-$Ww$ and $\beta\gamma$-$bc$ systems, the relevant field transformations are thus given by
\begin{eqnarray}
\label{78} 
v^1 & \to & v^1 \nonumber \\
v^2 & \to & v^2 \nonumber \\
V_1& \to &V_1'= V_1+ t{\partial v^2\over v^1v^2} \nonumber \\
V_2& \to &V_2'= V_2 - t{\partial v^1\over v^1v^2}  \\
b^1 & \to & b^1 \nonumber \\
b^2 & \to & b^2 \nonumber \\
c_1 & \to & c_1 \nonumber \\
c_2 & \to & c_2. \nonumber
\end{eqnarray}
Here $t$ is a complex parameter. We will see shortly that it is related to the level $k$ of the underlying $SU(2)$ WZW model. Hence, we obtain a family of sheaves of mirror CDR, parameterized by $t$, by declaring that  the fields undergo this transformation
from $U_1$ to $U_2$ when we glue the sheaves together.

Let us determine how some important operators behave under this deformation. Notice that the stress tensor $T=V_1\partial v^1+V_2\partial v^2 +W^1\partial w_1+ W^2\partial w_2$ is invariant. Hence, the deformed theory, for any value of $t$, has a stress tensor of $c=0$. This is in accordance with the fact that the $(2,2)$ model on $\bf{S}^3 \times \bf{S}^1$ is conformally invariant for all $k$, and that the difference between its left and right central charges is always 0. 

Let us now consider the $GL(1)$ current, which at $t=0$ (i.e. without considering the modulus) was defined to be $K=-{1\over 2}\left(V_1v^1+ W^1w_1 + V_2v^2 + W^2w_2 \right)$.   Under (\ref{78}),  we have
\be
K \to K-{t\over 2}\left({\partial v^2\over
v^2}-{\partial v^1\over v^1}\right) = {\widetilde K}.
\label{K to K'}   
\ee  
Note that the shift in $K$ under this transformation to $\widetilde K$ (in going from $U_1$ to $U_2$) is {\it not} an anomaly that spoils the existence of $K$ at $t \not=0$. The reason is  because in contrast to the situation encountered with the dimension 1 operator ${\cal J}(z)$ in the $\mathbb {CP}^1$ example, ${\widetilde K} - K$ can be expressed as a difference between a term (namely $t\,\partial v^1/2v^1$) that is holomorphic in $U_1$ and a term (namely $t\,\partial v^2/2v^2$)
that is holomorphic in $U_2$. 

Since we want to study how the current algebra will depend on $t$, it will be necessary for us to re-express the above globally-defined $GL(1)$ current generator $K$ in such a way that its explicit $t$ dependence is made manifest in $\it{both}$ $U_1$ and $U_2$. In order to be consistent with (\ref{K to K'}), we just need to ensure that the difference in the new  expressions of $K$ in $U_2$ and $U_1$ is given by 
\be
-{t\over 2}\left({\partial v^2\over v^2}-{\partial v^1\over v^1}\right).
\ee       
In addition, these expressions in $U_1$ and $U_2$ must be invariant under the geometrical symmetry $v^i \to \lambda v^i$ if $K$ is to be an allowable operator. Noting these requirements, we arrive at the following; in $U_1$, the current is represented by 
\be
{K^{[1]}}= {-{1\over 2}\left (V_1v^1+ W^1w_1 + V_2v^2 + W^2w_2 \right) - {t \over 2}{\partial v^1\over v^1}},
\ee  
while in $U_2$, it is represented by
\be
{K^{[2]}}= {-{1\over 2}\left (V_1v^1+ W^1w_1 + V_2v^2 + W^2w_2 \right) - {t \over 2}{\partial v^2\over v^2}}.
\ee  
Recalling that the original expression of $K$ given by $-{1\over 2}\left(V_1v^1+ W^1w_1 + V_2v^2 + W^2w_2 \right)$ is a global section of the sheaf of mirror CDR and is thus holomorphic in both $U_1$ and $U_2$, we see that ${K^{[1]}}$ and ${K^{[2]}}$ are holomorphic in $U_1$ and $U_2$ respectively. Moreover, as required, ${K^{[1]}}$ also transforms under (\ref{78} ) into $K^{[2]}$. Hence, for any value of $t$, the sheaf of mirror CDR has a global section $K$ that is represented in $U_1$ by $K^{[1]}$ and in $U_2$ by $K^{[2]}$.

Now we can compute the OPE singularity of $K$ for any $t$:
\be
{K(z)K(z')\sim - {t\over 2} {1\over (z-z')^2}.}
\ee 
To arrive at this result, we can either work in $U_1$, setting $K=K^{[1]}$ and
computing the OPE, or we can work in $U_2$, setting $K=K^{[2]}$ and
computing the OPE. The answer will come out the same in either case,
because the transformation (\ref{78}) is an automorphism of the CFT. Thus, the level of the $GL(1)$ current algebra is $-t$.

Likewise, we can work out the transformation of the $SL(2)$
currents under (\ref{78}).  The currents as defined at $t=0$, namely
$J_3=-{1 \over 2}(V_1v^1 + W^1w_1 -V_2v^2 - W^2 w_2)$, $J_+=-(V_2v^1 + W^2 w_1)$, $J_-= -(V_1v^2 + W^1w_2)$,
transform as
\begin{eqnarray}
J_3 & \to & J_3-{t\over 2}\left({\partial v^1\over v^1}+{\partial v^2\over v^2}\right)  \nonumber\\
J_+& \to & J_+ +t{\partial v^1\over v^2} \\
J_-&\to & J_- -t{\partial v^2\over v^1}. \nonumber
\end{eqnarray}
Similarly, the shifts in each current can be ``split'' as a difference of terms holomorphic in $U_1$ and $U_2$.  So the currents can be re-expressed to inherit $t$-dependent terms such that they can be defined at $t\not=0$. The new expressions of these currents which satisfy all the necessary requirements are given, in $U_1$ and $U_2$ respectively, by
\begin{eqnarray}
{J^{[1]}_3} &= & {-{1 \over 2}\left(V_1v^1 + W^1w_1 -V_2v^2 - W^2 w_2 \right)+{t \partial v^1}/ {2v^1}}\\
{J^{[2]}_3}&= & {-{1 \over 2}\left(V_1v^1 + W^1w_1 -V_2v^2 - W^2 w_2 \right) -{t \partial v^2}/ {2v^1}} 
\end{eqnarray}
together with
\begin{eqnarray}
{J^{[1]}_+}& = & -(V_2v^1 + W^2 w_1) \\
{J^{[2]}_+}& = & -(V_2v^1 + W^2 w_1) + {t \partial v^1}/ {v^2} 
\end{eqnarray} 
and
\begin{eqnarray}
{J^{[1]}_-}& = & -(V_1v^2 + W^1w_2) + {t \partial v^2}/ {v^1} \\
{J^{[2]}_-}& = & -(V_1v^2 + W^1w_2). 
\end{eqnarray} 
As required, $J^{[1]}_3$, $J^{[1]}_+$ and $J^{[1]}_-$ transform into $J^{[2]}_3$, $J^{[2]}_+$ and $J^{[2]}_-$ respectively under (\ref{78}). Hence, for any value of $t$, the sheaf of CDR also has global sections $J_3$, $J_+$ and $J_-$ that are represented in $U_1$ by $J^{[1]}_3$, $J^{[1]}_+$ and $J^{[1]}_-$, and in $U_2$ by $J^{[2]}_3$, $J^{[2]}_+$ and $J^{[2]}_-$.
    
We shall now compute the OPE's of these current operators, working in either $U_1$ or $U_2$, whichever proves to be more convenient. We obtain an $SL(2)$ current algebra at level $t$:
\begin{eqnarray}
J_3(z)J_3(z') &  \sim & {t \over 2}{1 \over {(z- z')^2}} \nonumber \\
J_3(z)J_{\pm}(z') & \sim & {\pm {J_{\pm}(z') \over {z-z'}}} \\
J_+(z)J_-(z') & \sim & {t \over 2}{1 \over {(z- z')^2}}+ {2J_3(z')\over z-z'}. \nonumber  
\end{eqnarray}

The $SL(2)$ and $GL(1)$ current algebras thus have levels $t$ and $-t$, in agreement with what we had anticipated from the half-twisted B-model if the level $k$ of the underlying $SU(2)$ WZW theory is related to the CDR parameter $t$ by $k=t-2$; indeed at $t=0$, the level of the $SL(2)$ and $GL(1)$ current algebras are the same at 0, where $k=-2$.  

Note  that the $Q_R$-cohomology of $\bf{S}^3 \times \bf{S}^1$ does not receive instanton corrections. For any
target space $X$, such corrections (because they are local on the Riemann surface $\Sigma$, albeit global in $X$) come only from
holomorphic curves in $X$ of genus zero. There is no such curve in $\bf{S}^3 \times \bf{S}^1$. Therefore, the above analysis of the $Q_R$-cohomology of $\bf{S}^3 \times \bf{S}^1$ is exact in the full theory.

\newsection{Relation to the Mirror Symmetry of Twisted Generalised Complex Manifolds}

\vspace{-0.5cm}

In this section, we will provide an $\it{a}$ $\it{priori}$ reason for the mysterious relation between the level $k$ and the complex parameter $t$ of the half-twisted B-model on $\bf{S}^3 \times \bf{S}^1$, i.e., $k=t-2$ - we will show that the relation is consistent with and therefore due to the mathematical results in \cite{ben} on the mirror symmetry of generalised complex manifolds. Since the origin of the relation rests upon the principle of mirror symmetry, we will be able to provide as well, an $\it{a}$ $\it{priori}$ reason for a similar relation revealed in the context of the half-twisted A-model on the same group manifold involving the sheaf of CDR.         

\newsubsection{Mirror Symmetry of the 2-torus and Strong-Weak Duality}    

Let us start by discussing the simplest form of mirror symmetry - that of a one-dimensional Calabi-Yau manifold given by a 2-torus. A 2-torus $\bf{T}^2 = S^1 \times S^1$ has both complex and K\"ahler moduli - if we let the radii of the two $\bf{S}^1$ circles be $R_1$ and $R_2$ respectively, the complex modulus will be given by $R_1/R_2$, and the K\"ahler modulus will be given by $R_1R_2$. By definition, mirror symmetry exchanges the complex and K\"ahler moduli of a Calabi-Yau manifold. In the case of a 2-torus, this is effected by $R_2 \leftrightarrow {1/R_2}$. Notice that this is simply a T-duality operation on the second $\bf{S}^1$ circle in $\bf{T}^2$. 

Note that for a sigma model on $\bf{T}^2$, one usually works with the complexified K\"ahler moduli and complex structure - the complex structure can be written as $\sigma = \sigma_1 + i \sigma_2$, where $\sigma_2 = {R_1 /R_2}$, and the K\"ahler modulus can be written as $\rho = \rho_1 + i \rho_2$, where $\rho_1 = (1/ {2\pi} )\int_{\Sigma} \phi^* (B)$ and $\rho_2 = R_1 R_2$. Mirror symmetry then exchanges $\sigma \leftrightarrow \rho$. 

The observation that mirror symmetry is T-duality is consistent with the fact that the (half-twisted) A-model on $\bf{T}^2$ is equivalent to the (half-twisted) B-model on ${\widetilde {\bf{T}}}^{\bf 2}$, where $\bf{T}^2$ and ${\widetilde {\bf{T}}}^{\bf 2}$ are mirror partners \footnote{Note that the mirror symmetric relation between the topological A- and B-models on mirror Calabi-Yau spaces can be applied to their half-twisted variants because the spectrum of local operators in either  variants is given by the cohomology with respect to the $\it{same}$ ${\overline Q}_+$ operator.} - recall that one can map between the A- and B-models by flipping the sign of the left-moving ghost number current $J(z)$, and this can be effected by a T-duality on one of the $\bf{S}^1$'s of the 2-torus target space.        

Last but not least, recall that the strength of the sigma model coupling is proportional to the inverse of the metric of the target space. Hence, since $R_2 \leftrightarrow {1/ R_2}$ under T-duality, we equivalently have a strong-weak duality of the worldsheet theory under mirror symmetry.

\newsubsection{An Equivalence of Theories as a Form of Twisted Generalised Mirror Symmetry}

Recall from the description of $\bf{S}^3 \times \bf{S}^1$ as a non-K\"ahler complex manifold with torsion in section 3.4, that it can be constructed as a non-trivial fibration of $\mathbb{CP}^1$ with fibres $E = \bf{T}^2$, where $\bf{T}^2$ is a genus one complex Riemann surface. The complex structure of $\bf{S}^3 \times \bf{S}^1$ will then given by the non-zero complex number $\lambda$ that has been made manifest in its other description as ${\mathbb {C}}^2 / {\mathbb Z}$. 

Now, consider the B-model on a 2-torus $\bf{T}^2$ at weak coupling. This is equivalent to the A-model on a mirror 2-torus ${\widetilde {\bf{T}}}^{\bf 2}$ at strong coupling. Let us fibre the $\bf{T}^2$ on the B-model side adiabatically\footnote{The condition of adiabaticity is not absolutely necessary. In fact, as reviewed in \cite{vafa}, there can even be singular fibres over various points on the base space, and the resulting physics will still be consistent.} over a $\mathbb {CP}^1$ base. Likewise, let us fibre the ${\widetilde {\bf{T}}}^{\bf 2}$ on the A-model side adiabatically over the $\it{same}$ $\mathbb {CP}^1$ base. Then, via the principle of fibrewise duality \cite{vafa}, and the above description of $\bf{S}^3 \times \bf{S}^1$ as a 2-torus fibration of $\mathbb {CP}^1$, we find that the B-model on $X = \bf{S}^3 \times \bf{S}^1$ at weak coupling, is $\it{equivalent}$ to the A-model on ${\widetilde X} = {\widetilde {\bf{S}}^{\bf 3}} \times {\widetilde{\bf{S}}^{\bf 1}}$ at strong coupling, where $X$ has fibres $E = \bf{T}^2$, while $\widetilde X$ has fibres ${\widetilde E} = {\widetilde {\bf{T}}}^{\bf 2}$. Alternatively, we find that the A-model on $X = \bf{S}^3 \times \bf{S}^1$ at weak coupling, is $\it{equivalent}$ to the B-model on ${\widetilde X} = {\widetilde {\bf{S}}^{\bf 3}} \times {\widetilde{\bf{S}}^{\bf 1}}$ at strong coupling. It is in this sense that $X$ and $\widetilde X$ define a twisted (i.e. non-zero torsion) generalised mirror pair. 

The above observation that $X$ and $\widetilde X$ should be mirror pairs, is mathematically consistent with the recent results by Ben-Bassat in \cite{ben} which furnishes an extension of the Strominger-Yau-Zaslow concept of T-duality as mirror symmetry for Calabi-Yau manifolds, to generalised complex manifolds with possibly non-zero torsion. This can be seen as follows. Firstly, in \cite{ben}, it is shown that for a real $n$-torus bundle with sections over an $n$-dimensional base such that one can define a flat connection over the total space, the mirror geometry will be given by a dual $n$-torus bundle with the same base but with dual torus fibres, where the dual torus fibre can be obtained from the orginal torus fibre via T-duality in an odd number of directions. Secondly, notice that $X$ and $\widetilde X$ are twisted generalised complex manifolds which can be viewed as real $2$-torus bundles with sections over the same $n$-dimensional base, where one can define a flat connection over the total space because $X$ and $\widetilde X$ are parallelised. Consequently, since the 2-torus fibres $E$ and $\widetilde E$ of $X$ and $\widetilde X$ are related by T-duality along one of the two circle directions, the twisted generalised mirror geometry of $X$ is indeed mathematically given by $\widetilde X$.

\newsubsection{The Relation $k=t -2$ and Twisted Generalised Mirror Symmetry}

In \cite{ben}, it is shown that from the flat connections on a pair of mirror $n$-torus bundles, one can define (semi-flat, generalised) complex structures on them. In addition, it is also shown that there is an explicit bijective correspondence between these complex structures. We shall now show that the relation $k=t-2$ obtained for the half-twisted B-model on $X$ involving the sheaves of mirror CDR, and the simlar relation ${\tilde k} = {\tilde t} -2$ obtained in \cite{MC} for the half-twisted A-model on $\widetilde X$ involving the sheaves of CDR, (where $\tilde k$ is the level of the underlying $SU(2)$ WZW model, and $\tilde t$ is a complex parameter associated with the modulus of CDR), is consistent with this bijective correspondence. 

Firstly, recall from section 3.3 and 3.4 that we have the three-form flux ${\cal H} = 2 i \partial \omega_T$ which generates a class in $H^1({\bf{S}^3 \times \bf{S}^1}, \Omega^{2,cl}) \cong \mathbb C$ that is associated with the moduli of mirror CDR, where the hermitian $(1,1)$-form $\omega_T$ can be explicitly written as ${\omega_T}= dt\wedge \zeta + \pi^*(\omega_0)$. Likewise, for the A-model on $\widetilde X$, we have the three-form flux ${\widetilde {\cal H}} = 2 i \partial {\widetilde \omega}_T$ which generates a class in $H^1({{\widetilde {\bf{S}}}^3 \times {\widetilde {\bf{S}}}^1}, \Omega^{2,cl}) \cong \mathbb C$ that is associated with the moduli of CDR, where the hermitian $(1,1)$-form ${\widetilde \omega}_T$ can be explicitly written as ${{\widetilde {\omega}}_T} = {\widetilde dt} \wedge {\widetilde \zeta} + \pi^*({\widetilde \omega}_0)$. 

As pointed out in section 3.4, $\omega_0$ is an $SO(3)$-invariant form on the $\bf{S}^2$ base space of $X$. Since $X$ and $\widetilde X$ are 2-torus fibrations of the $\it{same}$ $\bf{S}^2$ base space, it will mean that ${{\omega}}_T$ and ${\widetilde {\omega}}_T$ are distinct from each other only because of the terms $dt\wedge \zeta $ and ${\widetilde dt} \wedge {\widetilde \zeta}$, where $\zeta$ and $\widetilde \zeta$ are one-forms which can be kept fixed as the complex structures on $X$ and $\widetilde X$ are varied. 

As explained in section 3.4, the choices of the one-forms $dt$ and $\widetilde {dt}$ will determine the choices of the complex structures on $X$ and $\widetilde X$ and vice-versa. This means that the choices of the complex structures on $X$ and $\widetilde X$ will determine the choices of ${{\omega}}_T$ and ${\widetilde {\omega}}_T$, which in turn will determine $\cal H$ and $\widetilde {\cal H}$ respectively. Since $\cal H$ and $\widetilde {\cal H}$ generate the one-dimensional classes in $H^1({{{\bf{S}}}^3 \times {{\bf{S}}}^1}, \Omega^{2,cl})$ and $H^1({{\widetilde {\bf{S}}}^3 \times {\widetilde {\bf{S}}}^1}, \Omega^{2,cl})$ which are correlated with the choices of $t$ and $\tilde t$ (as shown in section 7.2 and \cite{MC}), it will mean from the relations $k = t-2$ and ${\tilde k} = {\tilde t} -2$, that a choice of the complex structure on $X$ or $\widetilde X$ will determine a choice of $k$ or $\tilde k$. 

Recall from section 3.4 that the value of $k$ or $\tilde k$ will determine the K\"ahler moduli of $E = \bf{T}^2$ or ${\widetilde E} = {\bf{\widetilde T}}^2$. From our earlier discussion in section 8.1 on the mirror symmetry of 2-torus's, we showed that the K\"ahler moduli of $E = \bf{T}^2$ will be given by the complex structure moduli of ${\widetilde E}= {\bf{\widetilde T}}^2$ and vice-versa. We also saw in section 3.4, from the two different constructions of $X$ and therefore $\widetilde X$, that the complex structure of $X$ or $\widetilde X$ is determined by the complex structure of $E$ or $\widetilde E$ respectively. Collectively, this means that a choice of the complex structure on $X$, will determine a choice of the complex structure on $\widetilde X$, and vice-versa. In other words, we have a bijective correspondence between the complex structure on $X = \bf{S}^3 \times \bf{S}^1$ and the complex structure on its twisted generalised mirror ${\widetilde X} = {\widetilde {\bf{S}}^{\bf 3}} \times {\widetilde{\bf{S}}^{\bf 1}}$,  in agreement with the mathematical results of \cite{MC}.

Hence, the existence of a bijective correspondence of the complex structures on mirror pairs of $2$-torus bundles, serves as an $\it{a}$ $\it{priori}$ reason for the once mysterious relations $k = t-2$ and ${\tilde k} = {\tilde t} -2$ uncovered in the context of the A and B-models on a mirror pair of $\bf{S}^3 \times \bf{S}^1$'s, whence the corresponding chiral algebras can be described by sheaves of CDR and mirror CDR respectively. Moreover, as explained at the end of section 8.1, since there is a strong-weak duality between the A-model on $X$ and the B-model on $\widetilde X$, and since there are no worldsheet instanton corrections to the analysis that lead up to these relations as explained in section 7.2,  we find that they will hold beyond the perturbative regime in the full theory.

\vspace{0.5cm}  
\hspace{-1.0cm}{\large \bf Acknowledgements:}\\
I would like to take this opportunity to thank  A. Adams, E. Sharpe and in particular F. Malikov, for helpful email correspondences. 
\newline

\end{document}